\documentclass[a4paper,12pt]{report}
\usepackage{epsf,epsfig,amsmath,amsfonts,a4wide,exscale}
\usepackage{graphicx}
\usepackage{latexsym,fancyhdr}
\renewcommand{\chaptermark}[1]{\markboth{\chaptername\ \thechapter. \ #1}{}}
\renewcommand{\sectionmark}[1]{\markright{\thesection\ #1}}
\fancypagestyle{plain}{\fancyhf{}\renewcommand{\headrulewidth}{0pt}\renewcommand{\footrulewidth}{0pt}}

\newcommand{\clearemptydoublepage}{\newpage{\pagestyle{empty}\cleardoublepage}}
\newtheorem{lem}{Lemma}[section]

\newtheorem{rem}{Remark}[section]

\newtheorem{prop}{Proposition}[section]

\newcommand{\half}{\mbox{$\textstyle \frac{1}{2}$}}
\newcommand{\rd}{\mbox{$\rm d$}}
\newcommand{\re}{\mbox{$\rm e$}}

\newcommand{\Q}{\mathbb{Q}}

\newcommand{\lb}{\left}
\newcommand{\rb}{\right}
\newcommand{\E}{\mathbb{E}}
\newcommand{\F}{\mathcal{F}}
\newcommand{\im}{\textrm{i}}
\newcommand{\e}{\textrm{e}}
\newcommand{\Var}{\textrm{Var}}
\newcommand{\nn}{\nonumber}

\newcommand{\R}{\mathbb{R}}
\newcommand{\N}{\mathbb{N}}

\begin{document}
\thispagestyle{empty}%
\null\vskip0.2in%
\begin{center}
\LARGE{{\bf An Information-Based Framework \\ for Asset Pricing:\\
$X$-Factor Theory and its Applications}}
\end{center}
\vfill
\begin{center}
{\Large {\bf by}}\\
\mbox{} \\
{\Large {\bf Andrea Macrina}}
\end{center}
\vfill
\begin{center}
\large{\bf{Department of Mathematics \\
King's College London \\
The Strand, London WC2R 2LS, United Kingdom}}
\end{center}
\vfill
\begin{center}
\large{\bf{Submitted to the University of London \\
for the degree of \\
Doctor of Philosophy}}
\end{center}
\vfill
\begin{center}
{\large\bf{24 October 2006}}
\end{center}

\newpage
\mbox{}\newline \vspace{20mm} \mbox{}\newline \LARGE
{\bf Abstract} \normalsize \vspace{5mm}

\noindent{\bf{An Information-Based Framework for Asset Pricing:
$X$-Factor Theory and its Applications}}. This thesis presents a
new framework for asset pricing based on modelling the information
available to market participants. Each asset is characterised by
the cash flows it generates. Each cash flow is expressed as a
function of one or more independent random variables called market
factors or ``$X$-factors". Each $X$-factor is associated with a
``market information process", the values of which become
available to market participants. In addition to true information
about the $X$-factor, the information process contains an
independent ``noise" term modelled here by a Brownian bridge. The
information process thus gives partial information about the
$X$-factor, and the value of the market factor is only revealed at
the termination of the process. The market filtration is assumed
to be generated by the information processes associated with the
$X$-factors. The price of an asset is given by the risk-neutral
expectation of the sum of the discounted cash flows, conditional
on the information available from the filtration. The thesis
develops the theory in some detail, with a variety of applications
to credit risk management, share prices, interest rates, and
inflation. A number of new exactly solvable models are obtained
for the price processes of various types of assets and derivative
securities; and a novel mechanism is proposed to account for the
dynamics of stochastic volatility and correlation. If the cash
flows associated with two or more assets share one or more
$X$-factors in common, then the associated price processes are
dynamically correlated in the sense that they share one or more
Brownian drivers in common. A discrete-time version of the
information-based framework is also developed, and is used to
construct a new class of models for the real and nominal interest
rate term structures, and the dynamics of the associated price
index.

\newpage

\mbox{}\newline \vspace{20mm} \mbox{}\newline \LARGE
{\bf Acknowledgements} \normalsize \vspace{5mm}

\noindent I am very grateful to Lane P. Hughston, my supervisor,
for his support and encouragement, and for introducing me to the
great intellectual world of mathematical finance. I would also
like to express my gratitude to other members of the financial
mathematics group (both present and former) at King's College,
including I. Buckley, G. Iori, A. L\"okka, M. Pistorius, W. Shaw,
and M. Zervos; and also A. L. Bronstein, J. Dear, A. Jack, M.
Jeannin, Z. Jiang, T. Johnson, A. Mehri, O. Precup, and A.
Rafailidis; as well as G. Di Graziano, A. Elizalde, S. Galiani, G.
Ha\l aj, M. Hinnerich, and R. Suzuki. I have benefitted from
interactions with many members of the London mathematical finance
community at Birkbeck College, Imperial College, and the London
School of Economics, both students and faculty members---too many
to name in full; but I would like to thank in particular my
co-author D. Brody, and also P. Barrieu, D. Becherer, M. Davis,
and L. Foldes. I have been lucky to have come to know many other
members of the international mathematical finance community, and I
mention T. Bielecki, M. Bruche, H. B\"uhlmann, P. Carr, R. Cont,
F. Delbaen, P. Embrechts, W. Farkas, B. Flesaker, H. F\"ollmer, M.
Grasselli, V. Henderson, A. Hirsa, D. Hobson, H. Hulley, T. Hurd,
P. Imkeller, F. Jamshidian, R. Jarrow, M. Jeanblanc, I. Karatzas,
H. Korezlioglu, D. Madan, A. McNeil, F. Mercurio, M. Monoyios, D.
Ocone, T. Ohmoto, M. Owen, R. Repullo, L. C. G. Rogers, W.
Runggaldier, M. Rutkowski, S. Shreve, M. Schweizer, P.
Sch\"onbucher, P. Spreij, L. Stettner, D. Taylor, S. Turnbull, A.
Wiese, and M. W\"uthrich, where helpful remarks and critical
observations have influenced this work, and also especially T.
Bj\"ork and H. Geman. I would as well like to thank D. Berger, R.
Curry, P. Friedlos, A. Gfeller, and I. Kouletsis. I acknowledge
financial support from EPSRC, Erziehungsdirektion des Kantons
Bern, Switzerland, and from the UK ORS scheme. Finally, as ever, I
express my thanks to my father, my mother, and my sister for their
unflagging support. Mamma, Pap\`a, Alessia, questi studi sono
stati un'impresa di famiglia. L'unione \`e sempre stata la nostra
forza. Grazie.

\newpage
\begin{center}
\phantom{} \vspace{8cm} \noindent The work presented in this
thesis is my own.\\ \vspace{4cm} Andrea Macrina
\end{center}

\newpage
\tableofcontents
\newpage
\pagestyle{fancy} \fancyhf{}
\fancyhead[RO]{\sffamily\small\thepage}
\fancyhead[LO]{\sffamily\small Introduction and summary}
\fancyhead[LE]{\sffamily\small }
\fancyhead[RE]{\sffamily\small\leftmark}
\fancypagestyle{plain}{\fancyhf{}\renewcommand{\headrulewidth}{0pt}\renewcommand{\footrulewidth}{0pt}}

\chapter{Introduction and summary}

When confronted with the task of developing a model for asset
pricing one soon faces questions of the following type: ``Which
asset classes are going to be considered?", ``How does one
distinguish between the various asset classes?", ``What types of
risk are involved?", ``In what ways do the various assets depend
upon one another?", ``In what ways do the various assets depend
upon the state of the economy?", and so on. All these questions
have to be kept in mind as one considers what features should be
incorporated into the mathematical design of an asset pricing
model. One general characteristic that we would like to include in
such models is flexibility, to ensure that a suitable variety of
market features can be captured. We also want tractability and
efficiency, to enable us to give precise answers even when complex
asset structures are being analysed. We need to give special
consideration to the relation between the level of sophistication
of the modelling framework and the fundamental requirements of
transparency and simplicity. The modelling framework has to offer
a degree of sophistication sufficiently high to ensure realistic
arbitrage-free prices and risk-management policies, even for
complex structures. Transparency and simplicity, on the other
hand, are needed to guarantee the consistency and integrity of the
models being developed, and that the results obtained are based on
sound mathematical foundations. To achieve a significant element
of success in satisfying these often conflicting requirements is a
serious challenge for any modelling framework.

Our view in what follows will be that asset pricing models should
be constructed in such a way that attention is focussed on the
cash flows generated by the assets under consideration, and on the
economic variables that determine these cash flows.

In particular, the approach pursued in this thesis is based on the
analysis of the information regarding market factors available to
market participants. We place special emphasis on the construction
of the market filtration. This point of view can be contrasted
with what is perhaps the more common modelling approach in
mathematical finance, where the market filtration is simply
``given". For example, in many studies the market consists of a
number of assets for which the associated price processes are
driven, collectively, by a multi-dimensional Brownian motion. The
underlying filtered probability space is then merely assumed to
have a sufficiently rich structure to support this
multi-dimensional Brownian motion. Alternatively, the filtration
is often assumed to be that generated by the multi-dimensional
Brownian motion---as described, for example, in Karatzas \& Shreve
1998. However no deeper ``economic foundation" for the
construction of the filtration is offered. More generally, the
system of asset price processes is sometimes assumed to be a
collection of semi-martingales with various specified properties;
but again, typically, little is said about the relevant filtration
apart from the general requirement that the various asset price
processes should be adapted to it. In the present approach, on the
other hand, we model the filtration explicitly in terms of the
information available to the market.

The contents of the thesis are adapted in part from the following
research papers: (i) Brody, Hughston \& Macrina (2007), henceforth
BHM1; (ii) Brody, Hughston \& Macrina (2006), henceforth BHM2; and
(iii) Hughston \& Macrina (2006), henceforth HM. In particular,
Chapters 2, 3, 4, 5, and 6, in which the details of the
information-based framework are developed and applied to a number
of examples involving credit and equity related products, contain
research associated with BHM1 and BHM2. In addition, a number of
further results are presented in these chapters, including the
material concerning the information-based Arrow-Debreu technique
appearing in Sections \ref{sec:3.5}, \ref{sec6.8}, and
\ref{sec6.9}, the material on the Black-Scholes model in Section
\ref{sec6.3a}, the material on correlated cash flows in Section
\ref{sec6.5}, and the material on the reduction theory for
dependent cash flows appearing in Sections \ref{sec6.6} and
\ref{sec6.7}. The results presented in Chapter 7, concerning
interest rates and inflation, are adapted from HM.

In the greater part of the thesis we concentrate on the
continuous-time formulation of the information-based framework and
the development of the associated $X$-factor theory. We work, in
general, in an incomplete-market setting with no arbitrage, and we
assume the existence of a fixed pricing kernel (or, equivalently,
the existence of a fixed pricing measure $\Q$). Explained in a
nutshell, the information-based approach can be summarised as
follows: First we identify the random cash flows occurring at the
pre-specified dates pertinent to the particular asset or group of
assets under consideration. Then we analyse the structure of the
cash flows in more detail by introducing an appropriate set of
$X$-factors, which are assumed to be independent of one another in
an appropriate choice of measure, typically the preferred pricing
measure, e.g., the risk-neutral measure. To each $X$-factor we
associate an information process that consists of two terms: a
signal component, and a noise component. The signal term embodies
``genuine" information about the possible outcomes of the
$X$-factor; while the ``noise" component is modelled here by a
Brownian bridge process, and plays the role of market speculation,
inuendo, gossip, and so on. The signal term and the Brownian
bridge term are assumed to be independent; in particular, the
Brownian bridge term carries no useful information about the value
of the relevant market factor. We assume that the information
processes collectively generate the market filtration. The price
of an asset is calculated by use of the standard risk-neutral
valuation formula; that is to say, the asset price is given by the
sum of discounted expected future cash flows, conditional on the
information supplied by the market filtration.

Chapter 2 begins with a simple model for credit-risky zero-coupon
bonds, where we have a single cash flow at the bond maturity. The
cash flow is modelled first by a binary random variable. Then in
Section \ref{sec2.3} we derive the price process of a defaultable
discount bond in the case for which the cash flow is modelled by a
random variable with a more general discrete spectrum. This allows
for a random recovery in the case of default. Here default is
defined in general as a failure to fully honour a required
payment, and hence as a cash flow of less than the contracted
value. In this section we also explore the properties of the
particular chosen form for the information process, showing that
it satisfies the Markov property. This in turn facilitates the
calculation of the price process of a defaultable bond, because
the expectation involved need only be conditional on the current
value of the information process. We are able to obtain a
closed-form expression for the price process of the bond. We then
proceed to analyse the dynamics of the price process of the bond
and find that the driver is given by a Brownian motion that is
adapted to the filtration generated by the market information
process. This construction indicates the sense in which the price
process of an asset can be viewed as an ``emergent'' phenomenon in
the present framework.

Since the resulting bond price at each moment of time is given by
a function of the value of the information process at that time,
we are able to show in Section \ref{sec2.7} that simulations for
the bond price trajectories are straightforward to generate by
simulating paths of the information process. The resulting plots
describe various scenarios, ranging from unexpected default events
to announced declines. In particular, the figures provide a means
to understand better the effects on asset prices due to large or
small values of the information flow rate parameter. This
parameter measures the rate at which ``genuine" information is
leaked into the market. We show in Section \ref{sec2.6} that the
model presented is in a certain sense invariant in its overall
form when it is updated or re-calibrated. We call this property
``dynamic consistency". In deriving these results we make use of
some special orthogonality properties of the Brownian bridge
process (Yor 1992, 1996), which also come into play in our
analysis of the dynamics of option prices.

In Chapter 3 we compute the prices of options on credit-risky
discount bonds. In particular, in the case of a defaultable binary
discount bond we obtain the price of a call option in analytical
form. The striking property of the resulting expression for the
option price is that it is very similar to the well-known
Black-Scholes price for stock options. In particular, we see that
the information flow rate parameter appearing in the definition of
the information process plays a role that is in many respects
analogous to that of the Black-Scholes volatility parameter.

This correspondence suggests that we should undertake a
sensitivity analysis of the option price with respect to different
values of the information flow parameter. By analogy with the
Black-Scholes greeks, in Section \ref{sec:3.3} we define
appropriate expressions for the vega and the delta, and we show
that the vega is positive. From this investigation we conclude
that bond options can in principle be used to calibrate the model.
In Section \ref{sec:3.4} we derive an expression for the price
process of a bond option, and conclude that a position in the
underlying bond market can serve as a hedge for a position in an
associated call option.

In Section \ref{sec:3.5} we present an alternative technique for
deriving the price processes of derivatives in the
information-based framework. Instead of performing a change of
measure, we introduce the concept of information-based
Arrow-Debreu securities. This concept, in turn, leads to the
notion of a new class of derivatives which we call ``information
derivatives". The information-based Arrow-Debreu technique is
considered further, and in greater generality, in Sections
\ref{sec6.8} and \ref{sec6.9}.

Complex credit-linked structures are investigated in Chapter 4,
where we begin with the consideration of defaultable coupon bonds.
Such bonds are analysed in greater depth in Section \ref{sec6.4},
once the theory of $X$-factors has been developed to a higher
degree of generality. We obtain an exact expression for the price
process of a coupon bond, taking advantage of the analytical
solution of the conditional expectations involved. In Sections
\ref{sec4.2}, \ref{sec4.3}, and \ref{sec4.4} we consider other
complex credit products, such as credit default swaps (CDSs) and
baskets of credit-risky bonds. One of the strengths of the
$X$-factor approach is that it allows for a great deal of
flexibility in the modelling of the correlation structures
involved with complex credit instruments.

We then leave the field of credit risk modelling as such, and step
to a more general domain of asset classes. In Chapter 5 we
consider cash flows described by continuous random variables. This
paves the way for the consideration of equity products. We begin
in Sections \ref{sec5.2} and \ref{sec5.3} with a discussion on how
we should model the cash flows associated with an asset that pays
a sequence of dividends---e.g., a stock. After defining the
information process appropriate for this type of financial
instrument, in Sections \ref{sec5.4} and \ref{sec5.5} we derive an
analytical formula for the price of a single-dividend-paying
asset. We are also able to price European-style options written on
assets with a single cash flow. In Section~\ref{sec5.6}, pricing
formulae are presented for the situation where the random variable
associated with the single cash flow has an exponential
distribution or, more generally, a gamma distribution.

The extension of this framework to assets generating multiple cash
flows is established in Section \ref{sec6.1}. We show that once
the relevant cash flows are decomposed in terms of a collection of
independent market factors, then a closed-form expression for the
asset price associated with a complex cash-flow structure can be
obtained. Moreover, by allowing distinct assets to share one or
more common market factors in the determination of one or more of
their respective cash flows, we obtain a natural correlation
structure for the associated asset price processes. This is
described in Section \ref{sec6.5}. This method for introducing
correlation in asset price movements contrasts with the
essentially \textit{ad hoc} approach adopted in most financial
modelling. Indeed, both for portfolio risk management and for
credit risk management there is a pressing need for a better
understanding of correlation modelling, and one of the goals of
the present work is to make a new contribution to this line of
investigation.

In Section \ref{sec6.3} we demonstrate that if two or more market
factors affect the future cash flows associated with an asset,
then the corresponding price process will exhibit unhedgeable
stochastic volatility. In particular, once two or more market
factors are involved, an option position cannot, in general, be
hedged with a position in the underlying. In this framework there
is therefore no need to introduce stochastic volatility into the
price process on an artificial basis. The $X$-factor theory makes
it possible to investigate the relationships holding between
classes of assets that are different in nature from one another,
and therefore have different types of risks. In Section
\ref{sec6.3a} we show how the standard geometric Brownian motion
model for asset price dynamics can be derived in an
information-based approach. In this case the relevant $X$-factor
is a Gaussian random variable.

The $X$-factor approach is developed further in the following
sections, where we discuss the issue of how to reduce a set of
dependent factors, which we call $Z$-factors, into a set of
independent $X$-factors. This is carried out in Sections
\ref{sec6.6} and \ref{sec6.7}, where we focus on dependent binary
random variables. In this way we illustrate a possible approach to
disentangling more general discrete and dependent random factors
by providing what we call a reduction algorithm. We conclude
Chapter 6 with a further development of the Arrow-Debreu theory
introduced in Section \ref{sec:3.5}. In Section \ref{sec6.8} we
extend the Arrow-Debreu theory to the case of a continuous
$X$-factor, and in Section \ref{sec6.9} we work out the price of a
bivariate intertemporal Arrow-Debreu security.

Up to this point the discussion has focussed on applications of
the information-based framework to credit-risky assets and equity
products in a continuous-time setting. In Chapter 7 we develop a
framework for the arbitrage-free dynamics of nominal and real
interest rates, and the associated price index. The goal of this
chapter is the development of a scheme for the pricing and risk
management of index-linked securities, in an information-based
setting. We begin with a general model for discrete-time asset
pricing, with the introduction in Section \ref{sec7.1} of two
axioms. The first axiom establishes the intertemporal relations
for dividend-paying assets. The second axiom specifies the
existence a positive non-dividend-paying asset with a positive
return. Armed with these axioms, we derive the price process of an
asset with limited liability, pinning down a transversality
condition. If this condition is satisfied, then the value of the
dividend-paying asset is dispersed in its dividends in the long
run. In Section \ref{sec7.2} we discuss the relationships that
hold between the nominal pricing kernel and the positive-return
asset. Nominal discount bonds are treated in Section \ref{sec7.3}
where we investigate the properties that follow from the theory
developed up to this stage in Chapter 7. By choosing a specific
form for the pricing kernel, we are in a position to create
interest rate models of the ``rational" type, including
discrete-time models with no immediate analogue in continuous
time. We also demonstrate that any nominal interest-rate model
consistent with the general scheme presented here admits a
discrete-time representation of the Flesaker-Hughston type.

In Section \ref{sec7.4} we embark on an analysis of the
money-market account process. We find that this process is a
special case of the positive-return process defined in Axiom B, if
the money-market account is defined as a previsible
strictly-positive non-dividend-paying asset. This property is
embodied in Axiom B$^{\ast}$, which can be used as an alternative
basis of the theory, and in Proposition 7.4.1 we show that the
original two Axioms A and B proposed in Section \ref{sec7.1} imply
Axiom B$^{\ast}$. Returning to the information-based framework, we
propose in Section \ref{sec7.5} to discretise the information
processes associated with the $X$-factors, and construct in this
way the filtration to which the pricing kernel is adapted. The
fact that we can generate explicit models for the pricing kernel
enables us to build explicit models for the discount bond price
process and for the associated money-market account process
described in Section \ref{sec7.4}.

The resulting nominal interest rate system is then embedded in a
wider system incorporating macroeconomic factors relating to the
money supply, aggregate consumption, and price level. In Section
\ref{sec7.6} we consider a representative agent who obtains
utility from real consumption and from the real liquidity benefit
of the money supply. We then calculate the maximised expected
utility of aggregate consumption and money supply liquidity
benefit, subject to a budget constraint, where utility is
discounted making use of a psychological discount factor. The fact
that the utility depends on the real liquidity benefit of the
money supply leads to fundamental links between the processes of
aggregate consumption, money supply, price level, and the nominal
liquidity benefit. The link between the nominal pricing kernel and
the money supply gives rise in a natural way both to inflationary
and deflationary scenarios. Using the same formulae we are then in
a position to price index-linked securities, and to consider the
pricing and risk management of inflation derivatives.

In concluding this introduction we make a few remarks to help the
reader place the material of this thesis in the broader context of
finance theory. The majority of the work that has been carried out
in mathematical finance, as represented in standard textbooks
(e.g., Baxter \& Rennie 1996, Bielecki \& Rutkowski 2002, Bj\"ork
2004, Duffie 2001, Hunt \& Kennedy 2004, Karatzas \& Shreve 1998,
Musiela \& Rutkowski 2004) operates at what one might call a
``macroscopic" level. That is to say, the price processes of the
``basic" assets are regarded as ``given", and no special attempts
are made to ``derive" these processes from deeper principles.
Instead, the emphasis is typically placed on the valuation of
derivatives, and various types of optimization problems. The
standard Black-Scholes theory is, for example, in this sense a
``macroscopic" theory.

On the other hand, there is also a well-developed literature on
so-called market microstructure which investigates how prices are
formed in markets (see, e.g., O'Hara 1995 and references cited
therein). We can regard this part of finance theory as referring
to the ``microscopic" level of the subject. Broadly speaking,
models concerning interacting agents, heterogeneous preferences,
asymmetric information, bid-ask spreads, statistical arbitrage,
and insider trading can be regarded as operating to a significant
extent at the ``microscopic" level.

The work in this thesis can be situated in between the
``macroscopic" and ``microscopic" levels, and therefore might
appropriately be called ``mesoscopic". Thus at the ``mesoscopic"
level we emphasize the modelling of cash flows and market
information; but we assume homogeneous preferences, and symmetric
information. Thus, there is a universal market filtration
modelling the development of available information; but we
construct the filtration, rather than taking it as given.
Similarly, an asset will have a well-defined price process across
the whole market; but we construct the price process, rather than
taking it as given. It should be emphasized nevertheless that the
``output" of our essentially mesoscopic analysis is a family of
macroscopic models; thus to that extent these two levels of
analysis are entirely compatible (see, e.g., Section
\ref{sec6.3a}).

\clearemptydoublepage
\pagestyle{fancy} \fancyhf{}
\renewcommand{\chaptermark}[1]{\markboth{\chaptername\ \thechapter. \ #1}{}}
\renewcommand{\sectionmark}[1]{\markright{\thesection\ #1}}
\fancyhead[RO]{\sffamily\small \thepage}
\fancyhead[LO]{\sffamily\small \rightmark}
\fancyhead[LE]{\sffamily\small \thepage}
\fancyhead[RE]{\sffamily\small \leftmark}
\fancypagestyle{plain}{\fancyhf{}\renewcommand{\headrulewidth}{0pt}\renewcommand{\footrulewidth}{0pt}}
\chapter{Credit-risky discount bonds}\label{chap2}
\section{The need for an information-based approach for credit-risk
modelling}\label{sec2.1}

Models for credit risk management and, in particular, for the
pricing of credit derivatives are usually classified into two
types: structural models and reduced-form models.  For some
general overviews of these approaches---see, e.g., Jeanblanc \&
Rutkowski 2000, Hughston \& Turnbull 2001, Bielecki \& Rutkowski
2002, Duffie \& Singleton 2003, Sch\"onbucher 2003, Bielecki {\it
et al}. 2004, Giesecke 2004, Lando 2004, and Elizalde 2005.

There are differences of opinion in the literature as to the
relative merits of the structural and reduced-form methodologies.
Both approaches have strengths, but there are also shortcomings in
each case. Structural models attempt to account at some level of
detail for the events leading to default---see, e.g., Merton 1974,
Black \& Cox 1976, Leland \& Toft 1996, Hilberink \& Rogers 2002,
and Hull \& White 2004a,b. One of the important problems of the
structural approach is its inability to deal effectively with the
multiplicity of situations that can lead to failure. For example,
default of a sovereign state, corporate default, and credit card
default would all require quite different treatments in a
structural model. For this reason the structural approach is
sometimes viewed as unsatisfactory as a basis for a practical
modelling framework.

Reduced-form models are more commonly used in practice on account
of their tractability, and on account of the fact that, generally
speaking, fewer assumptions are required about the nature of the
debt obligations involved and the circumstances that might lead to
default---see, e.g., Jarrow \& Turnbull 1995, Lando 1998, Flesaker
{\it et al}. 1994, Duffie {\it et al}. 1996, Jarrow {\it et al}.
1997,  Madan \& Unal 1998, Duffie \& Singleton 1999, Hughston \&
Turnbull 2000, Jarrow \& Yu 2001, and Chen \& Filipovi\'c 2005. By
a reduced-form model we mean a model that does not address
directly the issue of the ``cause" of the default. Most
reduced-form models are based on the introduction of a random time
of default, where the default time is typically modelled as the
time at which the integral of a random intensity process first
hits a certain critical level, this level itself being an
independent random variable. An unsatisfactory feature of such
intensity models is that they do not adequately take into account
the fact that defaults are typically associated with a failure in
the delivery of a promised cash flow---for example, a missed
coupon payment. It is true that sometimes a firm will be forced
into declaration of default on a debt obligation, even though no
payment has yet been missed; but this will in general often be due
to the failure of some other key cash flow that is vital to the
firm's business. Another drawback of the intensity approach is
that it is not well adapted to the situation where one wants to
model the rise and fall of credit spreads---which can in practice
be due in part to changes in the level of investor confidence.

The purpose of this chapter is to introduce a new class of
reduced-form credit models in which these problems are addressed.
The modelling framework that we develop is broadly speaking in the
spirit of the incomplete-information approaches of Kusuoka 1999,
Duffie \& Lando 2001, Cetin {\it et al} 2004, Gieseke 2004,
Gieseke \& Goldberg 2004, Jarrow \& Protter 2004, and Guo {\it et
al} 2005.

In our approach, no attempt is made as such to bridge the gap
between the structural and the intensity-based models. Rather, by
abandoning the need for an intensity-based approach we are able to
formulate a class of reduced-form models that exhibit a high degree
of intuitively natural behaviour.

For simplicity we assume in this chapter that the underlying
default-free interest rate system is deterministic. The cash flows
of the debt obligation---in the case of a coupon bond, the coupon
payments and the principal repayment---are modelled by a
collection of random variables, and default will be identified as
the event of the first such payment that fails to achieve the
terms specified in the contract. We shall assume that partial
information about each such cash flow is available at earlier
times to market participants. However, the information of the
actual value of the cash flow will be obscured by a Gaussian noise
process that is conditioned to vanish once the time of the
required cash flow is reached. We proceed under these assumptions
to derive an exact expression for the bond price process.

In the case of a defaultable discount bond admitting two possible
payouts---e.g., either the full principal, or some partial
recovery payment---we shall derive an exact expression for the
value of an option on the bond. Remarkably, this turns out to be a
formula of the Black-Scholes type. In particular, the parameter
$\sigma$ that governs the rate at which the true value of the
impending cash flow is revealed to market participants against the
background of the obscuring noise process turns out to play the
role of a volatility parameter in the associated option pricing
formula; this interpretation is reinforced with the observation
that the option price can be shown to be an increasing function of
this parameter, as will be shown in Section \ref{sec:3.3}.
\section{Simple model for defaultable discount bonds}\label{sec2.2}
The object in this chapter is to build an elementary modelling
framework in which matters related to credit are brought to the
forefront. Accordingly, we assume that the background default-free
interest-rate system is deterministic. This assumption serves the
purpose of allowing us to focus attention entirely on
credit-related issues; it also allows us to derive explicit
expressions for certain types of credit derivative prices. The
general philosophy is that we should try to sort out
credit-related matters first, before attempting to incorporate
stochastic default-free interest rates into the picture.

As a further simplifying feature we take the view that credit
events are directly associated with anomalous cash flows. Thus a
default (in the sense that we use the term) is not something that
happens in the abstract, but rather is associated with the failure
of some agreed contractual cash flow to materialise at the
required time.

Our theory will be based on modelling the flow of incomplete
information to market participants about impending debt obligation
payments. As usual, we model the unfolding of chance in the
financial markets with the specification of a probability space
$(\Omega, {\mathcal F}, {\mathbb Q})$ with filtration $\{{\mathcal
F}_t\}_{0\le t<\infty}$. The probability measure ${\mathbb Q}$ is
understood to be the risk-neutral measure, and the filtration
$\{{\mathcal F}_t\}$ is understood to be the market filtration.
Thus all asset-price processes and other information-providing
processes accessible to market participants are adapted to
$\{{\mathcal F}_t\}$. Our framework is, in particular, completely
compatible with the standard arbitrage-free pricing theory as
represented, for example, in Bj\"ork 2004, or Shiryaev 1999.

The real probability measure does not directly enter into the
present discussion. We assume the absence of arbitrage. The First
Fundamental Theorem then guarantees the existence of a (not
necessarily unique) risk-neutral measure. We assume, however, that
the market has chosen a fixed risk-neutral measure $\Q$ for the
pricing of all assets and derivatives. We assume further that the
default-free discount-bond system, denoted $\{P_{tT}\}_{0\le t\le
T<\infty}$, can be written in the form
\begin{equation}\label{}
    P_{tT}=\frac{P_{0T}}{P_{0t}},
\end{equation}
where the function $\{P_{0t}\}_{0\le t<\infty}$ is assumed to be
differentiable and strictly decreasing, and to satisfy
$0<P_{0t}\leq 1$ and $\lim_{t\rightarrow\infty}P_{0t}=0$. Under
these assumptions it follows that if the integrable random
variable $H_T$ represents a cash flow occurring at $T$, then its
value $H_t$ at any earlier time $t$ is given by
\begin{eqnarray}
H_t=P_{tT}{\mathbb E}\left[H_T|{\mathcal F}_t\right].
\end{eqnarray}

Now let us consider more specifically the case of a simple
credit-risky discount bond that matures at time $T$ to pay a
principal of $h_1$ dollars, if there is no default. In the event
of default, the bond pays $h_0$ dollars, where $h_0<h_1$. When
just two such payoffs are possible we shall call the resulting
structure a ``binary" discount bond. In the special case given by
$h_1=1$ and $h_0=0$ we call the resulting defaultable debt
obligation a ``digital" bond.

We shall write $p_1$ for the probability that the bond will pay
$h_1$, and $p_0$ for the probability that the bond will pay $h_0$.
The probabilities here are risk-neutral, and hence build in any
risk adjustments required in expectations needed in order to
deduce appropriate prices. Thus if we write $B_{0T}$ for the price
at time $0$ of the credit-risky discount bond then we have
\begin{eqnarray}
B_{0T}=P_{0T}(p_1 h_1+p_0 h_0).
\end{eqnarray}
It follows that, providing we know the market data $B_{0T}$ and
$P_{0T}$, we can infer the {\it{a priori}} probabilities $p_1$ and
$p_0$. In particular, we obtain
\begin{eqnarray}
p_0 = \frac{1}{h_1-h_0}\left(h_1-\frac{B_{0T}}{P_{0T}}\right),
\quad p_1 = \frac{1}{h_1-h_0}\left( \frac{B_{0T}}{P_{0T}}
-h_0\right) .
\end{eqnarray}

Given this setup we now proceed to model the bond-price process
$\{B_{tT}\}_{0\leq t\leq T}$. We suppose that the true value of
$H_T$ is not fully accessible until time $T$; that is, we assume
$H_T$ is ${\mathcal F}_T$-measurable, but not necessarily
${\mathcal F}_t$-measurable for $t<T$. We shall assume,
nevertheless, that {\it partial} information regarding the value
of the principal repayment $H_T$ is available at earlier times.
This information will in general be imperfect---one is looking
into a crystal ball, so to speak, but the image is cloudy and
indistinct. Our model for such cloudy information will be of a
simple type that allows for analytic tractability. In particular,
we would like to have a model in which information about the true
value of the debt repayment steadily increases over the life of
the bond, while at the same time the obscuring factors at first
increase in magnitude, and then eventually die away just as the
bond matures. More precisely, we assume that the following
$\{{\mathcal F}_t\}$-adapted process is accessible to market
participants:
\begin{eqnarray}
\xi_t=\sigma H_T t+\beta_{tT}. \label{eq:6}
\end{eqnarray}
We call $\{\xi_t\}$ a market information process. The process
$\{\beta_{tT}\}_{0\le t\le T}$ appearing in the definition of
$\{\xi_t\}$ is a standard Brownian bridge on the time interval
$[0,T]$. Thus $\{\beta_{tT}\}$ is a Gaussian process satisfying
$\beta_{0T}=0$ and $\beta_{TT}=0$, and such that ${\mathbb E}
[\beta_{tT}]=0$ and
\begin{equation}
{\mathbb E}\left[\beta_{sT}\beta_{tT}\right] = \frac{s(T-t)}{T}
\end{equation}
for all $s,t$ satisfying $0\le s\le t\le T$. We assume that
$\{\beta_{tT}\}$ is independent of $H_T$, and thus represents
``purely uninformative" noise. Market participants do not have
direct access to $\{\beta_{tT}\}$; that is to say,
$\{\beta_{tT}\}$ is not assumed to be adapted to $\{{\mathcal
F}_t\}$. We can thus think of $\{\beta_{tT}\}$ as representing the
rumour, speculation, misrepresentation, overreaction, and general
disinformation often occurring, in one form or another, in
connection with financial activity, all of which distort and
obscure the information contained in $\{\xi_t\}$ concerning the
value of $H_T$.

Clearly the choice (\ref{eq:6}) can be generalised to include a
wider class of models enjoying similar qualitative features. In
this thesis we shall primarily consider information processes of
the form (\ref{eq:6}) for the sake of definiteness and
tractability. Indeed, the ansatz $\{\xi_t\}$ defined by
(\ref{eq:6}) has many attractive features, and can be regarded as
a convenient ``standard" model for an information process.

The motivation for the use of a bridge process to represent the
noise is intuitively as follows. We assume that initially all
available market information is taken into account in the
determination of the price; in the case of a credit-risky discount
bond, the relevant information is embodied in the {\it a priori}
probabilities. After the passage of time, however, new rumours and
stories start circulating, and we model this by taking into
account that the variance of the Brownian bridge steadily
increases for the first half of its trajectory. Eventually,
however, the variance drops and falls to zero at the maturity of
the bond, when the outcome is realised.

The parameter $\sigma$ in this model represents the rate at which
the true value of $H_T$ is ``revealed" as time progresses. Thus,
if $\sigma$ is low, then the value of $H_T$ is effectively hidden
until very near the maturity date of the bond; on the other hand,
if $\sigma$ is high, then we can think of $H_T$ as being revealed
quickly. 
A rough measure for the timescale $\tau_D$ over which information
is revealed is given by
\begin{equation}
\tau_D = \frac{1}{\sigma^2(h_1-h_0)^2}.
\end{equation}
In particular, if $\tau_D\ll T$, then the value of $H_T$ is
typically revealed rather early in the history of the bond, e.g.,
after the passage of a few multiples of $\tau_D$. In this
situation, if default is ``destined" to occur, even though the
initial value of the bond is high, then this will be signalled by
a rapid decline in the value of the bond. On the other hand, if
$\tau_D\gg T$, then the value of $H_T$ will only be revealed at
the last minute, so to speak, and the default will come as a
surprise, for all practical purposes. It is by virtue of this
feature of the present modelling framework that the use of
inaccessible stopping times can be avoided.

To make a closer inspection of the default timescale we proceed as
follows. For simplicity, we assume in our model that the only
market information available about $H_T$ at times earlier than $T$
comes from observations of $\{\xi_t\}$. Let us denote by
${\mathcal F}^{\xi}_t$ the subalgebra of ${\mathcal F}_t$
generated by $\{\xi_s\}_{0\le s\le t}$. Then our simplifying
assumption is that
\begin{equation}
{\mathbb E}[H_T|{\mathcal F}_t]={\mathbb E} [H_T |{\mathcal
F}^{\xi}_t].
\end{equation}
With this assumption in place, we are now in a position to
determine the price-process $\{B_{tT}\}_{0\leq t\leq T}$ for a
credit-risky bond with payout $H_T$. In particular, we wish to
calculate
\begin{eqnarray}
B_{tT}=P_{tT}H_{tT}, \label{eq:230}
\end{eqnarray}
where $H_{tT}$ is the conditional expectation of the bond payout:
\begin{eqnarray}
H_{tT}={\mathbb E} \left[H_T\Big|{\mathcal F}_t\right].
\label{eq:x10}
\end{eqnarray}
It turns out that $H_{tT}$ can be worked out explicitly. The
result is given by the following expression:
\begin{eqnarray}\label{eq:HtT}
H_{tT}=\frac{p_0 h_0\exp\left[\frac{T}{T-t}\left(\sigma
h_0\xi_t-\tfrac{1}{2}\sigma^2 h_0^2 t\right)\right]+p_1
h_1\exp\left[\frac{T}{T-t}\left(\sigma h_1\xi_t
-\tfrac{1}{2}\sigma^2 h_1^2
t\right)\right]}{p_0\exp\left[\frac{T}{T-t}\left(\sigma h_0\xi_t
-\tfrac{1}{2}\sigma^2 h_0^2 t\right)\right]+ p_1\exp\left[
\frac{T}{T-t}\left(\sigma h_1\xi_t-\tfrac{1}{2}\sigma^2 h_1^2
t\right)\right]}. \label{eq:14}
\end{eqnarray}
We note, in particular, that there exists a function $H(x,y)$ of
two variables such that $H_{tT}=H(\xi_t,t)$. The fact that the
process $\{H_{tT}\}$ converges to $H_T$ as $t$ approaches $T$
follows from (\ref{eq:x10}) and the fact that $H_T$ is
$\F_T$-measurable. The details of the derivation of the formula
(\ref{eq:HtT}) will be given in the next section.

Since $\{\xi_t\}$ is given by a combination of the random bond
payout and an independent Brownian bridge, it is straightforward
to simulate trajectories for $\{B_{tT}\}$. Explicit examples of
such simulations are presented in Section \ref{sec2.7}.
\section{Defaultable discount bond price processes}\label{sec2.3}
Let us now consider the more general situation where the discount
bond pays out the possible values
\begin{equation}
H_T=h_i
\end{equation}
$(i=0,1,\ldots,n)$ with {\it a priori} probabilities
\begin{equation}
{\mathbb Q}[H_T=h_i]=p_i.
\end{equation}
For convenience we assume that
\begin{equation}
h_n>h_{n-1}>\dots>h_1>h_0.
\end{equation}
The case $n=1$ corresponds to the binary bond we have just
discussed. In the general situation we think of $H_T=h_n$ as the
case of no default, and all the other cases as various possible
degrees of recovery.

Although we consider, for simplicity, a discrete payout spectrum
for $H_T$, the case of a continuous recovery value in the event of
default can be formulated analogously. In that case we assign a
fixed {\it a priori} probability $p_1$ to the case of no default,
and a continuous probability distribution function
\begin{equation}
p_0(x)={\mathbb Q}[H_T<x]
\end{equation}
for values of $x$ less than or equal to $h$, satisfying
\begin{equation}
p_1+p_0(h)=1.
\end{equation}

Now defining the information process $\{\xi_t\}$ as before by
(\ref{eq:6}), we want to find the conditional expectation
(\ref{eq:x10}). We note, first, that the conditional probability
that the credit-risky bond pays out $h_i$ is given by
\begin{equation}
\pi_{it}=\Q\left(H_T=h_i\vert\F_t\right),
\end{equation}
or equivalently,
\begin{equation}
\pi_{it}=\E\left[{\bf 1}_{\{H_T=h_i\}}\vert\F_t\right].
\end{equation}
For $H_{tT}$ we can then write
\begin{equation}\label{H_tT}
H_{tT}=\sum_{i=0}^n h_i\pi_{it}.
\end{equation}
It follows, however, from the Markovian property of $\{\xi_t\}$,
which will be established in Proposition \ref{propinfoproc} below,
that to compute (\ref{eq:x10}) it suffices to take the conditional
expectation of $H_T$ with respect to the $\sigma$-subalgebra
generated by the random variable $\xi_t$ alone. Therefore, we have
\begin{equation}
H_{tT}={\mathbb E}[H_T|\xi_t],
\end{equation}
and also
\begin{equation}
\pi_{it}={\mathbb Q}(H_T=h_i|\xi_t),
\end{equation}
or equivalently,
\begin{equation}
\pi_{it} = {\mathbb E}\left[ {\bf 1}_{\{H_T=h_i\}}|\xi_t\right].
\end{equation}

\begin{prop}\label{propinfoproc}
The information process $\{\xi_t\}_{0\le t\le T}$ satisfies the
Markov property.
\end{prop}

\noindent{\bf Proof}. We need to verify that
\begin{eqnarray}
{\mathbb E}\left[f(\xi_t)\,\vert\,{\mathcal F}_s^\xi\right]=
{\mathbb E}\left[ f(\xi_t)\,\vert\,\xi_s\right]
\end{eqnarray}
for all $s,t$ such that $0\leq s\leq t\leq T$ and any measurable
function $f(x)$ with $\sup_x|f(x)|<\infty$. It suffices (see,
e.g., Liptser \& Shiryaev 2000, theorems 1.11 and 1.12) to show
that
\begin{eqnarray}
{\mathbb E}\left[ f(\xi_t)\,\vert\,\xi_s, \xi_{s_1}, \xi_{s_2},
\cdots, \xi_{s_k}\right] ={\mathbb
E}\left[f(\xi_t)\,\vert\,\xi_s\right]
\end{eqnarray}
for any collection of times $t,s,s_1,s_2,\ldots,s_k$ such that
\begin{equation}
T\ge t\ge s\ge s_1\ge s_2\ge\cdots\ge s_k\ge 0.
\end{equation}
First, we remark that for any times $t,s,s_1$ satisfying $t\ge
s\ge s_1$ the random variables $\beta_{tT}$ and
$\beta_{sT}/s-\beta_{s_1T}/s_1$ have vanishing covariance, and
thus are independent. More generally, for $s\ge s_1\ge s_2\ge s_3$
the random variables $\beta_{sT}/s-\beta_{s_1T} /s_1$ and
$\beta_{s_2T}/s_2-\beta_{s_3T}/s_3$ are independent. Next, we note
that
\begin{equation}
\frac{\xi_s}{s}-\frac{\xi_{s_1}}{s_1}=\frac{\beta_{sT}}{s} - \frac{\beta_{s_1T} }{s_1}.
\end{equation}
It follows that
\begin{align}
&{\mathbb E}\left[ f(\xi_t)\,\vert\,\xi_s, \xi_{s_1}, \xi_{s_2},
\cdots,
\xi_{s_k}\right]\nonumber\\
&={\mathbb E}\left[ f(\xi_t)\,\Big|\,\xi_s,
\frac{\xi_s}{s}-\frac{\xi_{s_1}}{s_1}, \frac{\xi_{s_1}}{s_1}-
\frac{\xi_{s_2}}{s_2}, \cdots, \frac{\xi_{s_{k-1}}}{s_{k-1}}-
\frac{\xi_{s_k}}{s_k}\right] \nonumber \\
&={\mathbb E}\left[f(\xi_t)\,\Big|\,\xi_s,
\frac{\beta_{sT}}{s}-\frac{\beta_{s_1T}}{s_1},
\frac{\beta_{s_1T}}{s_1}- \frac{\beta_{s_2T}}{s_2}, \cdots,
\frac{\beta_{s_{k-1}T}}{s_{k-1}}- \frac{\beta_{s_kT}}{s_k}\right].
\end{align}
However, since $\xi_s$ and $\xi_t$ are independent of
$\beta_{sT}/s - \beta_{s_1T} /s_1$, $\beta_{s_1T}/s_1 -
\beta_{s_2T} /s_2$, $\cdots$, $\beta_{s_{k-1}T}/s_{k-1}-
\beta_{s_kT}/s_k$, we see that the desired result follows
immediately.\hfill$\Box$ \\

Next we observe that the {\it a priori} probability $p_i$ and the
{\it a posteriori} probability $\pi_{it}$ at time $t$ are related
by the Bayes formula:
\begin{eqnarray}
{\mathbb Q}(H_T=h_i|\xi_t) = \frac{p_i \rho(\xi_t|H_T=h_i)}
{\sum_i p_i \rho(\xi_t|H_T=h_i)}. \label{eq:27}
\end{eqnarray}
Here the conditional density function $\rho(\xi|H_T=h_i)$,
$\xi\in\mathbb{R}$, for the random variable $\xi_t$ is defined by
the relation
\begin{eqnarray}
{\mathbb Q}\left(\xi_t\leq x|H_T=h_i\right)=\int_{-\infty}^x
\rho(\xi|H_T=h_i)\,\rd\xi,
\end{eqnarray}
and is given more explicitly by
\begin{eqnarray}\label{eq:28}
\rho(\xi|H_T=h_i)=\frac{1}{\sqrt{2\pi t(T-t)/T}} \exp\left(-\half
\frac{(\xi-\sigma h_it)^2}{t(T-t)/T}\right).
\end{eqnarray}
This expression can be deduced from the fact that conditional on
$H_T=h_i$ the random variable $\xi_t$ is normally distributed with
mean $\sigma h_i t$ and variance $t(T-t)/T$. As a consequence of
(\ref{eq:27}) and (\ref{eq:28}), we find that
\begin{eqnarray}
\pi_{it}=\frac{p_i\exp\left[\frac{T}{T-t}(\sigma h_i \xi_t-
\frac{1}{2} \sigma^2 h_i^2 t)\right]} {\sum_i\,p_i\exp\left[
\frac{T}{T-t}(\sigma h_i \xi_t- \frac{1}{2} \sigma^2 h_i^2
t)\right]}. \label{eq:25}
\end{eqnarray}
It follows then, on account of (\ref{H_tT}), that
\begin{eqnarray}\label{eq:25rev}
H_{tT} = \frac{\sum_i p_i h_i \exp\left[\frac{T}{T-t}\left(\sigma
h_i \xi_t-\frac{1}{2}\sigma^2 h_i^2 t\right)\right]}{\sum_i p_i
\exp \left[ \frac{T}{T-t}\left(\sigma h_i \xi_t-\frac{1}{2}
\sigma^2 h_i^2 t \right) \right]} .
\end{eqnarray}
This is the desired expression for the conditional expectation of
the bond payoff. In particular, for the binary case $i=0,1$ we
recover formula (\ref{eq:14}). The discount-bond price process
$\{B_{tT}\}$ is therefore given by (\ref{eq:230}), with $H_{tT}$
defined as in (\ref{eq:25rev}).
\section{Defaultable discount bond volatility}\label{sec2.4}
In this section we analyse the dynamics of the defaultable bond
price process $\{B_{tT}\}$ determined in the previous section. The
key relation we need for working out the dynamics of the bond
price is that the conditional probability process $\{\pi_{it}\}$
satisfies a stochastic equation of the form
\begin{eqnarray}
\rd\pi_{it}=\frac{\sigma T}{T-t}(h_i-H_{tT})\pi_{it}\,\rd W_t,
\label{eq:23}
\end{eqnarray}
for $0\le t<T$ where $H_{tT}$ is given by equation (\ref{H_tT}),
and the process $\{W_t\}_{0\leq t<T}$, defined by
\begin{eqnarray}\label{WINNOV}
W_t = \xi_t + \int_0^t \frac{1}{T-s}\,\xi_s\,\rd s - \sigma T
\int_0^t \frac{1}{T-s}\,H_{sT}\,\rd s, \label{eq:24}
\end{eqnarray}
is an $\{\F_t\}$-Brownian motion.

Perhaps the most direct way of obtaining (\ref{eq:23}) and
(\ref{eq:24}) is by appeal to the well-known
Fujisaki-Kallianpur-Kunita (FKK) filtering theory, the main
results of which we summarise below in an abbreviated form,
suppressing technicalities (see, e.g., Fujisaki {\it et al.} 1972,
or Liptser \& Shiryaev 2000, chapter 8, for a more complete
treatment). A probability space is given, with a background
filtration $\{\mathcal{G}_t\}$, on which we specify a pair of
processes $\{\xi_t\}$ (the ``observed" process) and $\{x_t\}$ (the
``unobserved" process). We assume that
\begin{equation}\label{obs}
\xi_t=\int^t_0\mu_s\rd s+Y_t
\end{equation}
and
\begin{equation}\label{unobs}
x_t=x_0+\int^t_0\vartheta_s\rd s+M_t,
\end{equation}
and that the processes $\{\xi_t\}$, $\{x_t\}$, $\{\mu_t\}$,
$\{\vartheta_t\}$, $\{Y_t\}$, and $\{M_t\}$ are
$\{\mathcal{G}_t\}$-adapted. We take $\{Y_t\}$ to be a
$\{\mathcal{G}_t\}$-Brownian motion, and $\{M_t\}$ to be a
$\{\mathcal{G}_t\}$-martingale which, for simplicity, we assume
here to be independent of $\{Y_t\}$. The idea is that the values
of $\{\xi_t\}$ are observable, and from this information we wish
to obtain information about the unobservable process $\{x_t\}$.
Let us write $\{\F_t\}$ for the filtration generated by the
observed process, and for any process $\{X_t\}$ let us write
$\hat{X}_t=\E[X_t\,\vert\,\F_t]$. Thus $\hat{X}_t$ represents the
best estimate of $X_t$, given the information of the observations
up to time $t$. Then the basic result of the FKK theory is that
the dynamics of $\{\hat{x}_t\}$ are given by
\begin{equation}\label{filter}
\rd\hat{x}_t=\hat{\vartheta}_t\rd t+\left[(\widehat{\mu
x})_t-\hat{\mu}_t\hat{x}_t\right]\rd W_t,
\end{equation}
where the so-called innovations process $\{W_t\}$, given by
\begin{equation}\label{innovW}
W_t=\xi_t-\int^t_0\hat{\mu}_s\rd s,
\end{equation}
turns out to be an $\{\F_t\}$-Brownian motion.

Now let us see how the FKK theory can be used to obtain the
dynamics of $\{\pi_{it}\}$. The link to the FKK theory is
established by letting  the market information process $\{\xi_t\}_{0\le t\le T}$ be the observed process, and by
letting the dynamically constant process ${\bf 1}\{H_T=h_i\}$ be
the unobserved process.

To obtain an appropriate expression for the dynamics of
$\{\xi_t\}$ in the form (\ref{obs}), we recall (Karatzas \& Shreve
1991) that a standard Brownian bridge $\{\beta_{tT}\}_{0\le t\le
T}$ satisfies a stochastic differential equation of the form
\begin{equation}
\rd\beta_{tT}=-\frac{\beta_{tT}}{T-t}\rd t+\rd Y_t
\end{equation}
for $0\le t<T$, where $\{Y_t\}$ is a standard Brownian motion.
Then if we set $\xi_t=\sigma H_T t+\beta_{tT}$, a short calculation shows that
\begin{eqnarray}
\rd\xi_t=\frac{\sigma H_T T-\xi_t}{T-t}\rd t+\rd Y_t.
\end{eqnarray}
Thus in equation (\ref{obs}) we can set
\begin{equation}\label{mu}
\mu_t=\frac{\sigma H_T T-\xi_t}{T-t},
\end{equation}
and it follows that
\begin{equation}
\hat{\mu}_t=\frac{\sigma H_{tT}T-\xi_t}{T-t},
\end{equation}
where $H_{tT}=\E[H_T\,\vert\,\F_t]$. Inserting this expression for
$\hat{\mu}_t$ into (\ref{innovW}), we are then led to expression
(\ref{WINNOV}) for the innovation process.

To work out the dynamics of $\{\pi_{it}\}$ we need expressions for
$\hat{x}_t$ and $(\widehat{\mu x})_t$, with $x_t={\bf
1}\{H_T=h_i\}$ and $\mu_t$ as given by (\ref{mu}). Thus
$\hat{x}_t=\pi_{it}$, and
\begin{equation}
(\widehat{\mu x})_t=\pi_{it}\frac{\sigma h_i T-\xi_t}{T-t}.
\end{equation}
Inserting the expressions that we have obtained for $\hat{\mu}_t$,
$\hat{x}_t$, and $(\widehat{\mu x})_t$ into (\ref{filter}), and
noting that $\vartheta_t=0$ (since the unobserved process is
dynamically constant in the present context), we are then led to
(\ref{eq:23}), as desired. The point here is that the FKK theory
allows us to deduce equation (\ref{eq:23}) and tells us that
$\{W_t\}$ is an $\{F_t\}$-Brownian motion.

Alternatively, we can derive the dynamics of $\{\pi_{it}\}$ from
(\ref{eq:25}). This is achieved by applying Ito's lemma and using
the fact that $(\rd\xi_t)^2=\rd t$. The fact that $\{W_t\}$, as
defined by (\ref{eq:24}), is an $\{\F_t\}$-Brownian motion can
then be verified by use of the L\'evy criterion. In particular one
needs to show that $\{W_t\}$ is an $\{\F_t\}$-martingale and that
$(\rd W_t)^2=\rd t$. To prove that $\{W_t\}$ is an $\{{\mathcal
F}_t\}$-martingale we need to show that ${\mathbb
E}[W_u\vert{\mathcal F}_t]=W_t$, for $0\le t\le u<T$. First we
note that it follows from (\ref{eq:24}) and the Markov property of
$\{\xi_t\}$ that
\begin{eqnarray}\label{eq:24a}
{\mathbb E}[W_u\vert{\mathcal F}_t]&=&W_t+{\mathbb
E}\left[(\xi_u-\xi_t)\vert\xi_t\right]+{\mathbb E}\left[
\int^u_t\frac{1}{T-s}\,\xi_s\,\rd s\bigg\vert\xi_t\right]
\nonumber\\
&&-\sigma T\,{\mathbb E}\left[\int^u_t\frac{1}{T-s}\,H_{sT}\,\rd
s\bigg\vert\xi_t\right].
\end{eqnarray}
This expression can be simplified if we recall that
$H_{sT}={\mathbb E}[H_T\vert\xi_s]$ and use the tower property in
the last term on the right. Inserting the definition (\ref{eq:6})
into the second and third terms on the right we then have:
\begin{eqnarray}\label{eq:24b}
{\mathbb E}[W_u\vert{\mathcal F}_t] &=& W_t+{\mathbb E}[\sigma H_T
u+\beta_{uT}\vert\xi_t]-{\mathbb E}[\sigma H_T
t+\beta_{tT}\vert\xi_t]+\sigma{\mathbb E}[H_T\vert\xi_t]
\int^u_t\frac{s}{T-s}\,\rd s \nonumber \\ && + {\mathbb
E}\left[\int^u_t\frac{1}{T-s}\,\beta_{sT}\,\rd
s\bigg\vert\xi_t\right]-\sigma{\mathbb E}[H_T\vert\xi_t]
\int^u_t\frac{T}{T-s}\,\rd s.
\end{eqnarray}
Taking into account the fact that
\begin{equation}
\int^u_t\frac{s}{T-s}\,\rd s=t-u+\int^u_t\frac{T}{T-s}\,\rd s,
\label{eq:24bb}
\end{equation}
we see that all terms involving the random variable $H_T$ cancel
each other in (\ref{eq:24b}). This leads us to the following
relation:
\begin{equation}
{\mathbb E}[W_u\vert{\mathcal F}_t]=W_t+{\mathbb E}
[\beta_{uT}\vert\xi_t]-{\mathbb E}
[\beta_{tT}\vert\xi_t]+\int^u_t\frac{1}{T-s}\,{\mathbb E}
[\beta_{sT}\vert\xi_t]\,\rd s. \label{eq:24d}
\end{equation}
Next we use the tower property and the independence of
$\{\beta_{tT}\}$ and $H_T$ to deduce that
\begin{equation}
{\mathbb E}[\beta_{uT}\vert\xi_t]={\mathbb E}[{\mathbb
E}[\beta_{uT}\vert H_T,\beta_{tT}]\vert\xi_t]={\mathbb E}
[{\mathbb E}\left[\beta_{uT}\vert\beta_{tT}]\vert\xi_t\right].
\label{eq:24e}
\end{equation}
To calculate the inner expectation ${\mathbb E} [\beta_{uT}
\vert\beta_{tT}]$ we use the fact that the random variable
$\beta_{uT}/(T-u)-\beta_{tT}/(T-t)$ is independent of the random
variable $\beta_{tT}$. This can be checked by calculating their
covariance, and using the relation ${\mathbb E}
[\beta_{uT}\beta_{tT}]=t(T-u)/T$. We conclude after a short
calculation that
\begin{equation}
{\mathbb E}[\beta_{uT}\vert\beta_{tT}]
=\frac{T-u}{T-t}\,\beta_{tT}. \label{eq:24g}
\end{equation}
Inserting this result into (\ref{eq:24e}) we obtain the following
formula:
\begin{equation}
{\mathbb E}[\beta_{uT}\vert\xi_t]=\frac{T-u}{T-t}\,{\mathbb
E}[\beta_{tT}\vert\xi_t]. \label{eq:24h}
\end{equation}
Applying this formula to the second and fourth terms on the right
side of (\ref{eq:24d}), we deduce that ${\mathbb
E}[W_u\vert{\mathcal F}_t]=W_t$. That establishes that $\{W_t\}$
is an $\{{\mathcal F}_t\}$-martingale. Now we need to show that
$(\rd W_t)^2=\rd t$. This follows if we insert (\ref{eq:6}) into
the definition of $\{W_t\}$ above and use again the fact that
$(\rd\beta_{tT})^2=\rd t$. Hence, by Levy's criterion $\{W_t\}$ is
an $\{{\mathcal F}_t\}$-Brownian motion.

The $\{\F_t\}$-Brownian motion $\{W_t\}$, the existence of which
we have established, can be regarded as part of the information
accessible to market participants. We note in particular that,
unlike $\beta_{tT}$, the value of $W_t$ contains ``real"
information relevant to the outcome of the bond payoff. It follows
from (\ref{H_tT}) and (\ref{eq:23}) that for the discount bond
dynamics we have
\begin{eqnarray}
\rd B_{tT}=r_t B_{tT}\,\rd t+\Sigma_{tT}\, \rd W_t.
\end{eqnarray}
Here the expression
\begin{equation}
r_t=-\frac{\partial \ln P_{0t}}{\partial t}
\end{equation}
is the (deterministic) short rate at $t$, and the absolute
bond volatility $\Sigma_{tT}$ is given by
\begin{eqnarray}
\Sigma_{tT} =  \frac{\sigma T}{T-t}P_{tT}V_{tT}, \label{eq:26}
\end{eqnarray}
where $V_{tT}$ is the conditional variance of the
terminal payoff $H_T$, defined by:
\begin{eqnarray}
V_{tT}=\sum_i\,(h_i-H_{tT})^2\pi_{it}. \label{eq:3.12}
\end{eqnarray}
We thus observe that as the maturity date is approached the
absolute discount bond volatility will be high unless the
conditional probability has its mass concentrated around
the ``true" outcome; this ensures that the correct level is
eventually reached.

\begin{prop}\label{propCondVar}
The process $\{V_{tT}\}_{0\le t<T}$ for the conditional variance
of the terminal payoff $H_T$ is a supermartingale.
\end{prop}

\noindent{\bf Proof}. This follows directly from the fact that
$\{V_{tT}\}_{0\le t\le T}$ can be expressed as the difference
between a martingale and a submartingale:
\begin{equation}\label{condvar}
V_{tT}=\E_t\left[H^2_T\right]-\left(\E_t\left[H_T\right]\right)^2.
\end{equation}
\hfill$\Box$

The interpretation of this result is that in the filtration
$\{\F_t\}$ generated by $\{\xi_t\}$ one on average ``gains
information" about $H_T$. In other words, the uncertainty in the
conditional estimate of $H_T$ tends to reduce.

Given the result of Proposition \ref{propCondVar}, we shall now
derive an expression for the volatility of the volatility. The
``second order" volatility is of interest in connection with
pricing models for options on realised volatility. In the present
example it turns out that the ``vol-of-vol" has a particularly
simple form. Starting with (\ref{condvar}), we see that the
conditional variance can be written in the form
\begin{equation}
V_{tT}=\sum_{i=0}^n h_i^2\pi_{it}-H_{tT}^2.
\end{equation}
By use of Ito's formula, for the dynamics of $\{V_{tT}\}$ we thus
obtain
\begin{equation}
\rd V_{tT}=\sum_{i=0}^n h_i^2\rd\pi_{it}-2H_{tT}\rd H_{tT}-(\rd
H_{tT})^2.
\end{equation}
For the dynamics of $H_{tT}$, on the other hand, we have
\begin{equation}
\rd H_{tT}=\frac{\sigma T}{T-t}V_{tT}\rd W_t,
\end{equation}
from which it follows that
\begin{equation}
(\rd H_{tT})^2=\left(\frac{\sigma T}{T-t}\right)^2 V_{tT}^2\rd t.
\end{equation}
Combining these relations with (\ref{eq:23}) we then obtain
\begin{equation}
\rd V_{tT}=-\sigma^2\left(\frac{T}{T-t}\right)^2 V_{tT}^2\rd
t+\frac{\sigma T}{T-t}K_{tT}\rd W_t,
\end{equation}
where
\begin{equation}
K_{tT}=\sum^n_{i=0}\left(h_i-H_{tT}\right)^3\pi_{it}
\end{equation}
is the conditional skewness (third central moment) of terminal
payoff. Writing $V_{0T}$ for the {\it a priori} variance of $H_T$,
we thus have the expression
\begin{equation}
V_{tT}=V_{0T}-\sigma^2\int^t_0\left(\frac{T}{T-s}\right)^2V_{sT}^2\rd
s+\sigma\int^t_0\frac{T}{T-s}K_s\rd W_s
\end{equation}
for the ``risk" associated with the payoff $H_T$.

To calculate the vol-of-vol of the defaultable discount bond, we
need to work out the dynamics of $\{\Sigma_{tT}\}$, given the
dynamics of $\{V_{tT}\}$. It should be evident therefore that the
second-order absolute volatility $\Sigma^{(2)}_{tT}$, i.e. the
vol-of-vol of $\{B_{tT}\}$, is given by
\begin{equation}
\Sigma^{(2)}_{tT}=\left(\frac{\sigma T}{T-t}\right)^2
P_{tT}K_{tT}.
\end{equation}

It is interesting to observe that, as a consequence of equation
(\ref{eq:24}), the market information process $\{\xi_t\}$
satisfies the following stochastic differential equation:
\begin{eqnarray}
\rd \xi_t = \frac{1}{T-t}\left( \sigma T H(\xi_t, t) -\xi_t\right)
\rd t + \rd W_t.
\end{eqnarray}
We see that $\{\xi_t\}$ is a diffusion process; and since
$H(\xi_t,t)$ is monotonic in its dependence on $\xi_t$, we deduce
that $\{B_{tT}\}$ is also a diffusion process. To establish that
$H(\xi_t,t)$ is monotonic in $\xi_t$, and thus that $B_{tT}$ is
increasing in $\xi_t$ we note that
\begin{equation}
P_{tT}H^\prime(\xi_t,t)=\Sigma_{tT},
\end{equation}
where $H^\prime(\xi,t)= \partial H(\xi,t)/\partial\xi$. It follows
therefore that, in principle, instead of ``deducing" the dynamics
of $\{B_{tT}\}$ from the arguments of the previous sections, we
might have simply ``proposed" on an {\it ad hoc} basis the
one-factor diffusion process described above, noting that it leads
to the correct default dynamics. This line of reasoning shows that
the information formalism can be viewed, if desired, as leading to
purely ``classical" financial models, based on observable price
processes. In that sense the information-based approach adds an
additional layer of interpretation and intuition to the classical
framework, without altering any of its fundamental principles.
\section{Digital bonds, and binary bonds with partial recovery}\label{sec2.5}
It is interesting to ask, incidentally, whether in the case of a
binary bond with partial recovery, with possible payoffs
$\{h_0,h_1\}$, the price process admits the representation
\begin{eqnarray}
B_{tT}=P_{tT} h_0 + D_{tT}(h_1-h_0). \label{eq:240}
\end{eqnarray}
Here $D_{tT}$ denotes the value of a ``digital" credit-risky bond
that pays at maturity a unit of currency with probability $p_1$
and zero with probability $p_0=1- p_1$. Thus $h_0$ is the amount
guaranteed, whereas $h_1-h_0$ is the part that is ``at risk". It
is well known that such a relation can be deduced in
intensity-based models (Lando 1994, 1998). The problem is thus to
find a process $\{D_{tT}\}$ consistent with our scheme such that
(\ref{eq:240}) holds. It turns out that this can be achieved as
follows. Suppose we consider a digital payoff structure $D_T
\in\{0,1\}$ for which the parameter $\sigma$ is replaced by
\begin{equation}
{\bar\sigma}=\sigma(h_1-h_0).
\end{equation}
In other words, in establishing the appropriate dynamics for
$\{D_{tT}\}$ we ``renormalise" $\sigma$ by replacing it with
${\bar\sigma}$. The information available to market participants
in the case of the digital bond is represented by the process
$\{{\bar\xi}_t\}$ defined by
\begin{eqnarray}
{\bar\xi}_t={\bar\sigma}D_T t+\beta_{tT}.
\end{eqnarray}
It follows from (\ref{eq:25rev}) that the corresponding digital
bond price is given by
\begin{eqnarray}
D_{tT} = P_{tT} \frac{p_1\exp\left[\frac{T}{T-t}\left({\bar
\sigma}{\bar\xi}_t-\frac{1}{2}{\bar\sigma}^2t\right)\right]}
{p_0+p_1\exp\left[\frac{T}{T-t}\left({\bar\sigma}{\bar\xi}_t
-\frac{1}{2}{\bar\sigma}^2t\right)\right]}. \label{eq:Q}
\end{eqnarray}
A short calculation making use of (\ref{eq:14}) then allows us to
confirm that (\ref{eq:240}) holds, where $D_{tT}$ is given by
(\ref{eq:Q}). Thus even though at first glance the general binary
bond process (\ref{eq:14}) does not appear to admit a
decomposition of the form (\ref{eq:240}), in fact it does, once a
suitably renormalised value for the market information parameter
has been introduced.

A slightly more general result is the following. Let $H_T$ be a
random payoff and let $c_T$ be a constant payoff. Write
$\xi_t=\sigma H_T t+\beta_{tT}$ for the information process
associated with $H_T$, and $\xi'_t=\sigma(H_T+c_T)t+\beta_{tT}$
for the information process associated with the combined payoff
$H_T+c_T$. Then a straightforward calculation shows that
\begin{equation}
\E\left[H_T+c_T\,\big\vert\,\F_t^{\xi'}\right]=\E\left[H_T\,\big\vert\,\F^{\xi}_t\right]+c_T.
\end{equation}
There is no reason, on the other hand, to suppose that such
``linearity" holds more generally.
\section{Dynamic consistency and market re-calibration}\label{sec2.6}
The technique of ``renormalising" the information flow rate has
another useful application. It turns out that the model under
consideration exhibits a property that might appropriately be
called ``dynamic consistency".

Loosely speaking, the question is as follows: if the information
process is given as described, but then we ``update" or
re-calibrate the model at some specified intermediate time, is it
still the case that the dynamics of the model moving forward from
that intermediate time can be represented by an information
process?

To answer this question we proceed as follows. First, we define a
standard Brownian bridge over the interval $[t,T]$ to be a
Gaussian process $\{\gamma_{uT}\}_{t\leq u\leq T}$ satisfying
$\gamma_{tT}=0$, $\gamma_{TT}=0$, ${\mathbb E}[\gamma_{uT}]=0$ for
all $u\in[t,T]$, and
\begin{equation}
{\mathbb E}[\gamma_{uT}\gamma_{vT}]=\frac{(u-t)(T-v)}{(T-t)}
\end{equation}
for all $u,v$ such that $t\leq u\leq v\leq T$. Then we make note
of the following result.
\begin{lem}
Let $\{\beta_{tT}\}_{0\leq t\leq T}$ be a standard Brownian bridge
over the interval $[0,T]$, and define the process
$\{\gamma_{uT}\}_{t\leq u\leq T}$ by
\begin{eqnarray}
\gamma_{uT} = \beta_{uT} - \frac{T-u}{T-t} \,\beta_{tT} .
\end{eqnarray}
Then $\{\gamma_{uT}\}_{t\leq u\leq T}$ is a standard Brownian
bridge over the interval $[t,T]$, and is independent of
$\{\beta_{sT}\}_{0\leq s\leq t}$.
\end{lem}

\noindent{\bf Proof}. The lemma is easily established by use of
the covariance relation
\begin{equation}
{\mathbb E}[\beta_{tT}\beta_{uT}]=\frac{t(T-u)}{T}.
\end{equation}
We need to recall also that a necessary and sufficient condition
for a pair of Gaussian random variables to be independent is that
their covariance should vanish. Now let the information process
$\{\xi_s\}_{0\leq s\leq T}$ be given, and fix an intermediate time
$t\in(0,T)$. Then for all $u\in[t,T]$ let us define the process
$\{\eta_u\}_{0\leq u\leq T}$ by
\begin{eqnarray}
\eta_u = \xi_u- \frac{T-u}{T-t}\,\xi_t.
\end{eqnarray}
We claim that $\{\eta_u\}$ is an information process over the time
interval $[t,T]$. In fact, a short calculation establishes that
\begin{eqnarray}
\eta_u = {\tilde\sigma}H_T (u-t) + \gamma_{uT},
\end{eqnarray}
where $\{\gamma_{uT}\}_{t\leq u\leq T}$ is a standard Brownian
bridge over the interval $[t,T]$, and the new information flow
rate is given by
\begin{equation}
{\tilde\sigma}=\frac{\sigma T}{(T-t)}.
\end{equation}\hfill$\Box$

The interpretation of these results is as follows. The ``original"
information process proceeds from time 0 up to time $t$. At that
time we can re-initialise the model by taking note of the value of
the random variable $\xi_t$, and introducing the re-initialised
information process $\{\eta_u\}$. The new information process
depends on $H_T$; but since the value of $\xi_t$ is supplied, the
\textit{a priori} probability distribution for $H_T$ now changes
to the appropriate \textit{a posteriori} distribution consistent
with the information gained from the knowledge of $\xi_t$ at time
$t$.

These interpretive remarks can be put into a more precise form as
follows. Let $0\le t\le u<T$. What we want is a formula for the
conditional probability $\pi_{iu}$ expressed in terms of the
information $\eta_u$ and the ``new" {\it a priori} probability
$\pi_{it}$. Such a formula indeed exists, and is given as follows:
\begin{equation}\label{postcond}
\pi_{iu}=\frac{\pi_{it}\exp\left[\frac{T-t}{T-u}(
{\tilde\sigma}h_i\eta_u-\tfrac{1}{2}{\tilde\sigma}^2h_i^2(u-t))
\right]}{\Sigma_{i}\pi_{it}\exp\left[\frac{T-t}{T-u}
({\tilde\sigma}h_i\eta_u-\tfrac{1}{2}{\tilde\sigma}^2h_i^2
(u-t))\right]}.
\end{equation}
This relation can be verified by substituting the given
expressions for $\pi_{it}$, $\eta_{u}$, and $\bar{\sigma}$ into
the right-hand side of  (\ref{postcond}). But (\ref{postcond}) has
the same structure as the original formula (\ref{eq:25}) for
$\pi_{it}$, and thus we see that the model exhibits dynamic consistency.
\section{Simulation of defaultable bond-price processes}\label{sec2.7}
The model introduced in the previous sections allows for a simple
simulation methodology for the dynamics of defaultable bonds. In
the case of a defaultable discount bond all we need is to simulate
the dynamics of $\{\xi_t\}$. For each ``run" of the simulation we
choose at random a value for $H_T$ (in accordance with the {\it a
priori} probabilities), and a sample path for the Brownian bridge.
That is to say, each simulation corresponds to a choice of
$\omega\in\Omega$, and for each such choice we simulate the path
\begin{equation}
\xi_t(\omega)=\sigma t H_T(\omega)+ \beta_{tT}(\omega)
\end{equation}
for $t\in[0,T]$. One convenient way to simulate a Brownian bridge
is to write
\begin{equation}
\beta_{tT}=B_t-\frac{t}{T}B_T,
\end{equation}
where $\{B_t\}$ is a standard Brownian motion. It is
straightforward to verify that if $\{\beta_{tT}\}$ is defined this
way then it has the correct auto-covariance. Since the bond price
at time $t$ is expressed as a function of the random variable
$\xi_t$, this means that a path-wise simulation of the bond price
trajectory is feasible for any number of recovery levels.

The parameter $\sigma$ governs the ``speed" with which the bond
price converges to its terminal value. This can be seen as
follows. We return to the case of a binary discount bond with the
possible payoffs $h_0$ and $h_1$. Suppose, for example in a given
``run" of the simulation, the ``actual" value of the payout is
$H_T=h_0$. In that case we have
\begin{equation}
\xi_t=\sigma h_0 t +\beta_{tT},
\end{equation}
and thus by use of expression (\ref{eq:HtT}) for $\{H_{tT}\}$ we
obtain
\begin{align}
&H_{tT}=\frac{p_0 h_0\exp\left[\frac{T}{T-t}(\sigma h_0\beta_{tT}+
\frac{1}{2}\sigma^2 h_0^2 t)\right]+p_1h_1\exp\left[\frac{T}{T-t}
(\sigma h_1\beta_{tT} + \sigma h_0 h_1 t - \frac{1}{2} \sigma^2
h_1^2 t)\right]}{p_0\exp\left[\frac{T}{T-t} (\sigma h_0 \beta_{tT}
+ \frac{1}{2} \sigma^2 h_0^2 t)\right]+p_1 \exp\left[\frac{T}{T-t}
(\sigma h_1\beta_{tT} + \sigma h_0 h_1 t - \frac{1}{2} \sigma^2
h_1^2 t)\right]}.\nn\\
&\phantom{x}\label{eq:2.8}
\end{align}
Next, we divide the numerator and the denominator of this formula
by the coefficient of $p_0 h_0$. After some re-arrangement of
terms we get
\begin{eqnarray}
H_{tT}=\frac{p_0 h_0+p_1h_1\exp\left[-\frac{T}{T-t} \left(
\frac{1}{2} \sigma^2 (h_1-h_0)^2 t - \sigma (h_1-h_0)\beta_{tT}
\right)\right]}{p_0 +p_1 \exp\left[-\frac{T}{T-t} \left(
\frac{1}{2} \sigma^2 (h_1-h_0)^2 t - \sigma (h_1-h_0)\beta_{tT}
\right)\right]}.
\end{eqnarray}
Inspection of this expression shows that the convergence of the
bond price to the value $h_0$ is {\it exponential}. We note,
further, in line with the heuristic arguments in Section
\ref{sec2.2} concerning $\tau_D$, that the parameter
$\sigma^2(h_1-h_0)^2$ governs the speed at which the defaultable
discount bond converges to its destined terminal value. In
particular, if the {\it a priori} probability of no default is
high (say, $p_1\approx1$), and if $\sigma$ is very small, and if
$H_T=h_0$, then it will only be when $t$ is close to $T$ that
serious decay in the value of the bond price will set in.

In the figures that follow shortly, we present some sample
trajectories of the defaultable bond price process for various
values of $\sigma$ (I am grateful to I. Buckley for assistance
with the preparation of the figures). Each simulation is composed
of ten sample trajectories where the sample of the underlying
Brownian motion is the same for all paths. For all simulations we
have chosen the following values: the defaultable bond's maturity
is five years, the default-free interest rate system has a
constant short rate of 0.05, and the {\it a priori} probability of
default is set at 0.2. The object of these simulations is to
analyse the effect on the price process of the bond when the
information flow rate is increased. Each set of four figures shows
the trajectories for a range of information flow rates from a low
rate ($\sigma=0.04$) up to a high rate ($\sigma=5$).

The first four figures relate to the situation where two
trajectories are destined to default $(H_T=0)$ and the other eight
refer to the no-default case $(H_T=1)$. Figure 2.1 shows the case
where market investors have very little information
($\sigma=0.04$) about the future cash flow $H_T$ until the end of
the bond contract. Only in the last year or so, investors begin to
obtain more and more information when the noise process dies out
as the maturity is approached. In this simulation we see that
default comes as a surprise, and that investors have no chance to
anticipate the default.

In Figure 2.3, by way of contrast, the
information flow rate is rather high ($\sigma=1$) and already
after one year the bond price process starts to react strongly to
the high rate of information release. The interpretation is that
investors adjust their positions in the bond market according to
the amount of genuine information accessible to them and as a
consequence the volatility of the price process increases until
the signal term in the definition of the information process
dominates the noise produced by the bridge process.

In Figures 2.5-2.8 and Figures 2.9-2.12 we separate the
trajectories destined to not to default ($H_T=1$) from those that
will end in a state of default ($H_T=0$). As long as the
information rate is kept low the price process keeps its
stochasticity. If $\sigma$ is high, as in Figure 2.7 and in Figure
2.8, the trajectories become increasingly deterministic.

This is an expected phenomenon if one recalls that the default-free term
structure $P_{tT}$ is assumed to be deterministic. In other words,
the market participants, with much genuine information about the
future cash flow $H_T$ defining the credit-risky asset, will in
this case trade the defaultable bond similarly to the
credit-risk-free discount bond, making the price of the
defaultable bond approach that of a credit-risk-free bond.

The simulations referring to the case where the cash flow $H_T$ is
zero at the bond maturity manifest rather interesting features and
scenarios that are very much linked to episodes occurred in
financial markets. For instance, Figure 2.9 can be associated with
the crises at Parmalat and Swissair. Both companies had the
reputation to be reliable and financially robust until, rather as
a surprise, it was announced that they were not able to honour
their debts. Investors had very little genuine information about
the payoffs connected to the two firms, and the asset prices
reacted only at the last moment with a large drop in value.

An example in which there were earlier omens that a default might be
imminent is perhaps Enron's. The company seemed to be doing well for
quite some time until it became apparent that a continuous and
gradual deterioration in the company's finances had arisen that
eventually led to a state of default. This example would correspond
more closely with Figure 2.10 where the bond price is stable for the
first three and a half years but then commences to fluctuate,
reaching a very high volatility following the augmented amount of
information related to the increased likelihood of the possibility
of a payment failure.

The even more dramatic case in Figure 2.12 can be associated with the example of a new credit card holder who, very soon after receipt of his card, is not able
to pay his loan back, perhaps due to irresponsibly high
expenditures during the previous month.

\begin{figure}[H]
\begin{center}
\includegraphics[width=13cm]{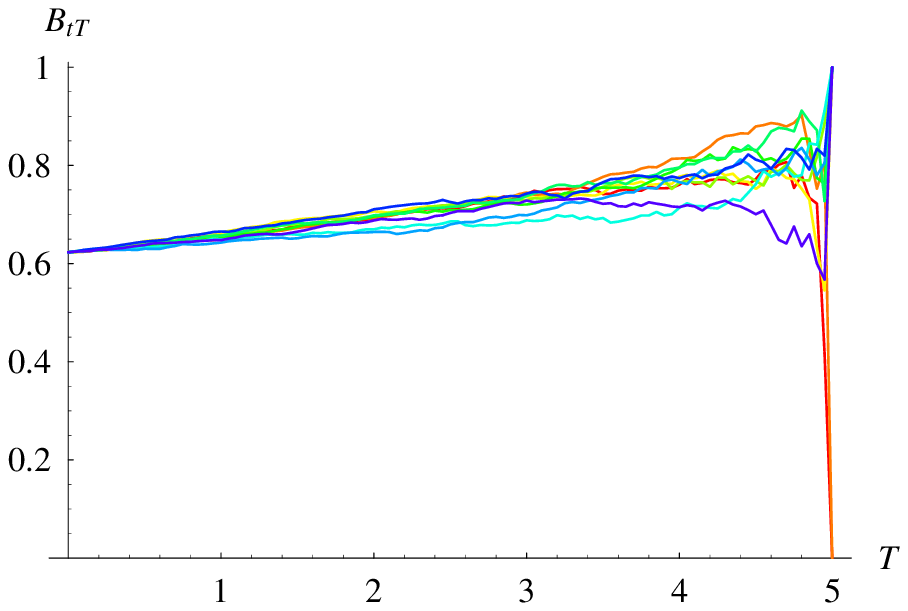}
\end{center}
\caption{Bond price process: $\sigma=0.04$. Two paths are
conditional on default $(H_T=0)$ and eight paths are conditional
on no-default $(H_T=0)$. The maturity of the bond is five years,
the default-free interest rate is constant at 5\%, the {\it a
priori} probability of default is 20\%, and the information flow
rate is 0.04.}
\begin{center}
\includegraphics[width=13cm]{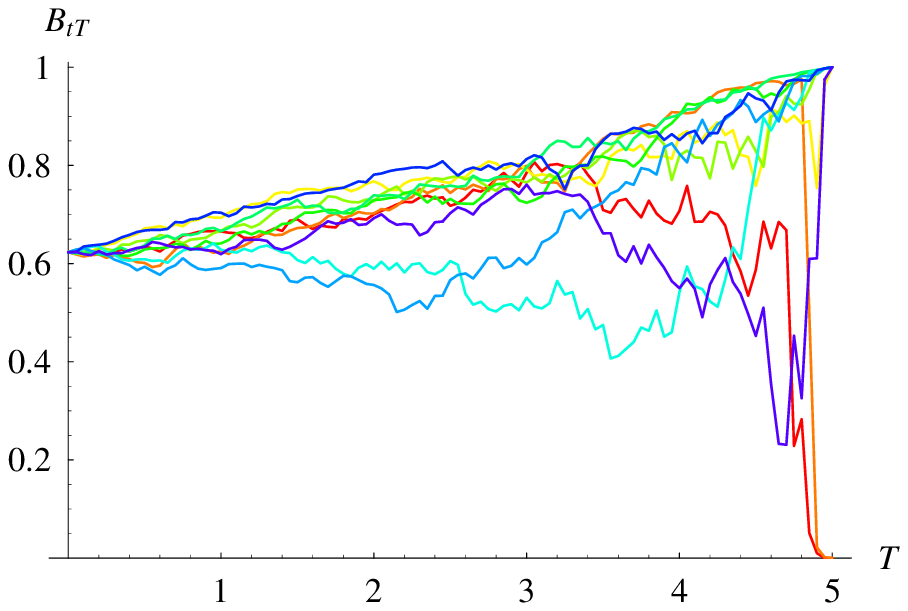}
\end{center}
\caption{Bond price process: $\sigma=0.2$. Two paths are
conditional on default $(H_T=0)$ and eight paths are conditional
on no-default $(H_T=0)$. The maturity of the bond is five years,
the default-free interest rate is constant at 5\%, the {\it a
priori} probability of default is 20\%, and the information flow
rate is 0.2.}
\end{figure}
\begin{figure}[H]
\begin{center}
\includegraphics[width=13cm]{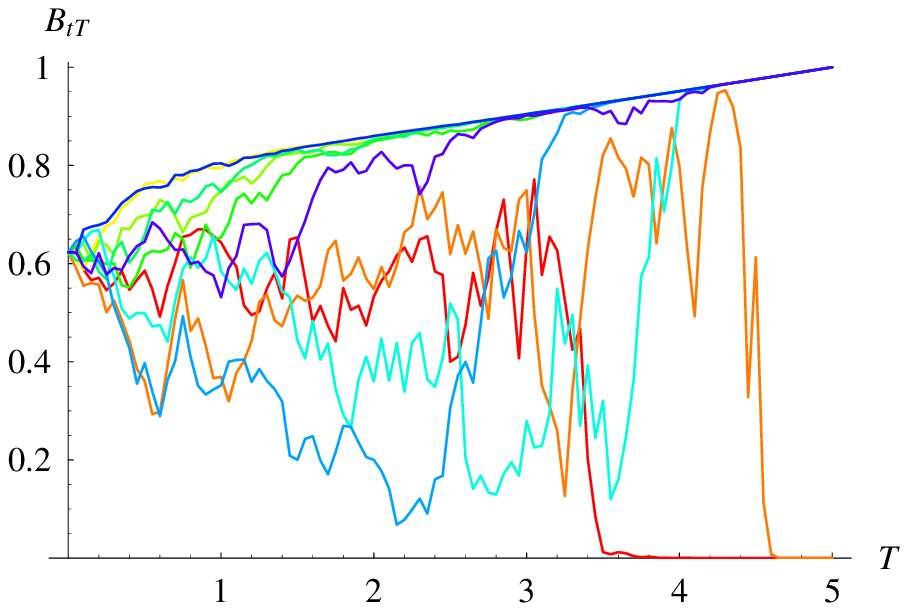}
\end{center}
\caption{Bond price process: $\sigma=1$. Two paths are conditional
on default $(H_T=0)$ and eight paths are conditional on no-default
$(H_T=0)$. The maturity of the bond is five years, the
default-free interest rate is constant at 5\%, the {\it a priori}
probability of default is 20\%, and the information flow rate is
1.0.}
\begin{center}
\includegraphics[width=13cm]{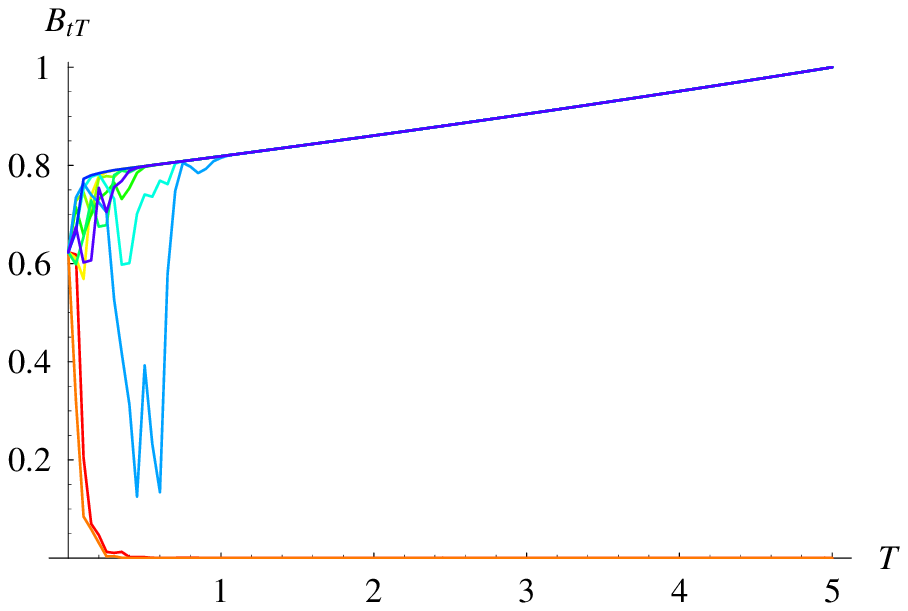}
\end{center}
\caption{Bond price process: $\sigma=5$. Two paths are conditional
on default $(H_T=0)$ and eight paths are conditional on no-default
$(H_T=0)$. The maturity of the bond is five years, the
default-free interest rate is constant at 5\%, the {\it a priori}
probability of default is 20\%, and the information flow rate is
5.0.}
\end{figure}
\begin{figure}[H]
\begin{center}
\includegraphics[width=13cm]{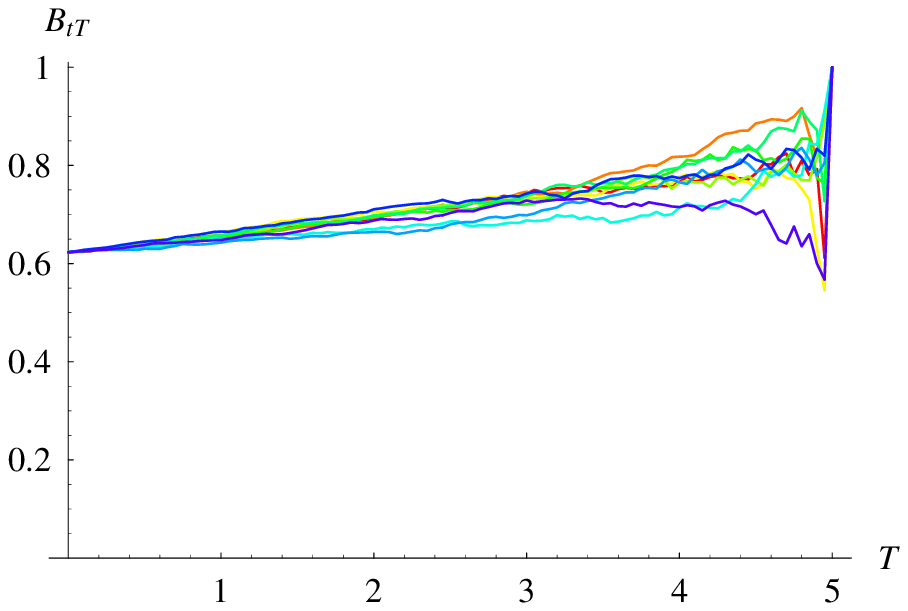}
\end{center}
\caption{Bond price process: $\sigma=0.04$. All paths are
conditional on no-default $(H_T=1)$. The maturity of the bond is
five years, the default-free interest rate is constant at 5\%, the
{\it a priori} probability of default is 20\%, and the information
flow rate is 0.04.}
\begin{center}
\includegraphics[width=13cm]{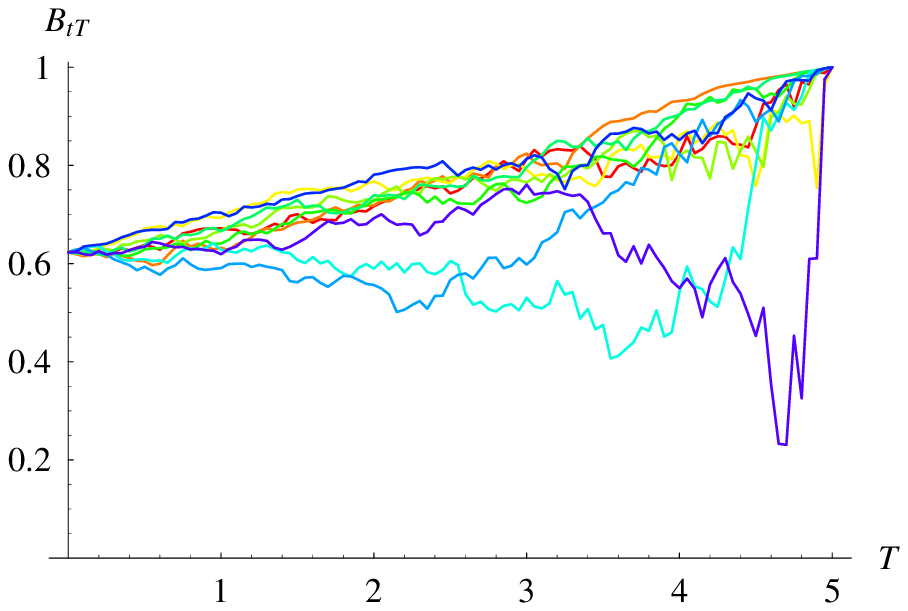}
\end{center}
\caption{Bond price process: $\sigma=0.2$. All paths are
conditional on no-default $(H_T=1)$. The maturity of the bond is
five years, the default-free interest rate is constant at 5\%, the
{\it a priori} probability of default is 20\%, and the information
flow rate is 0.2.}
\end{figure}
\begin{figure}[H]
\begin{center}
\includegraphics[width=13cm]{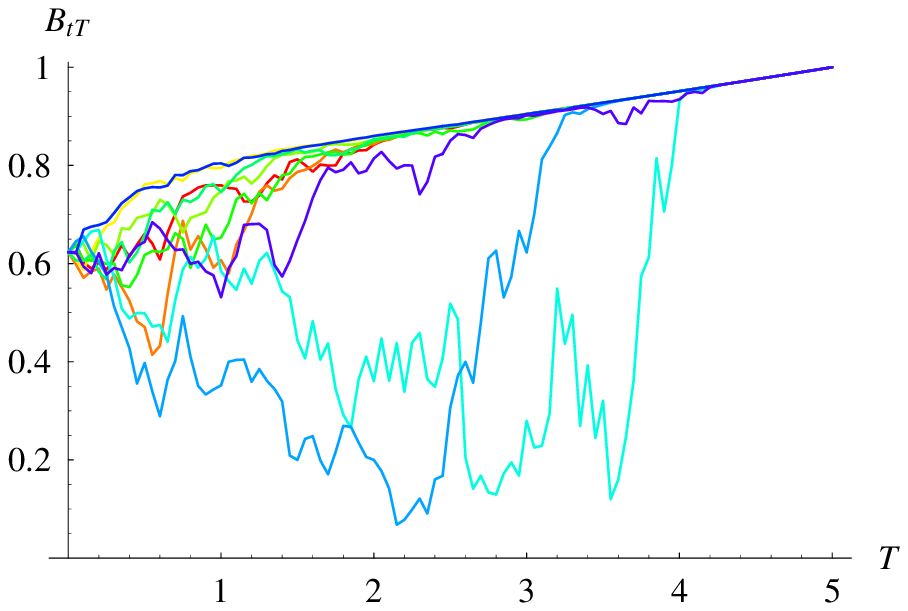}
\end{center}
\caption{Bond price process: $\sigma=1$. All paths are conditional
on no-default $(H_T=1)$. The maturity of the bond is five years,
the default-free interest rate is constant at 5\%, the {\it a
priori} probability of default is 20\%, and the information flow
rate is 1.0.}
\begin{center}
\includegraphics[width=13cm]{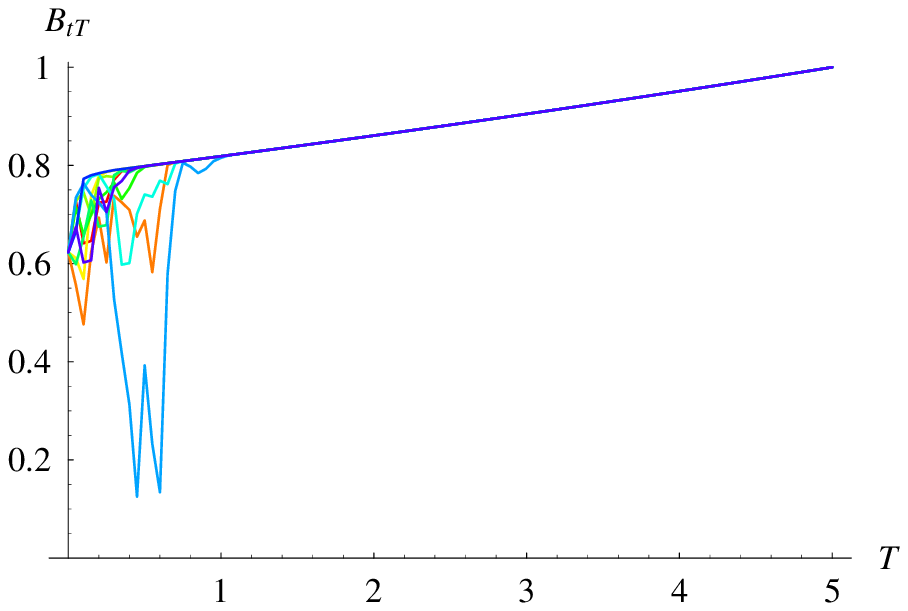}
\end{center}
\caption{Bond price process: $\sigma=5$. All paths are conditional
on no-default $(H_T=1)$. The maturity of the bond is five years,
the default-free interest rate is constant at 5\%, the {\it a
priori} probability of default is 20\%, and the information flow
rate is 5.0.}
\end{figure}
\begin{figure}[H]
\begin{center}
\includegraphics[width=13cm]{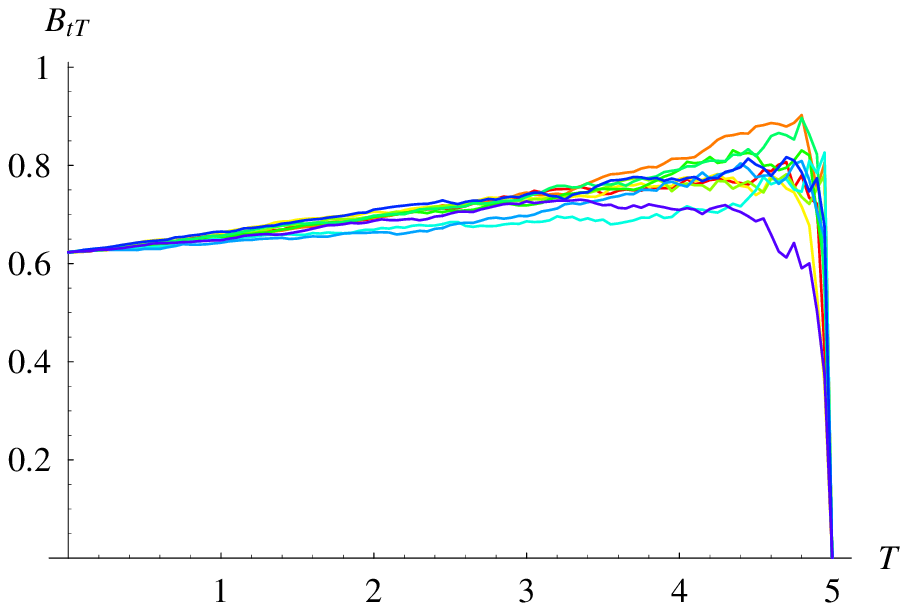}
\end{center}
\caption{Bond price process: $\sigma=0.04$. All paths are
conditional on default $(H_T=0)$. The maturity of the bond is five
years, the default-free interest rate is constant at 5\%, the {\it
a priori} probability of default is 20\%, and the information flow
rate is 0.04.}
\begin{center}
\includegraphics[width=13cm]{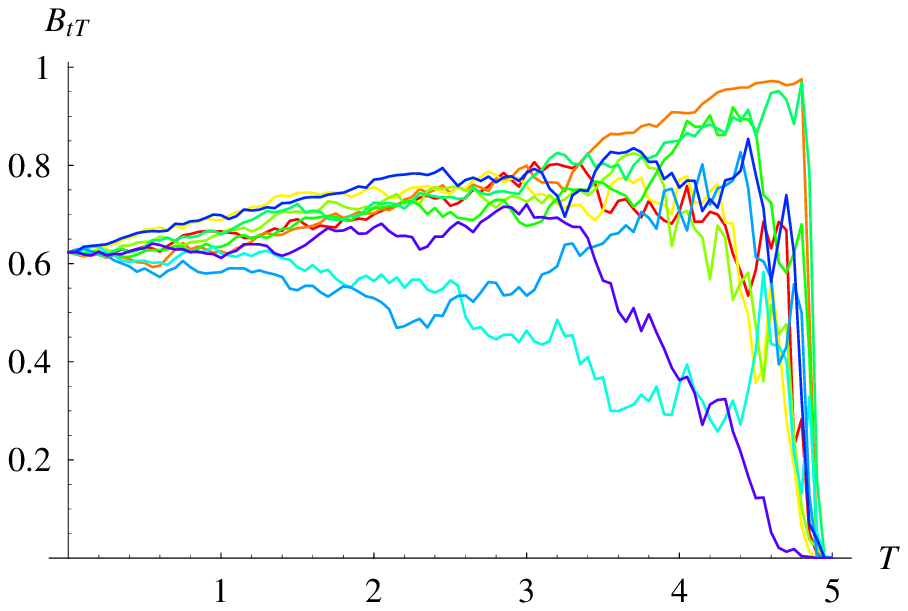}
\end{center}
\caption{Bond price process: $\sigma=0.2$. All paths are
conditional on default $(H_T=0)$. The maturity of the bond is five
years, the default-free interest rate is constant at 5\%, the {\it
a priori} probability of default is 20\%, and the information flow
rate is 0.2.}
\end{figure}
\begin{figure}[H]
\begin{center}
\includegraphics[width=13cm]{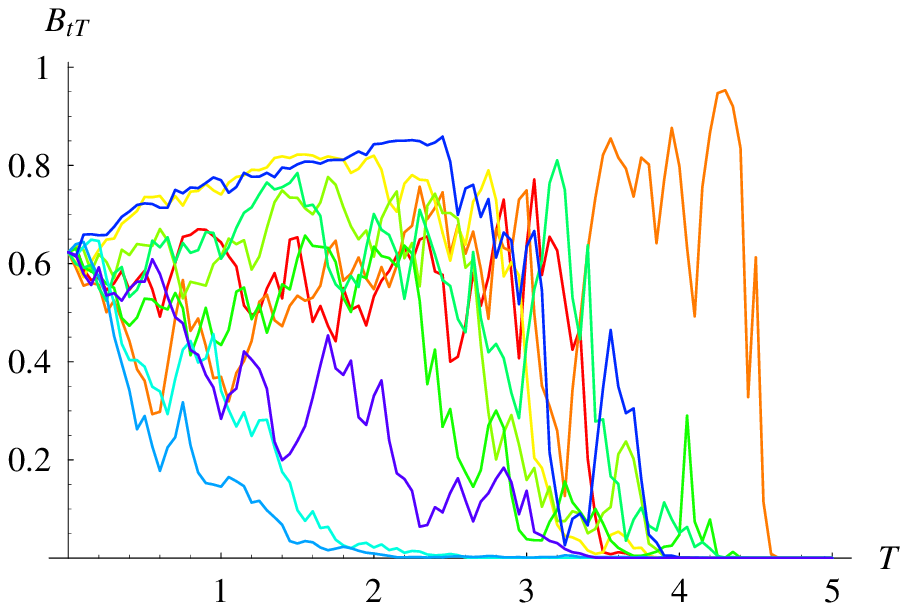}
\end{center}
\caption{Bond price process: $\sigma=1$. All paths are conditional
on default $(H_T=0)$. The maturity of the bond is five years, the
default-free interest rate is constant at 5\%, the {\it a priori}
probability of default is 20\%, and the information flow rate is
1.0.}
\begin{center}
\includegraphics[width=13cm]{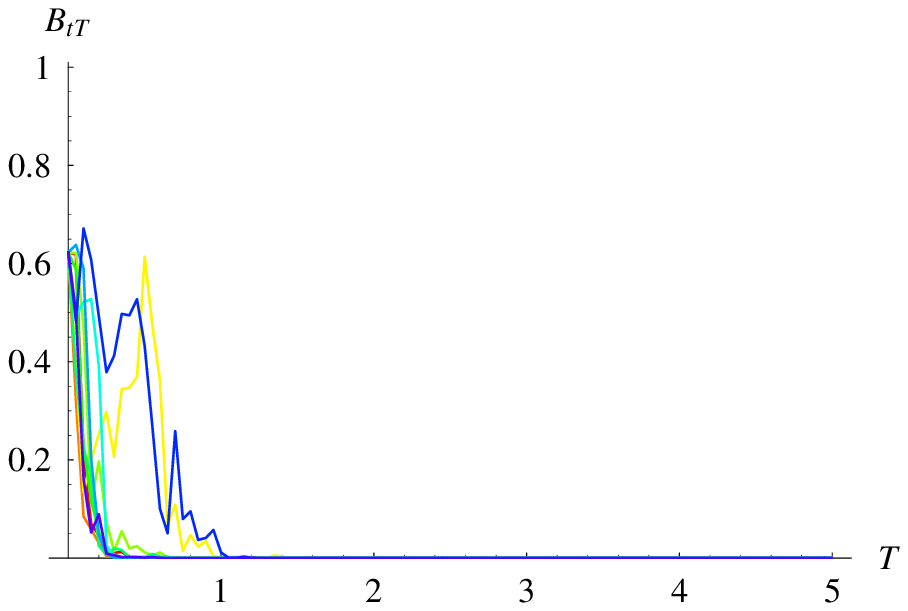}
\end{center}
\caption{Bond price process: $\sigma=5$. All paths are conditional
on default $(H_T=0)$. The maturity of the bond is five years, the
default-free interest rate is constant at 5\%, the {\it a priori}
probability of default is 20\%, and the information flow rate is
5.0.}
\end{figure}

\chapter{Options on credit-risky discount bonds}\label{chap3}
%
\section{Pricing formulae for bond options}\label{sec:3.2}
In this chapter we consider the pricing of options on credit-risky
bonds. In particular, we look at credit-risky discount bonds. As
we shall demonstrate, in the case of a binary bond there is an
exact solution for the valuation of European-style vanilla
options. The resulting expression for the option price exhibits a
structure that is strikingly analogous to that of the
Black-Scholes option pricing formula.

We consider the value at time 0 of an option that is exercisable
at a fixed time $t>0$ on a credit-risky discount bond that matures
at time $T>t$. The value $C_0$ of a call option at time 0 is given
by
\begin{eqnarray}
C_0=P_{0t}{\mathbb E}\left[(B_{tT}-K)^+\right], \label{eq:4.1}
\end{eqnarray}
where $B_{tT}$ is the bond price on the option maturity date and
$K$ is the strike price. Inserting formula (\ref{eq:230}) for
$B_{tT}$ into the valuation formula (\ref{eq:4.1}) for the option,
and making use of (\ref{eq:25}), we obtain
\begin{eqnarray}
C_0&=&P_{0t}\,{\mathbb E} \left[ \left( P_{tT}H_{tT}-K \right)^+
\right] \nonumber \\ &=& P_{0t}\,{\mathbb E} \left[ \left(
\sum_{i=0}^n P_{tT} \pi_{it} h_i-K \right)^+\right],
\end{eqnarray}
which after some further re-arrangement can be written in the form
\begin{eqnarray}
C_0&=& P_{0t}\,{\mathbb E} \left[ \left(\frac{1}{\Phi_t}
\sum_{i=0}^n P_{tT}p_{it} h_i-K\right)^+\right] \nonumber \\ &=&
P_{0t}\,{\mathbb E} \left[ \frac{1}{\Phi_t} \left( \sum_{i=0}^n
\Big(P_{tT} h_i-K\Big) p_{it}\right)^+\right]. \label{eq:30}
\end{eqnarray}
Here the quantities $p_{it}$ $(i=0,1,\ldots,n)$ are the
``unnormalised" conditional probabilities, defined by
\begin{eqnarray}
p_{it}=p_i\exp\left[\frac{T}{T-t}\left(\sigma h_i \xi_t
-\tfrac{1}{2}\sigma^2 h_i^2 t\right)\right]. \label{eq:pi}
\end{eqnarray}
Then for the ``normalised" conditional probabilities we have
\begin{equation}
\pi_{it}=\frac{p_{it}}{\Phi_t}
\end{equation}
where
\begin{equation}
\Phi_t=\sum_i p_{it},
\end{equation}
or, more explicitly,
\begin{eqnarray}
\Phi_t=\sum_{i=0}^n p_i\exp\left[\frac{T}{T-t}\left(\sigma h_i
\xi_t -\tfrac{1}{2}\sigma^2 h_i^2 t\right)\right] . \label{eq:365}
\end{eqnarray}
Our plan now is to use the factor $\Phi^{-1}_t$ appearing in
(\ref{eq:30}) to make a change of probability measure on
$(\Omega,\F_t)$. To this end, we fix a time horizon $u$ beyond the
option expiration but before the bond maturity, so $t\le u<T$. We
define a process $\{\Phi_t\}_{0\le t\le u}$ by use of the
expression (\ref{eq:365}), where now we let $t$ vary in the range
$[0,u]$. We shall now work out the dynamics of $\{\Phi_t\}$

Let $\{p_{it}\}$, $i=1,\ldots,n$, be the un-normalised probability
density processes given by (\ref{eq:pi}). The dynamics of the
probability density process $\{p_{it}\}$ can be obtained by
applying Ito's Lemma to the expression (\ref{eq:pi}).
\begin{eqnarray}\label{pi_dyn}
\frac{\rd p_{it}}{p_{it}}&=&\left[\frac{T}{(T-t)^2}\left(\sigma h_i\xi_t-\frac{1}{2}\sigma^2 h_i^2 t\right)-\frac{1}{2}\frac{T}{T-t}\sigma^2 h_i^2\right]\rd t\nn\\
&+&\frac{T}{T-t}\sigma h_i\rd\xi_t+\frac{1}{2}\frac{T^2}{(T-t)^2}\sigma^2 h_i^2(\rd\xi_t)^2.
\end{eqnarray}
Recalling that for the dynamics of the information process
$\{\xi_t\}$ we have
\begin{equation}
\rd\xi_t=\frac{1}{T-t}(\sigma T H_{tT}-\xi_t)\rd t+\rd W_t,
\end{equation}
we see that $(\rd\xi_t)^2=\rd t$. Inserting these expressions for
$\rd\xi_t$ and $(\rd\xi_t)^2$ into the equation (\ref{pi_dyn}) we
find that
\begin{equation}
\rd p_{it}=\sigma^2\left(\frac{T}{T-t}\right)^2 h_i
p_{it}H_{tT}\rd t+\sigma\frac{T}{T-t}h_i p_{it}\rd W_t.
\end{equation}
Now we sum over $i$, recalling that
\begin{equation}
\sum_{i=0}^n h_i p_{it}=\Phi_t H_{tT}
\end{equation}
and
\begin{equation}
\sum_{i=0}^n p_{it}=\Phi_t.
\end{equation}
We thus obtain
\begin{equation}
\rd\Phi_t=\sigma^2\left(\frac{T}{T-t}\right)^2H_{tT}^2\Phi_t\rd t+\sigma\frac{T}{T-t}H_{tT}\Phi_t\rd W_t.
\end{equation}
It follows then by Ito's Lemma that
\begin{equation}
\rd\Phi^{-1}_t=-\sigma\frac{T}{T-t}H_{tT}\Phi^{-1}_t\rd W_t,
\end{equation}
and hence that
\begin{equation}\label{invPhi_Dyn}
\Phi^{-1}_t=\exp\left(-\sigma T\int^t_0\frac{1}{T-s}H_{sT}\rd W_s-\frac{1}{2}\sigma^2 T^2\int^t_0\frac{1}{(T-s)^2}H_{sT}^2\rd s\right).
\end{equation}
Since $\{H_{sT}\}$ is bounded, and $t\le u<T$, we see that the
process $\{\Phi^{-1}_t\}_{0\le t\le u}$ is a martingale. In
particular, since $\Phi_0=1$, we deduce that
$\E\left[\Phi^{-1}_t\right]=1$, where $t$ is the option maturity
date, and hence that the factor $\Phi^{-1}_t$ appearing in
(\ref{eq:365}) can be used to effect a change of measure on
$(\Omega,\F_t)$ as we had earlier indicated. Writing ${\mathbb
B}_T$ for the new probability measure thus defined, we have
\begin{eqnarray}
C_0=P_{0t}\,{\mathbb E}^{{\mathbb B}_T}\left[ \left( \sum_{i=0}^n
\Big(P_{tT}h_i-K\Big) p_{it}\right)^+\right]. \label{eq:345}
\end{eqnarray}
We shall call ${\mathbb B}_T$  the ``bridge" measure because it
has the property that it makes $\{\xi_s\}_{0\leq s\leq t}$ a
Gaussian process with mean zero and covariance
\begin{equation}
{\mathbb E}^{{\mathbb B}_T}[\xi_r \xi_s]=\frac{r(T-s)}{T}
\end{equation}
for $0\leq r\leq s\leq t$. In other words, with respect to the
measure ${\mathbb B}_T$, and over the interval $[0,t]$, the
information process has the law of a standard Brownian bridge over
the interval $[0,T]$. Armed with this fact, we can then proceed to
calculate the expectation in (\ref{eq:345}).

The proof that $\{\xi_s\}_{0\leq s\leq t}$ has the claimed
properties in the measure ${\mathbb B}_T$ is as follows. For
convenience we introduce a process
$\{W_t^*\}_{0\leq t\leq u}$ which we define as a Brownian motion
with drift in the ${\mathbb Q}$-measure:
\begin{eqnarray}
W_t^* = W_t + \sigma T \int_0^t \frac{1}{T-s}\, H_{sT}\, \rd s.
\end{eqnarray}
It is straightforward to check that on $(\Omega,\F_u)$ the process
$\{W_t^{\ast}\}_{0\le t\le u}$ is a Brownian motion with respect
to the measure defined by use of the density martingale
$\{\Phi_t^{-1}\}_{0\le t\le u}$ given by (\ref{invPhi_Dyn}). It
then follows from the definition of $\{W_t\}$ given in equation
(\ref{eq:24}) that
\begin{eqnarray}
W_t^* = \xi_t + \int_0^t \frac{1}{T-s}\, \xi_{s}\, \rd s .
\label{eq:39}
\end{eqnarray}
Taking the stochastic differential of each side of this relation,
we deduce that
\begin{eqnarray}
\rd\xi_t = -\frac{1}{T-t}\,\xi_t\, \rd t + \rd W_t^*.
\label{eq:36}
\end{eqnarray}
We note, however, that (\ref{eq:36}) is the stochastic
differential equation satisfied by a Brownian bridge (see, e.g.,
Karatzas \& Shreve 1991, and Protter 2004) over the interval
$[0,T]$. Thus we see that in the measure $\mathbb{B}^T$ defined on
$(\Omega, \F_t)$ the process $\{\xi_s\}_{0\le s\le t}$ has the
properties of a standard Brownian bridge over the interval
$[0,T]$, restricted to the period $[0,t]$. For the transformation
back from $\mathbb{B}^{T}$ to $\Q$ on $(\Omega,\F_u)$, the
appropriate density martingale $\{\Phi_t\}_{0\le t\le u}$ with
respect to $\mathbb{B}^T$ is given by:
\begin{eqnarray}
\Phi_t = \exp\left( \sigma T \int_0^t \frac{1}{T-s}\, H_{sT} {\rd}
\,W_s^* - \half \sigma^2 T^2 \int_0^t \frac{1}{(T-s)^2}\, H_{sT}^2
\,{\rd}s \right) . \label{eq:4.100}
\end{eqnarray}

The crucial point that follows from this analysis is that the
random variable $\xi_t$ is ${\mathbb B}_T$-Gaussian. In the case
of a binary discount bond, therefore, the relevant expectation for
determining the option price can be carried out by standard
techniques, and we are led to a formula of the Black-Scholes type.
In particular, for a binary bond, equation (\ref{eq:345}) reads
\begin{eqnarray}
C_0 = P_{0t} {\mathbb E}^{{\mathbb B}_T} \left[ \Big(
(P_{tT}h_1-K)p_{1t} + (P_{tT}h_0-K)p_{0t}\Big)^+ \right],
\label{eq:999}
\end{eqnarray}
where $p_{0t}$ and $p_{1t}$ are given by
\begin{equation}
p_{0t}=p_0 \exp\left[ {\textstyle\frac{T}{T-t}}\left(\sigma h_0
\xi_t- \half \sigma^2 h_0^2 t\right) \right],
\end{equation}
and
\begin{equation}
p_{1t}=p_1\exp\left[ {\textstyle\frac{T}{T-t}} \left(\sigma h_1
\xi_t- \half \sigma^2 h_1^2 t\right) \right].
\end{equation}
To compute the value of (\ref{eq:999}) there are essentially three
different cases that have to be considered:
\begin{align}
&\hspace{-5cm}(1)\qquad P_{tT}h_1>P_{tT}h_0>K \nn\\
&\hspace{-5cm}(2)\qquad K>P_{tT}h_1>P_{tT}h_0 \nn\\
&\hspace{-5cm}(3)\qquad P_{tT}h_1>K>P_{tT}h_0 \nn
\end{align}

In case (1) the option is certain to expire in the money. Thus,
making use of the fact that $\xi_t$ is ${\mathbb B}_T$-Gaussian
with mean zero and variance $t(T-t)/T$, we see that ${\mathbb
E}^{{\mathbb B}_T}[p_{it}]=p_i$; hence in case (1) we have
$C_0=B_{0T}-P_{0t}K$.

In case (2) the option expires out of the money, and thus $C_0=0$.

In case (3) the option can expire in or out of the money, and
there is a ``critical" value of $\xi_t$ above which the argument
of (\ref{eq:999}) is positive. This is obtained by setting the
argument of (\ref{eq:999}) to zero and solving for $\xi_t$.
Writing ${\bar\xi}_t$ for the critical value, we find that
${\bar\xi}_t$ is determined by the relation
\begin{eqnarray}
\frac{T}{T-t}\sigma(h_1-h_0){\bar\xi}_t = \ln \left[
\frac{p_0(P_{tT}h_0-K)}{p_1(K-P_{tT}h_1)}\right]+\half\sigma^2
(h_1^2-h_0^2)\tau,
\end{eqnarray}
where
\begin{equation}
\tau=\frac{tT}{(T-t)}.
\end{equation}
Next we note that since $\xi_t$ is ${\mathbb B}_T$-Gaussian with
mean zero and variance $t(T-t)/T$, for the purpose of computing
the expectation in (\ref{eq:999}) we can set
\begin{equation}
\xi_t=Z\sqrt{\frac{t(T-t)}{T}},
\end{equation}
where $Z$ is ${\mathbb B}_T$-Gaussian with zero mean and unit
variance. Then writing ${\bar Z}$ for the corresponding critical
value of $Z$, we obtain
\begin{eqnarray}
{\bar Z} = \frac{\ln \left[ \frac{p_0(K-P_{tT}h_0)}
{p_1(P_{tT}h_1-K)} \right]+\half\sigma^2 (h_1^2-h_0^2)\tau}{\sigma
\sqrt{\tau}(h_1-h_0)}.
\end{eqnarray}
With this expression at hand, we calculate the expectation in
(\ref{eq:999}). The result is:
\begin{eqnarray}\label{Bridgecallprice}
C_0=P_{0t}\Big[ p_1(P_{tT}h_1-K)N(d^+)-p_0 (K-P_{tT}h_0) N(d^-)
\Big].
\end{eqnarray}
Here $d^+$ and $d^-$ are defined by
\begin{eqnarray}
d^{\pm}=\frac{\ln\left[\frac{p_1(P_{tT}h_1-K)}
{p_0(K-P_{tT}h_0)}\right]\pm\tfrac{1}{2}\sigma^2
(h_1-h_0)^2\tau}{\sigma\sqrt{\tau}(h_1-h_0)}.
\end{eqnarray}

It is interesting to note that the information flow-rate parameter
$\sigma$ plays a role like that of the volatility parameter in the
Black-Scholes model. The more rapidly information is ``leaked out"
about the ``true" value of the bond repayment, the higher the
volatility.

We remark that in the more general case for which there are
multiple recovery levels, a semi-analytic result can be obtained
that, for practical purposes, can be regarded as fully tractable.
In particular, starting from (\ref{eq:345}) we consider the case
where the strike price $K$ lies in the range
\begin{equation}
P_{tT}h_{k+1} > K > P_{tT}h_k
\end{equation}
for some value of $k\in\{0,1,\ldots,n\}$. It is an exercise to
verify that there exists a unique critical value of $\xi_t$ such
that the summation appearing in the argument of the $\max(x,0)$
function in (\ref{eq:345}) vanishes. Writing ${\bar \xi}_t$ for
the critical value, which can be obtained by numerical methods, we
define the scaled critical value ${\bar Z}$ as before by setting
\begin{equation}
{\bar\xi}_t={\bar Z}\sqrt{\frac{t(T-t)}{T}}.
\end{equation}
A calculation then shows that the option price is given by the
following formula:
\begin{eqnarray}
C_0 = P_{0t} \sum_{i=0}^n p_i\left( P_{tT}h_i-K\right)\, N(\sigma
h_i \sqrt{\tau}-{\bar Z}).
\end{eqnarray}
\section{Model input sensitivity analysis}\label{sec:3.3}
It is straightforward to verify that the option price has a
positive ``vega", i.e. that $C_0$ is an increasing function of
$\sigma$. This means in principalthat we can use bond option
prices (or, equivalently, the prices of caps and floors) to back
out an implied value for $\sigma$, and hence to calibrate the
model. Normally the term ``vega" is used in the Black-Scholes
theory to characterise the sensitivity of the option price to a
change in volatility; here we use the term analogously to denote
sensitivity with respect to the information flow-rate parameter.
Thus, writing
\begin{equation}
{\mathcal V}=\frac{\partial C_0}{\partial\sigma}
\end{equation}
for the option vega, after a calculation we obtain the following
positive expression:
\begin{eqnarray}\label{eq:vega}
{\mathcal V}=\frac{1}{\sqrt{2\pi}}\, \re^{-rt-\frac{1}{2}A}
(h_1-h_0) \sqrt{\tau p_0p_1(P_{tT}h_1-K)(K-P_{tT}h_0)},
\end{eqnarray}
where
\begin{eqnarray}
A=\frac{1}{\sigma^2\tau(h_1-h_0)^2} \ln^2\left[ \frac{p_1
(P_{tT}h_1-K)}{p_0(K-P_{tT}h_0)}\right] + \frac{1}{4}\sigma^2\tau
(h_1-h_0)^2.
\end{eqnarray}

Another interesting and important feature of this model is the
possibility to hedge an option position against moves in the
underlying asset by holding a position in the credit-risky bond.
The number of bond unites needed to hedge a short position in a
call option is given by the option delta, which is defined by
\begin{equation}\label{eq33}
\Delta=\frac{\partial C_0}{\partial B_{0T}}.
\end{equation}
To calculate the option delta at time zero we need to express the
initial call option value
\begin{eqnarray}\label{eq34}
C_0=P_{0t}\Big[ p_1(P_{tT}h_1-K)N(d^+)-p_0 (K-P_{tT}h_0)
N(d^-)\Big]
\end{eqnarray}
in terms of the initial value of the binary bond
\begin{equation}
B_{0T}=P_{0T}(p_0 h_0+p_1 h_1).
\end{equation}
For this purpose we substitute the {\it a priori} probabilities
$p_0$ and $p_1$ in the expression of the call option (\ref{eq34})
by setting
\begin{equation}\label{eq35}
p_0 = \frac{1}{h_1-h_0}\left(h_1-\frac{B_{0T}}{P_{0T}}\right),
\quad p_1 = \frac{1}{h_1-h_0}\left( \frac{B_{0T}}{P_{0T}}
-h_0\right).
\end{equation}
The call price now reads:
\begin{eqnarray}\label{eq37}
\nonumber C_0&=&P_{0t}\Bigg[
\frac{1}{h_1-h_0}\left(\frac{B_{0T}}{P_{0T}}-h_0\right)(P_{tT}h_1-K)N(d^+)\\
&-&\frac{1}{h_1-h_0}\left(h_1-\frac{B_{0T}}{P_{0T}}\right)
(K-P_{tT}h_0) N(d^-)\Bigg].
\end{eqnarray}
To obtain the option delta we need to differentiate (\ref{eq37})
with respect to $B_{0T}$, taking into account also the dependence
of $d^+$ and $d^-$ on $B_{0T}$. The result is given as follows:
\begin{equation}\label{eq36}
\Delta=\frac{(P_{tT}h_1-K)N(d^+)+(K-P_{tT}h_0)N(d^-)}{P_{tT}(h_1-h_0)}.
\end{equation}

\section{Bond option price processes}\label{sec:3.4}
In the Section \ref{sec:3.2} we obtained the initial value $C_0$
of an option on a binary credit-risky bond. In the present section
we extend this calculation to determine the price process of such
an option. We fix the bond maturity $T$ and the option maturity
$t$. Then the price $C_s$ of a call option at time $s\leq t$ is
given by
\begin{eqnarray}
C_s &=& P_{st}\,{\mathbb E}\left[(B_{tT}-K)^+ |{\mathcal F}_s
\right] \nonumber \\ &=& \frac{P_{st}}{\Phi_s}\,{\mathbb
E}^{{\mathbb B}_T} \left[\Phi_t (B_{tT}-K)^+ | {\mathcal F}_s
\right] \nonumber \\ &=& \frac{P_{st}}{\Phi_s}\,{\mathbb
E}^{{\mathbb B}_T}\left[\left. \left( \sum_{i=0}^n \Big(P_{tT}h_i-
K\Big) p_{it}\right)^+ \right| {\mathcal F}_s\right] .
\label{eq:option}
\end{eqnarray}
We recall that $p_{it}$, defined in (\ref{eq:pi}), is a function
of $\xi_t$. The calculation can thus be simplified by use of the
fact that $\{\xi_t\}$ is a ${\mathbb B}_T$-Brownian bridge. To
determine the conditional expectation (\ref{eq:option}) we note
that the ${\mathbb B}_T$-Gaussian random variable $Z_{st}$ defined
by
\begin{eqnarray}
Z_{st} = \frac{\xi_t}{T-t}-\frac{\xi_s}{T-s}
\end{eqnarray}
is independent of $\{\xi_u\}_{0\leq u\leq s}$. We can then express
$\{p_{it}\}$ in terms of $\xi_s$ and $Z_{st}$ by writing
\begin{eqnarray}
p_{it}=p_i\exp\left[\frac{T}{T-s}\sigma h_i T\xi_s - \half \frac{
T}{T-t}\sigma^2 h_i^2 t  +\sigma h_i Z_{st}T\right].
\label{eq:pi2}
\end{eqnarray}
Substituting (\ref{eq:pi2}) into (\ref{eq:option}), we find that
$C_s$ can be calculated by taking an expectation involving the
random variable $Z_{st}$, which has mean zero and variance
$v_{st}^2$ given by
\begin{eqnarray}
v_{st}^2 = \frac{t-s}{(T-t)(T-s)}.
\end{eqnarray}

In the case of a call option on a binary discount bond that pays
$h_0$ or $h_1$, we can obtain a closed-form expression for
(\ref{eq:option}). In that case the option price at time $s$ is
given by the following expectation:
\begin{eqnarray}
C_s = \frac{P_{st}}{\Phi_s}\,{\mathbb E}^{{\mathbb B}_T} \Big[
\big( (P_{tT}h_0-K)p_{0t} + (P_{tT}h_1-K)p_{1t}
\big)^+\,\vert\,\F_s \Big]. \label{eq:option2}
\end{eqnarray}
Substituting (\ref{eq:pi2}) in (\ref{eq:option2}) we find that the
expression in the expectation is positive only if the inequality
$Z_{st}>{\bar Z}$ is satisfied, where
\begin{eqnarray}
{\bar Z} = \frac{\ln\left[ \frac {\pi_{0s}(K-P_{tT}h_0)}
{\pi_{1s}(P_{tT}h_1-K)} \right]+\half \sigma^2 (h_1^2-h_0^2)
v_{st}^2 T}{\sigma(h_1-h_0)v_{st}T}.
\end{eqnarray}
It will be convenient to set
\begin{equation}
Z_s=v_{st} Z,
\end{equation}
where $Z$ is a ${\mathbb B}_T$-Gaussian random variable with zero
mean and unit variance. The computation of the expectation in
(\ref{eq:option2}) then reduces to a pair of Gaussian integrals,
and we obtain the following result:

\begin{prop}
Let $\{C_s\}_{0\le s\le t}$ denote the price process of a
European-style call option on a defaultable bond. Let $t$ denote
the option expiration date, let $K$ denote the strike price, and
let $T$ denote the bond maturity date. Then the option price at
time $s\in[0,t]$ is given by:
\begin{eqnarray}
C_s = P_{st}\Big[ \pi_{1s}\left( P_{tT}h_1-K\right) N(d_s^+) -
\pi_{0s}\left(K- P_{tT}h_0\right) N(d_s^-) \Big],
\end{eqnarray}
where the conditional probabilities $\{\pi_{is}\}$ are as defined
in \textrm{(\ref{eq:25})}, and
\begin{eqnarray}
d_s^{\pm}=\frac{\ln\left[\frac{\pi_{1s}\left(P_{tT}h_1 -K\right)}
{\pi_{0s}\left(K-P_{tT}h_0\right)} \right] \pm \tfrac{1}{2}
\sigma^2 v_{st}^2 T^2(h_1-h_0)^2} {\sigma v_{st} T(h_1-h_0)}.
\end{eqnarray}
\end{prop}

\begin{rem}
We note that $d_s^+ = d_s^- + \sigma v_{st} T(h_1-h_0)$, and that
$d_0^\pm = d^\pm$.
\end{rem}

One particularly attractive feature of the model worth pointing
out in the present context is that delta-hedging is possible. This
is because the option price process and the underlying bond price
process are one-dimensional diffusions driven by the same Brownian
motion. Since $C_t$ and $B_{tT}$ are both monotonic in their
dependence on $\xi_t$, it follows that $C_t$ can be expressed as a
function of $B_{tT}$; the delta of the option can then be defined
in the conventional way as the derivative of the option price with
respect to the value of the underlying. At time 0 this reduces to
the expression we developed earlier.

This brings us to another interesting point. For certain types of
instruments it may be desirable to model the occurrence of credit
events taking place at some time that precedes a cash-flow date.
In particular, we may wish to consider contingent claims based on
such events. In the present framework we can regard such
contingent claim as derivative structures for which the payoff is
triggered by the level of $\xi_t$. For example, it may be that a
credit event is established if $B_{tT}$ drops below some specific
level, or if the credit spread widens beyond some threshold. For
that reason we see that the consideration of a barrier option
becomes an important issue, where both the payoff and the barrier
level are expressed in terms of the information process.
\section{Arrow-Debreu technique and information derivatives}\label{sec:3.5}
In this section we consider an alternative method for option
pricing in the information-based framework. We concentrate on the
case where the underlying asset pays a single cash flow at the
maturity date $T$. We sketch the main ideas behind this
alternative method, which is based on the concept of an
Arrow-Debreu security. In section \ref{sec6.8} we then present an
extension of this technique to the case of an underlying asset for
which the dividends can be regarded as continuous random
variables.

In the present section we introduce also a new class of contingent
claims, which we call ``information derivatives". This type of
security is defined by a payoff that is a function of the value of
the information process. In other words the underlying of such a
derivative is the information available to market participants.
The price of an information derivative is given by the
risk-neutral pricing formula, that is:
\begin{equation}\label{infoderiv}
S_0=P_{0t}\E^{\Q}\left[f(\xi_t)\right],
\end{equation}
where $P_{0t}$ is the discount function, and $f(\xi_t)$ is the
payoff function. Here the derivative's maturity is denoted $t$. To
begin with we consider the elementary information security defined
by the following payoff:
\begin{equation}\label{eispayoff}
    f(\xi_t)=\delta(\xi_t-x).
\end{equation}
Let us write $A_{0t}(x)$ for the price at time 0 of such a
contract. Without the introduction of a great deal of additional
mathematics, the treatment of distribution-valued random variables
will have to be somewhat heuristic in what follows; but in
practice this causes no problems. In order to work out the price
of an elementary information security we use the standard Fourier
representation of the delta function, namely:
\begin{equation}\label{deltadef}
    \delta(\xi_t-x)=\frac{1}{2\pi}\int^{\infty}_{-\infty}\textrm{e}^{\textrm{i}(\xi_t-x)\kappa}\,\rd\kappa.
\end{equation}
As with all distributional expressions, this formula acquires its
meaning from the context in which it is used. Here we recall that,
conditional on a specific value of the random variable $H_T$, the
information process $\{\xi_t\}$ is normally distributed. This
ensures, along with the fact that the random variable $H_T$ can
take only a finite number of different states, that the integral
in (\ref{deltadef}) and the expectation in (\ref{infoderiv}) can
be in effect interchanged. Thus we get,
\begin{eqnarray}\label{eisprice}
A_{0t}(x)&=&P_{0t}\,\E^{\Q}[\delta(\xi_t-x)]\nn\\
         &=&P_{0t}\,\frac{1}{2\pi}\int^{\infty}_{-\infty}\e^{-\textrm{i}x\kappa}\E^{\Q}\left[\e^{\textrm{i}\xi_t\kappa}\right]\rd\kappa.
\end{eqnarray}
A standard calculation involving the computation of the moment
generating function of $\xi_t$ yields the value of the expectation
in the equation above, that is:
\begin{equation}\label{infoprocmgf}
    \E^{\Q}\left[\e^{\textrm{i}\xi_t\kappa}\right]=\sum^n_{j=0}p_j\exp\left[\textrm{i}\sigma
    h_j\,t\kappa-\frac{t(T-t)}{2T}\kappa^2\right].
\end{equation}
Inserting this intermediate result into (\ref{eisprice}) and
completing the square, we obtain
\begin{align}
&A_{0t}(x)\nn\\
&=P_{0t}\sum^n_{j=0}p_j\frac{1}{2\pi}\sqrt{\frac{t(T-t)}{T}}\int^{\infty}_{-\infty}\exp\left[
-\frac{1}{2}\left(\sqrt{\frac{t(T-t)}{T}}\,\kappa-\sqrt{\frac{(\sigma
h_j\,t-x)^2 T}{t(T-t)}}\,\textrm{i}\right)^2\right]\rd\kappa.
\end{align}
Carrying out the integration, we thus have:
\begin{equation}\label{finalEISprice}
A_{0t}(x)=P_{0t}\sum^n_{j=0}p_j\sqrt{\frac{T}{2\pi
t(T-t)}}\exp\left[-\tfrac{1}{2}\frac{(\sigma h_j t -x)^2
T}{t(T-t)}\right].
\end{equation}

The price $A_{0t}(x)$ of a security paying a delta function at the
derivative's maturity $t$ can be viewed also from another angle.
Let us introduce a general payoff function $g(\xi_t)$ defining an
exotic payout depending on the value of the information process
$\{\xi_t\}$ at time $t$. Then we decompose the function $g(\xi_t)$
in infinitesimal parts by use of the delta function by writing
\begin{equation}\label{defdeltadistr}
    g(x_0)=\int^{\infty}_{-\infty}\delta(x-x_0)g(x)\rd x.
\end{equation}
But this definition can be exploited to express the payoff
function $g(\xi_t)$ as a superposition of elementary information
securities. Thus we have:
\begin{eqnarray}\label{geninfoderiv}
    g(\xi_t)=\int^{\infty}_{-\infty}\delta(\xi_t-x)g(x)\rd x.
\end{eqnarray}
This decomposition in terms of elementary securities is in line
with the concept of a so-called Arrow-Debreu security. In fact,
the price (\ref{finalEISprice}) of the elementary information
security defined by the payoff (\ref{eispayoff}) can be regarded
as an example of an Arrow-Debreu price. Thus we can now express
the price of a general information derivative as a weighted
integral, where the elementary information securities with the
Arrow-Debreu price $A_{0t}(x)$ play the role of the weights.
Following the calculation of the price of the elementary
information security we hence have for the price of a general
exotic information derivative the following result:
\begin{equation}\label{PriceGenInfoDeriv}
  V_0=P_{0t}\E^{\Q}[g(\xi_t)]=P_{0t}\E^{\Q}\left[\int^{\infty}_{-\infty}\delta(\xi_t-x)g(x)\rd x\right]=\int^{\infty}_{-\infty}A_{0t}(x)\,g(x)\rd x.
\end{equation}

We can now treat a European call option on a credit-risky discount
bond as a further example of an exotic information derivative, and
apply the Arrow-Debreu technique just shown. We refer the reader
to Section \ref{sec6.8} for a derivation of the call option price,
using this alternative technique, in the case of an underlying
asset with a continuous cash flow function.

The price of a European call option terminating at time $t$
written on a defaultable discount bond with a discrete payoff
function and maturity $T$ is given by
\begin{equation}
C_0=P_{0t}\E^{\Q}\left[(B_{tT}-K)^+\right],
\end{equation}
where $K$ is the strike, and $\{B_{tT}\}$ is the price of the
bond, given by
\begin{equation}
B_{tT}=P_{tT}\frac{\sum_i p_i h_i
\exp\left[\frac{T}{T-t}\left(\sigma h_i \xi_t-\frac{1}{2}\sigma^2
h_i^2 t\right)\right]}{\sum_i p_i \exp \left[
\frac{T}{T-t}\left(\sigma h_i \xi_t-\frac{1}{2} \sigma^2 h_i^2 t
\right) \right]} . \label{eq:29.5}
\end{equation}
We note that since $B_{tT}$ is a function of $\xi_t$, the payoff
function
\begin{equation}\label{callpayoff}
g(\xi_t)=(B_{tT}-K)^+
\end{equation}
can be regarded as an example of an exotic information derivative
and hence we can apply the Arrow-Debreu technique. We write
\begin{eqnarray}
C_0=\int^{\infty}_{-\infty}A_{0t}(x)g(x)\rd x,
\end{eqnarray}
where $A_{0t}(x)$ is given as in (\ref{finalEISprice}) and the
payoff function $g(x)$ by formula (\ref{callpayoff}), with $\xi_t$
replaced by the variable $x$. In the case that the defaultable
bond has a binary payoff we recover, by following the calculation
in Section \ref{sec6.8}, the same expression as in
(\ref{Bridgecallprice}). In particular, if we use the Arrow-Debreu
technique there is no need to change from the risk-neutral measure
to the bridge measure. Hence we have a useful alternative
technique to price derivatives, in which the fundamental step is
the decomposition of the payoff function into elementary
information securities.

\chapter{Complex credit-linked structures}\label{chap4}
%
%
\section{Coupon bonds}\label{sec4.1}
The discussion so far has focused on simple structures such as
discount bonds and options on discount bonds. One of the
advantages of the present approach, however, is that its
tractability extends to situations of a more complex nature. In
this section we consider the case of a credit-risky coupon bond.
One should regard a coupon bond as being a rather complicated
instrument from the point of view of credit risk management, since
default can occur at any of the coupon dates. The market will in
general possess partial information concerning all of the future
coupon payments, as well as the principal payment.

As an illustration, we consider a bond with two payments
remaining---a coupon $H_{T_1}$ at time $T_1$, and a coupon plus
the principal totalling $H_{T_2}$ at time $T_2$. We assume that if
default occurs at $T_1$, then no further payment is made at $T_2$.
On the other hand, if the $T_1$-coupon is paid, default may still
occur at $T_2$. We model this by setting
\begin{equation}
H_{T_1}={\bf c}X_{T_1},\qquad H_{T_2}= ({\bf c}+{\bf
p})X_{T_1}X_{T_2},
\end{equation}
where $X_{T_1}$ and $X_{T_2}$ are independent random variables
taking the values $\{0,1\}$, and the constants ${\bf c},{\bf
p}$ denote the coupon and principal payments.
Let us write $\{p^{(1)}_0, p^{(1)}_1\}$ for the {\it a priori}
probabilities that $X_{T_1}=\{0,1\}$, and
$\{p^{(2)}_0,p^{(2)}_1\}$ for the {\it a priori} probabilities
that $X_{T_2}=\{0,1\}$. We introduce a pair of information
processes
\begin{eqnarray}
\xi^{(1)}_t=\sigma_1X_{T_1}t+\beta^{(1)}_{tT_1} \quad {\rm and}
\quad \xi^{(2)}_t=\sigma_2X_{T_2}t+\beta^{(2)}_{tT_2},
\end{eqnarray}
where $\{\beta^{(1)}_{tT_1}\}$ and $\{\beta^{(2)}_{tT_2}\}$ are
independent Brownian bridges, and $\sigma_1$ and $\sigma_2$ are
parameters. Then for the credit-risky coupon-bond price process we
have
\begin{eqnarray}
B_{tT_2}={\bf c}P_{tT_1}{\mathbb E} \left[X_{T_1}\big| \xi^{(1)}_t
\right]+({\bf c}+{\bf p}) P_{tT_2}{\mathbb
E}\left[X_{T_1}\big\vert \xi^{(1)}_t \right]{\mathbb
E}\left[X_{T_2}\big\vert\xi^{(2)}_t\right]. \label{eq:coupon1}
\end{eqnarray}
The two conditional expectations appearing in this formula can be
worked out explicitly using the techniques already described. The
result is:
\begin{eqnarray}
{\mathbb E} \left[X_{T_i}\big|\xi^{(i)}_t \right] = \frac{
p_1^{(i)}\exp\left[\frac{T_i}{T_i-t}\left( \sigma_i \xi_t^{(i)}-
\frac{1}{2} \sigma_i^2t\right)\right]} {p_0^{(i)}+p_1^{(i)} \exp
\left[ \frac{T_i}{T_i-t} \left(\sigma_i\xi^{(i)}_t -\frac{1}{2}
\sigma_i^2t\right)\right]}, \label{eq:coupon2}
\end{eqnarray}
for $i=1,2$. It should be evident that in the case of a bond with
two payments remaining we obtain a ``two-factor"
model---the factors (i.e.,~the Brownian drivers) being the two
innovation processes arising in connection with the
information processes $\{\xi_t^{(i)}\}_{i=1,2}$.

Similarly, if there are $n$ outstanding coupons, we model the
payments by
\begin{equation}
H_{T_k}={\bf c}X_{T_1} \cdots X_{T_k}
\end{equation}
for $k\leq n-1$ and
\begin{equation}
H_{T_n}=({\bf c}+ {\bf p}) X_{T_1} \cdots X_{T_n},
\end{equation}
and introduce the market information processes
\begin{eqnarray}
\xi^{(i)}_t=\sigma_iX_{T_i}t+\beta^{(i)}_{tT_i} \qquad
(i=1,2,\ldots,n).
\end{eqnarray}
The case of $n$ outstanding payments gives rise in general to an
$n$-factor model. The independence of the random variables
$\{X_{T_i}\}_{i=1,2,\ldots,n}$ implies that the price of a
credit-risky coupon bond admits a closed-form expression analogous
to that obtained in (\ref{eq:coupon1}).

With a slight modification of these expressions we can 
consider the case when there is recovery in the event of
default. In the two-coupon example discussed above, for instance,
we can extend the model by saying that in the event of default on
the first coupon the effective recovery rate (as a percentage of
coupon plus principal) is $R_1$; whereas in the case of default on
the final payment the recovery rate is $R_2$. Then we have
\begin{eqnarray}
H_{T_1}&=&{\bf c}X_{T_1}+R_1({\bf c}+{\bf p})(1- X_{T_1}),\\
H_{T_2}&=&({\bf c}+{\bf p})X_{T_1}X_{T_2}+R_2({\bf c}+ {\bf
p})X_{T_1}(1-X_{T_2}).
\end{eqnarray}
A further extension of this line of reasoning allows for the
introduction of random recovery rates.
\section{Credit default swaps}\label{sec4.2}
Swap-like structures can be treated in a similar way. For example,
in the case of a basic credit default swap we have a series of
premium payments, each of the amount ${\bf g}$, made to the seller
of protection. The payments continue until the failure of a coupon
payment in the reference bond, at which point a lump-sum payment
${\bf n}$ is made to the buyer of protection.

As an illustration, suppose we consider two reference coupons,
letting $X_{T_1}$ and $X_{T_2}$ be the associated independent
random variables, following the pattern of the previous example.
We assume for simplicity that the default-swap premium payments
are made immediately after the bond coupon dates. Then the value
of the default swap, from the point of view of the seller of
protection, is given by the following expression:
\begin{eqnarray}
V_t &=& {\bf g} P_{tT_1}{\mathbb E} \left[ X_{T_1}\big\vert
\xi^{(1)}_t \right] -{\bf n}P_{tT_1}{\mathbb
E}\left[1-X_{T_1}\big\vert\xi^{(1)}_t\right] \nonumber \\ &+& {\bf
g} P_{tT_2}{\mathbb E} \left[ X_{T_1} \big \vert
\xi^{(1)}_t\right]{\mathbb E} \left[X_{T_2}\big\vert \xi^{(2)}_t
\right]-{\bf n}P_{tT_2}{\mathbb E}\left[ X_{T_1}\big\vert
\xi^{(1)}_t \right]{\mathbb
E}\left[1-X_{T_2}\big\vert\xi^{(2)}_t\right].
\end{eqnarray}
After some rearrangement of terms, this can be expressed more
compactly as follows:

\begin{prop}
Let $\{V_t\}_{0\le T_2}$ be the price process of a credit default
swap. Let ${\bf g}$ denote the premium payment, and let ${\bf n}$
denote the payment made to the buyer of the protection in the
event of default. Then the price of a default swap written on a
reference defaultable two-coupon bond is given by
\begin{eqnarray}
V_t &=& -{\bf n}P_{tT_1}+\left[({\bf g}+{\bf n})P_{tT_1}-{\bf
n}P_{tT_2}\right]{\mathbb E}
\left[X_{T_1}\big\vert\xi^{(1)}_t\right] \nonumber \\ && +({\bf
g}+{\bf n}) P_{tT_2}{\mathbb E}
\left[X_{T_1}\big\vert\xi^{(1)}_t\right]{\mathbb E}
\left[X_{T_2}\big\vert\xi^{(2)}_t\right].
\end{eqnarray}
\end{prop}

A similar approach can be adapted in the multi-name credit
situation. The importance of multi-credit correlation modelling
has been emphasised by many authors---see e.g. Davis \& Lo 2001,
Duffie \& Garleaunu 2001, Frey \& McNeil 2003, and Hull \& White
2004a. The point that we would like to emphasise here is that in
the information-based framework there is a good deal of
flexibility available in the manner in which the various
cash-flows can be modelled to depend on one another, and in many
situations tractable expressions emerge that can be used as the
basis for the modelling of complex multi-name credit instruments.
\section{Baskets of credit-risky bonds}\label{sec4.3}
We consider now the valuation problem for a basket of bonds in the
situation for which there are correlations in the payoffs. We
shall demonstrate how to obtain a closed-form expression for the
value of a basket of defaultable bonds with various different
maturities.

For definiteness we consider a set of digital bonds each with two
possible payoffs $\{0,1\}$. It will be convenient to label the
bonds in chronological order with respect to their maturities.
Therefore, we let $H_{T_1}$ denote the payoff of the bond that
expires first; we let $H_{T_2}$ $(T_2\geq T_1)$ denote the payoff
of the first bond that matures after $T_1$; and so on. In general
the various bond payouts will not be independent.

We propose to model this set of dependent random variables in
terms of an underlying set of independent random variables. To
achieve this we let $X$ denote the random variable associated with
the payoff of the first bond: $H_{T_1}=X$. The random variable $X$
takes on the values $\{1,0\}$ with {\it a priori} probabilities
$\{p,1-p\}$. The payoff of the second bond $H_{T_2}$ can then be
represented in terms of three independent random variables:
$H_{T_2}= XX_1+(1-X)X_0$. Here $X_0$ takes the values $\{1,0\}$
with the probabilities $\{p_0,1-p_0\}$, and $X_1$ takes the values
$\{1,0\}$ with the probabilities $\{p_1,1-p_1\}$. Clearly, the
payoff of the second bond is unity if and only if the random
variables $(X,X_0,X_1)$ take the values $(0,1,0)$, $(0,1,1)$,
$(1,0,1)$, or $(1,1,1)$. Since these random variables are
independent, the {\it a priori} probability that the second bond
does not default is $p_0+p(p_1-p_0)$, where $p$ is the {\it a
priori} probability that the first bond does not default.

To represent the payoff of the third bond we introduce four
additional independent random variables:
\begin{eqnarray}
H_{T_3}= XX_1X_{11}+X(1-X_1)X_{10}+(1-X)X_0X_{01}
+(1-X)(1-X_0)X_{00}.
\end{eqnarray}
Here the random variables $\{X_{ij}\}_{i,j=0,1}$ take on the
values $\{1,0\}$ with the probabilities $\{p_{ij},1-p_{ij}\}$. It
is a matter of combinatorics to determine the {\it a priori}
probability that $H_{T_3}=1$ in terms of $p$, $\{p_i\}$, and
$\{p_{ij}\}$.

The scheme above can be extended to represent the payoff of a
generic bond in the basket with an expression of the following
form:
\begin{eqnarray}
H_{T_{n+1}} = \sum_{\{k_j\}=1,0} X^{\omega(k_1)}
X_{k_1}^{\omega(k_2)} X_{k_1k_2}^{\omega(k_3)}\cdots
X_{k_1k_2\cdots k_{n-1}}^{\omega(k_n)} X_{k_1k_2\cdots
k_{n-1}k_n}.
\end{eqnarray}
Here, for any random variable $X$ we define $X^{\omega(0)}=1-X$
and $X^{\omega(1)}=X$. The point is that if we have a basket of
$N$ digital bonds with arbitrary {\it a priori} default
probabilities and arbitrary {\it a priori} correlation, then we
can introduce $2^N-1$ independent digital random variables to
represent the $N$ correlated random variables associated with the
bond payoffs. The scheme above provides a convenient way of
achieving this.

One advantage of the decomposition into independent random
variables is that we retain analytical tractability for the
pricing of the basket. In particular, since the random variables
$\{X_{k_1k_2\cdots k_n}\}$ are independent, it is natural to
introduce a set of $2^N-1$ independent Brownian bridges to
represent the noise that hides the values of the independent
random variables:
\begin{eqnarray}
\xi^{k_1k_2\cdots k_n}_t=\sigma_{k_1k_2\cdots k_n}X_{k_1k_2\cdots
k_n}t + \beta^{k_1k_2\cdots k_n}_{tT_{n+1}} .
\end{eqnarray}
The number of independent factors in general grows rapidly with
the number of bonds in the portfolio. As a consequence, a market
that consists of correlated bonds is in general highly incomplete.
This, in turn, provides a justification for the creation
of products such as CDSs and CDOs that enhance the ``hedgeability"
of such portfolios.
\section{Homogeneous baskets}\label{sec4.4}

In the case of a ``homogeneous" basket the number of independent
random variables characterising the payoff of the portfolio can be
reduced. We assume for simplicity that the basket contains $n$
defaultable discount bonds, each maturing at time $T$, and each
paying $0$ or $1$, with the same {\it a priori} probability of
default. This is an artificial situation, but is of
interest as a first step in the analysis of the more general
setup.

The goal is to model default correlations in the portfolio, and in
particular to model the flow of market information concerning
default correlation. Let us write $H_T$ for the payoff at time $T$
of the homogeneous portfolio, and set
\begin{eqnarray}
H_T = n-Z_1-Z_1Z_2-Z_1Z_2Z_3-\cdots-Z_1Z_2\ldots Z_n,
\end{eqnarray}
where the random variables $\{Z_j\}_{j=1,2,\ldots,n}$, each taking
the values $\{0,1\}$, are assumed to be independent. Thus if
$Z_1=0$, then $H_T=n$; if $Z_1=1$ and $Z_2=0$, then $H_T=n-1$;  if
$Z_1=1$, $Z_2=1$, and $Z_3=0$, then $H_T=n-2$; and so on.

Now suppose we write $p_j={\mathbb Q}(Z_j=1)$ and $q_j={\mathbb
Q}( Z_j=0)$ for $j=1,2,\ldots,n$. Then ${\mathbb Q}(H_T=n)=q_1$,
${\mathbb Q}(H_T=n-1)=p_1q_2$, ${\mathbb Q}(H_T=n-2)=p_1p_2q_3$,
and so on. More generally, we have ${\mathbb
Q}(H_T=n-k)=p_1p_2\ldots p_kq_{k+1}$. Thus if $p_1\ll1$ but
$p_2,p_3,\ldots,p_k$ are large, then we are in a situation of low
default probability and high default correlation; that is to say,
the probability of a default occurring in the portfolio is small,
but conditional on at least one default occurring, the probability
of several defaults is high.

The market will take a view on the likelihood of various numbers
of defaults occurring in the portfolio. We model this by
introducing a set of independent market information processes
$\{\xi^j_{t}\}$ defined by
\begin{eqnarray}
\xi_{t}^j=\sigma_jX_jt+\beta_{tT}^j,
\end{eqnarray}
where $\{\sigma_j\}_{j=1,2,\ldots,n}$ are parameters, and
$\{\beta^j_{tT}\}_{j=1,2,\ldots,n}$ are independent Brownian
bridges. The market filtration $\{{\mathcal F}_t\}$ is taken to be
that generated collectively by $\{\xi^j_t\}_{j=1,2,\ldots,n}$, and
for the portfolio value $H_t=P_{tT}{\mathbb E}\left[ H_T|
{\mathcal F}_t\right]$ we have
\begin{eqnarray}
H_t = P_{tT}\Big[n- {\mathbb E}_t[X_1]- {\mathbb E}_t[X_1]
{\mathbb E}_t[X_2]-\cdots- {\mathbb E}_t[X_1] {\mathbb E}_t
[X_2]\ldots{\mathbb E}_t [X_n]\Big].
\end{eqnarray}
The conditional expectations appearing here can be calculated by
means of formulae established earlier in the paper. The resulting
dynamics for $\{H_t\}$ can thus be used to describe the evolution
of correlations in the portfolio.

For example, if ${\mathbb E}_t[X_1]$ is low and ${\mathbb
E}_t[X_2]$ is high, then the conditional probability at time $t$
of a default at time $T$ is small; whereas if ${\mathbb E}_t[X_1]$
were to increase suddenly, then the conditional probability of two
or more defaults at $T$ would rise as a consequence. Thus, the
model is sufficiently rich to admit a detailed account of the
correlation dynamics of the portfolio. The losses associated with
individual tranches can be identified, and derivative structures
associated with such tranches can be defined.

For example, a digital option that pays out in the event that
there are three or more defaults has the payoff structure
$H_T^{(3)}=X_1X_2X_3$. The homogeneous portfolio model has the
property that the dynamics of equity-level and mezzanine-level
tranches involve a relatively small number of factors. The market
prices of tranches can be used to determine the {\it a priori}
probabilities, and the market prices of options on tranches can be
used to fix the information-flow parameters.

In summary, we see that the information-based framework for
default dynamics introduced in this work is applicable to the
analysis of both single-name and multi-name credit products.

\chapter{Assets with general cash-flow structures}\label{chap5}
\section{Asset pricing: general overview of information-based framework}\label{sec5.0}
In the pricing of derivative securities, the starting point is
usually the specification of a model for the price process of the
underlying asset. Such models tend to be of an \textit{ad
hoc} nature. For example, in the Black-Scholes theory, the
underlying asset has a geometric Brownian motion as its price
process which, although very useful as a mathematical model is
nevertheless widely agreed to be in some respects artificial. More generally,
but more or less equally arbitrarily, the economy is often
modelled by a probability space equipped with the filtration
generated by a multi-dimensional Brownian motion, and it is
assumed that asset prices are Ito processes that are adapted to
this filtration. This particular example is of course the
``standard'' model within which a great deal of financial
engineering has been carried out.

The basic methodological problem with the standard model (and the
same applies to various generalisations thereof) is that the
market filtration is fixed once and for all, and little or no
comment is offered on the issue of ``where it comes from''. In
other words, the filtration, which represents the unfolding of
information available to market participants, is modelled first,
in an \textit{ad hoc} manner, and then it is assumed that the
asset price processes are adapted to it. But no indication is
given about the nature of this ``information"; and it is not at
all obvious, \textit{a priori}, why the Brownian filtration, for
example, should be regarded as providing information rather than
simply noise.

In a complete market there is a sense in which the
Brownian filtration provides all of the relevant information, and
no irrelevant information. That is, in a complete market
based on a Brownian filtration the asset price movements precisely
reflect the information content of the filtration. Nevertheless,
the notion that the market filtration should be ``prespecified" is
an unsatisfactory one in financial modelling.

The usual intuition behind the ``prespecified-filtration''
approach is to imagine that the filtration  represents the
unfolding in time of a succession of random events that
``influence'' the markets, thus causing prices to change. For
example, a spell of bad weather in South America results in a
decrease in the supply of coffee beans and hence an increase in
the price of coffee.

The idea is that one then ``abstractifies'' these various
influences in the form of a prespecified background filtration to
which asset price processes are adapted. What is unsatisfactory
about this is that so little structure is given to the filtration:
price movements behave as though they were spontaneous. In
reality, we expect the price-formation process to exhibit more
structure. It would be out of place, in this thesis, to attempt a
complete account of the process of price formation or to address
the literature of market microstructure in a systematic way.
Nevertheless, we can try to improve on the ``prespecified''
approach. In that spirit we proceed as follows.

We note that price changes arise from two rather distinct sources.
The first source of price change is that resulting from changes in
market-agent preferences---that is to say, changes in the pricing
kernel. Movements in the pricing kernel are associated with
(a)~changes in investor attitudes towards risk, and (b)~changes in
investor ``impatience'', i.e.,~the subjective discounting of
future cash flows. But equally important, if not more so, are
those changes in price resulting from the revelation to market
agents of information about the future cash flows derivable from
possession of a given asset.

When a market agent decides to buy or sell an asset, the decision
is made in accordance with the information available to the agent
concerning the likely future cash flows associated with the asset.
A change in the information available to the market agent about a
future cash flow will typically have an effect on the price at
which they are willing to buy or sell, even if the agent's
preferences remain unchanged.

Let us consider, for example, the situation where one is thinking
of purchasing an item at a price that seems attractive. But then,
by chance, one reads a newspaper article pointing out some
undesirable feature of the product. After some reflection, one
decides that the price is not so attractive. As a result, one
decides not to buy, and eventually---possibly because other
individuals have read the same report---the price drops.

The movements of the price of an asset should, therefore, be
regarded as constituting an emergent phenomenon. To put the matter
another way, the price process of an asset should be viewed as the
output of the various decisions made relating to possible
transactions in the asset, and these decisions in turn should be
understood as being induced primarily by the flow of information
to market participants.

Taking into account these elementary observations, we are now in a
position in this chapter to propose the outlines of a general
framework for asset pricing based on modelling of the flow of
market information. The information will be that concerning the
values of the future cash flows associated with the given assets.
For example, if the asset represents a share in a firm that will
make a single distribution at some pre-agreed date, then there is
a single cash flow corresponding to the random amount of the
distribution. If the asset is a credit-risky discount bond, then
the future cash flow is the payout of the bond at the maturity
date. In each case, based on the information available relating to
the likely payouts of the given financial instrument, market
participants determine, as best as they can, estimates for the
value of the right to the impending cash flows. These estimates,
in turn, lead to decisions concerning transactions, which then
trigger movements in the price.

In this chapter we present a simple class of models capturing the
essence of the scenario described above. As we remarked in the
introduction of the thesis, in building this framework we have
several criteria in mind that we would like to see satisfied:
\begin{itemize}
\item The first of these is that our model for the flow of market
information should be intuitively appealing, and should allow for
a reasonably sophisticated account of aggregate investor
behaviour. \item At the same time, the model should be simple
enough to allow one to derive explicit expressions for the asset
price processes thus induced, in a suitably rich range of
examples, as well as for various associated derivative price
processes. \item The framework should also be flexible enough to
allow for the modelling of assets having complex cash-flow
structures. \item Furthermore, it should be suitable for practical
implementation, with the property that calibration and pricing can
be carried out swiftly and robustly, at least for more elementary
structures. \item We would like the framework to be mathematically
sound, and to be manifestly arbitrage-free.
\end{itemize}
In what follows we shall attempt to make some headway with these
diverse criteria.
\section{The three ingredients}\label{sec5.2}
In asset pricing we require three basic ingredients, namely, (a)
the cash flows, (b) the investor preferences, and (c) the flow of
information available to market participants. Translated into
somewhat more mathematical language, these ingredients amount to
the following: (a$'$) cash flows are modelled as random variables;
(b$'$) investor preferences are modelled with the determination of
a pricing kernel; and (c$'$) the market information flow is
modelled with the specification of a filtration. As we have
indicated above, asset pricing theory conventionally attaches more
weight to (a) and (b) than to (c). In this paper, however, we
emphasise the importance of ingredient (c).

Our theory will be based on modelling the flow of information
accessible to market participants concerning the future cash flows
associated with the possession of an asset, or with a position in
a financial contract. The idea that information should play a
foundational role in asset pricing has been long
appreciated---see, e.g., Back 1992, Back \& Baruch 2004, and
references cited therein. Our contribution to this area will
involve an explicit technique for modelling the filtration. We
start by setting the notation and introducing the assumptions
employed in this paper. We model the financial markets with the
specification of a probability space $(\Omega,{\mathcal F},
{\mathbb Q})$ on which a filtration $\{{\mathcal F}_t\}_{0\leq t <
\infty}$ will be constructed. The probability measure ${\mathbb
Q}$ is understood to be the risk-neutral measure, and the
filtration $\{{\mathcal F}_t\}$ is understood to be the market
filtration. All asset-price processes and other
information-providing processes accessible to market participants
will be adapted to $\{{\mathcal F}_t\}$. We do not regard
$\{{\mathcal F}_t\}$ as something handed to us on a platter.
Instead, it will be modelled explicitly.

Several simplifying assumptions will be made, so that we can
concentrate our efforts on the problems associated with the flow
of market information. The first of these assumptions is the use
of the risk-neutral measure.  The ``real'' probability measure
does not enter into the present investigation.

We leap over that part of the economic analysis that determines
the pricing measure. More specifically, we assume the absence of
arbitrage and the existence of an established pricing kernel (see,
e.g., Cochrane 2005, and references cited therein). With these
conditions the existence of a unique risk-neutral pricing measure
${\mathbb Q}$ is ensured, even though the markets we consider
will, in general, be incomplete. For a discussion of the issues
associated with pricing in incomplete markets see, e.g., Carr {\it
et al.} 2001.

The second assumption is that we take the default-free system of
interest rates to be deterministic. The view is that we should
first develop our framework in a simplified setting, where certain
essentially macroeconomic issues are put to one side; then, once
we are satisfied with the tentative framework, we can attempt to
generalise it in such a way as to address these issues. We
therefore assume a deterministic default-free discount bond
system. The absence of arbitrage implies that the corresponding
system of discount functions $\{P_{tT}\}_{0\le t \leq T<\infty}$
can be written in the form $P_{tT}= P_{0T}/P_{0t}$ for $t \leq T$,
where $\{P_{0t}\}_{0\le t<\infty}$ is the initial discount
function, which we take to be part of the initial data of the
model. The function $\{P_{0t}\}_{0\leq t< \infty}$ is assumed to
be differentiable and strictly decreasing, and to satisfy
$0<P_{0t}\leq 1$ and $\lim_{t\rightarrow\infty}P_{0t}=0$. These
conditions can be relaxed somewhat for certain applications. A
method for extending in the information-based framework to a
background stochastic interest rate environment is considered in
Rutkowski \& Yu 2005, using a forward measure technique (Geman
{\it et al.} 1995).

We also assume, for simplicity, that all cash flows occur at
pre-determined dates. Now clearly for some purposes we would like
to allow for cash flows occurring effectively at random times---in
particular, at stopping times associated with the market
filtration. But in the present exposition we want to avoid the
idea of a ``prespecified'' filtration with respect to which
stopping times are defined. We take the view that the market
filtration is a ``derived'' notion, generated by information about
impending cash flows, and by the actual values of cash flows when
they occur. In the present paper we regard a ``randomly-timed''
cash flow as being a set of random cash flows occurring at various
times---and with a joint distribution function that ensures only
one of these flows is non-zero. Hence in our view the ontological
status of a cash flow is that its timing is definite, only the
amount is random---and that cash flows occurring at different
times are, by their nature, different cash flows.
\section{Modelling the cash flows}\label{sec5.3}
First we consider the case of a single isolated cash flow
occurring at time $T$, represented by a random variable $D_T$. We
assume that $D_T\ge 0$. The value $S_t$ of the cash flow at any
earlier time $t$ in the interval $0\le t< T$ is then given by the
discounted conditional expectation of $D_T$:
\begin{eqnarray}
S_t=P_{tT}{\mathbb E}^{{\mathbb Q}}\left[D_T\vert{\mathcal
F}_t\right]. \label{eq:2-1}
\end{eqnarray}
In this way we model the price process $\{S_t\}_{0\le t<T}$ of a
limited-liability asset that pays the single dividend $D_T$ at
time $T$. The construction of the price process here is carried
out in such a way as to guarantee an arbitrage-free market if
other assets are priced by the same method---see Davis 2004 for a
related point of view. With a slight abuse of terminology we shall
use the terms ``cash flow'' and ``dividend'' more or less
interchangeably. If a more specific use of one of these terms is
needed, then this will be evident from the context. We adopt the
convention that when the dividend is paid the asset price goes
``ex-dividend'' immediately. Hence in the example above we have
$\lim_{t\to T}S_{t}=D_T$ and $S_T=0$.

In the case that the asset pays a sequence of dividends $D_{T_k}$
$(k=1,2,\ldots,n)$ on the dates $T_k$ the price (for values of $t$
earlier than the time of the first dividend) is
\begin{eqnarray}
S_t=\sum^n_{k=1}P_{tT_k}{\mathbb E}^{{\mathbb Q}} \left[ D_{T_k}
\vert{\mathcal F}_t\right]. \label{eq:2-3}
\end{eqnarray}
More generally, taking into account the ex-dividend behaviour, we
have
\begin{eqnarray}
S_t=\sum^n_{k=1}{\bf 1}_{\{t<T_k\}}P_{tT_k}{\mathbb E}^{{\mathbb
Q}} \left[D_{T_k}\vert{\mathcal F}_t\right].
\end{eqnarray}

It turns out to be useful  if we adopt the convention that a
discount bond also goes ex-dividend on its maturity date. In the
case of a discount bond we assume that the price of the bond is
given, for dates earlier than the maturity date, by the product of
the principal and the relevant discount factor. But at maturity
(when the principal is paid) the value of the bond drops to zero.
In the case of a coupon bond, there is a downward jump in the
price of the bond at the time a coupon is paid (the value lost may
be captured back in the form of an ``accrued interest'' payment).
In this way we obtain a consistent treatment of the
``ex-dividend'' behaviour of the asset price processes under
consideration. With this convention it follows that all price
processes have the property that they are right continuous with
left limits.
\section{Construction of the market information flow}\label{sec5.4}
Now we present a simple model for the flow of market information.
We consider first the case of a single distribution, occurring at
time $T$, and assume that market participants have only partial
information about the upcoming cash flow $D_{T}$. The information
available in the market about the cash flow is assumed to be
contained in a process $\{\xi_t\}_{0\leq t\leq T}$ defined by:
\begin{eqnarray}\label{eq:2-21}
\xi_t=\sigma D_{T} t+\beta_{tT}.
\end{eqnarray}
We call $\{\xi_t\}$ the market information process. The
information process is composed of two parts. The term $\sigma
D_{T}t$ contains the ``true information'' about the upcoming
dividend. This term grows in magnitude as $t$ increases.

The process $\{\beta_{tT}\}_{0\leq t\leq T}$ is a standard
Brownian bridge over the time interval $[0,T]$. Thus
$\beta_{0T}=0$, $\beta_{TT}=0$, and at time $t$ the random
variable $\beta_{tT}$ has mean zero and variance $t(T-t)/T$; the
covariance of $\beta_{sT}$ and $\beta_{tT}$ for $s \leq t$ is
$s(T-t)/T$. We assume that $D_{T}$ and $\{\beta_{tT}\}$ are
independent. Thus the information contained in the bridge process
is ``pure noise''. The information contained in $\{\xi_t\}$ is
clearly unchanged if we multiply $\{\xi_t\}$ by some overall scale
factor.

An earlier well-known example of the use of a Brownian bridge
process in the context of interest rate modelling can be found in
Ball \& Torous 1983; they are concerned, however, with the
default-free interest rate term structure, and their model is
unrelated to the approach presented in this thesis.

We assume that the market filtration $\{{\mathcal F}_t\}$ is
generated by the market information process. That is to say, we
assume that $\{{\mathcal F}_t\}=\{{\mathcal F}^{\xi}_t\}$, where
$\{{\mathcal F}^{\xi}_t\}$ is the filtration generated by
$\{\xi_t\}$. The dividend $D_T$ is therefore ${\mathcal
F}_T$-measurable, but is not ${\mathcal F}_t$-measurable for
$t<T$. Thus the value of $D_T$ becomes ``known'' at time $T$, but
not earlier. The bridge process $\{\beta_{tT}\}$ is not adapted to
$\{{\mathcal F}_t\}$ and thus is not directly accessible to market
participants. This reflects the fact that until the dividend is
paid the market participants cannot distinguish the ``true
information" from the ``noise" in the market.

The introduction of the Brownian bridge models the fact that
market perceptions, whether valid or not, play a role in
determining asset prices. Initially, all available information is
used to determine the \textit{a priori} risk-neutral probability
distribution for $D_T$. Then after the passage of time rumours,
speculations, and general disinformation start circulating,
reflected in the steady increase in the variance of the Brownian
bridge. Eventually the variance drops and falls to zero at the
time the distribution to the share-holders is made. The parameter
$\sigma$ represents the rate at which information about the true
value of $D_T$ is revealed as time progresses. If $\sigma$ is low,
the value of $D_T$ is effectively hidden until very near the time
of the dividend payment; whereas if $\sigma$ is high, then the
value of the cash flow is for all practical purposes revealed very
quickly.

In the example under consideration we have made some simplifying
assumptions concerning our choice for the market information
structure. For instance, we assume that $\sigma$ is constant. We
have also assumed that the random dividend $D_T$ enters directly
into the structure of the information process, and enters
linearly. As we shall indicate later, a more general and in some
respects more natural setup is to let the information process
depend on a random variable $X_T$ which we call a ``market
factor''; then the dividend is regarded as a function of the
market factor. This arrangement has the advantage that it easily
generalises to the situation where a cash flow might depend on
several independent market factors, or indeed where cash flows
associated with different financial instruments have one or more
market factors in common. But for the moment we regard the single
cash flow $D_T$ as being the relevant market factor, and we assume
the information-flow rate to be constant.

With the market information structure described above for a single
cash flow in place, we proceed to construct the associated price
dynamics. The price process $\{S_t\}$ for a share in the firm
paying the specified dividend is given by formula (\ref{eq:2-1}).
It is assumed that the \textit{a priori} probability distribution
of the dividend $D_T$ is known. This distribution is regarded as
part of the initial data of the problem, which in some cases can
be calibrated from knowledge of the initial price of the asset,
possibly along with other price data.

The general problem of how the {\it a priori} distribution is
obtained is an important one---any asset pricing model has to
confront some version of this issue---which we defer for later
consideration. The main point is that the initial distribution is
not to be understood as being ``absolutely'' determined, but
rather represents the ``best estimate'' for the distribution given
the data available at that time, in accordance with what one might
call a Bayesian point of view. We recall the fact that the
information process $\{\xi_t\}$ is Markovian, which we showed
earlier. Making use of this property of the information process
together with the fact that $D_T$ is ${\mathcal F}_T$-measurable
we deduce that
\begin{eqnarray}
S_t={ \bf 1}_{\{t<T\}}P_{tT}{\mathbb E}^{{\mathbb Q}}
\left[D_T\vert\xi_t\right]. \label{eq:2-7}
\end{eqnarray}
If the random variable $D_T$ that represents the payoff has a
continuous distribution, then the conditional expectation in
(\ref{eq:2-7}) can be expressed in the form
\begin{eqnarray}
{\mathbb E}^{{\mathbb Q}}\left[D_T\vert\xi_t\right]=
\int_0^{\infty} x \pi_t(x)\,\rd x.
\end{eqnarray}
Here $\pi_t (x)$ is the conditional probability density for the
random variable $D_T$:
\begin{eqnarray}
\pi_t(x)=\frac{\rd}{\rd x}\,{\mathbb Q}(D_T\le x\vert\xi_t).
\end{eqnarray}
We implicitly assume appropriate technical conditions on the
distribution of the dividend that will suffice to ensure the
existence of the expressions under consideration. Also, for
convenience we use a notation appropriate for continuous
distributions, though corresponding results can easily be inferred
for discrete distributions, or more general distributions, by
slightly modifying the stated assumptions and conclusions.

Bearing in mind these points, we note that the conditional
probability density process for the dividend can be worked out by
use of a form of the Bayes formula:
\begin{eqnarray}
\pi_t(x) = \frac{p(x)\rho(\xi_t|D_T=x)}{\int_0^\infty p(x)
\rho(\xi_t|D_T=x){\rm d}x}. \label{eq:08}
\end{eqnarray}
Here $p(x)$ denotes the {\it a priori} probability density for
$D_T$, which we assume is known as an initial condition, and
$\rho(\xi_t|D_T=x)$ denotes the conditional density function for
the random variable $\xi_t$ given that $D_T=x$. Since $\beta_{tT}$
is a Gaussian random variable with variance $t(T-t)/T$, the
conditional probability density for $\xi_t$ is
\begin{eqnarray}
\rho(\xi_t|D_T=x) = \sqrt{{\frac{T}{2\pi t(T-t)}}} \exp\left(
-\frac{(\xi_t-\sigma t x)^2T}{2t(T-t)}\right) . \label{eq:4.13}
\end{eqnarray}
Inserting this expression into the Bayes formula, we get
\begin{equation}
\pi_t(x)= \frac{p(x)\exp\left[ \frac{T}{T-t}(\sigma
x\xi_t-\tfrac{1}{2}\sigma^2 x^2 t)\right]}{\int^{\infty}_0
p(x)\exp\left[\frac{T}{T-t}(\sigma x\xi_t-\tfrac{1}{2}\sigma^2 x^2
t)\right]\rd x}.
\end{equation}
We thus obtain the following result for the asset price:

\begin{prop}
The information-based price process $\{S_t\}_{0\le t\le T}$ of a
limited-liability asset that pays a single dividend $D_T$ at time
$T$ with distribution
\begin{equation}\label{}
{\mathbb Q}(D_T\leq y)=\int_0^y p(x)\, {\rm d}x
\end{equation}
is given by
\begin{eqnarray}
S_t = { \bf 1}_{\{t<T\}} P_{tT}\frac{\int^{\infty}_0 x
p(x)\exp\left[ \frac{T}{T-t}(\sigma x\xi_t-\tfrac{1}{2}\sigma^2
x^2 t)\right]\rd x}{\int^{\infty}_0
p(x)\exp\left[\frac{T}{T-t}(\sigma x\xi_t-\tfrac{1}{2}\sigma^2 x^2
t)\right]\rd x}, \label{eq:2-11}
\end{eqnarray}
where $\xi_t=\sigma D_T t+\beta_{tT}$ is the market information.
\end{prop}
\section{Asset price dynamics in the case of a single random cash flow}\label{sec5.5}
In order to analyse the properties of the price process deduced
above, and to be able to compare it with other models, we need to
work out the dynamics of $\{S_t\}$. One of the advantages of the
model under consideration is that we have a completely explicit
expression for the price process at our disposal. Thus in
obtaining the dynamics we need to find the stochastic differential
equation of which $\{S_t\}$ is the solution. This turns out to be
an interesting exercise because it offers some insights into what
we mean by the assertion that market price dynamics should be
regarded as constituting an ``emergent phenomenon''. The basic
mathematical tool that we make use of here is nonlinear filtering
theory---see, e.g., Bucy \& Joseph 1968, Kallianpur \& Striebel
1968, Davis \& Marcus 1981, and Liptser \& Shiryaev 2000. The
specific applications that we make of the theory here are
original.

To obtain the dynamics associated with the price process $\{S_t\}$
of a single-dividend-paying asset, let us write
\begin{eqnarray}\label{eq:3.13a}
D_{tT}={\mathbb E}^{{\mathbb Q}} [D_T\vert\xi_t]
\end{eqnarray}
for the conditional expectation of $D_T$ with respect to the
market information $\xi_t$. Evidently, $D_{tT}$ can be expressed
in the form $D_{tT}=D(\xi_t,t)$, where the function $D(\xi,t)$ is
defined by
\begin{eqnarray}
D(\xi,t)=\frac{\int^{\infty}_0 x
p(x)\exp\left[\frac{T}{T-t}(\sigma x\xi-\tfrac{1}{2}\sigma^2 x^2
t)\right]\rd x}{\int^{\infty}_0 p(x)\exp\left[\frac{T}{T-t}(\sigma
x\xi-\tfrac{1}{2}\sigma^2 x^2 t)\right]\rd x}.
\end{eqnarray}
A straightforward calculation making use of the Ito rules shows
that the dynamical equation for the conditional expectation
$\{D_{tT}\}$ is given by
\begin{eqnarray}
\rd D_{tT} = \frac{\sigma T}{T-t} V_t \left[ \frac{1}{T-t}
\Big(\xi_t - \sigma T D_{tT} \Big) \rd t + \rd \xi_t \right].
\end{eqnarray}
Here $V_t$ is the conditional variance of the dividend:
\begin{eqnarray}
V_t=\int^{\infty}_0 x^2 \pi_t(x)\,\rd x - \left(\int^{\infty}_0 x
\pi_t(x)\,\rd x\right)^2.
\end{eqnarray}
Therefore, if we define a new process $\{W_t\}_{0\le t<T}$ by
setting
\begin{eqnarray}
W_t=\xi_t-\int^t_0\,\frac{1}{T-s}\Big(\sigma T D_{sT}
-\xi_s\Big)\rd s, \label{eq:2-16}
\end{eqnarray}
we find, after some rearrangement of terms, that
\begin{eqnarray}
\rd D_{tT} =\frac{\sigma T}{T-t}V_t  \rd W_t. \label{eq:2-17}
\end{eqnarray}
For the dynamics of the asset price process we thus have
\begin{eqnarray}
\rd S_t=r_t S_t\rd t+\Gamma_{tT}\rd W_t, \label{eq:2-18}
\end{eqnarray}
where the short rate $r_t$ is given by $r_t=-\rd\ln P_{0t}/\rd t$,
and the absolute price volatility $\Gamma_{tT}$ is given by
\begin{eqnarray}
\Gamma_{tT}=P_{tT}\frac{\sigma T}{T-t}V_t. \label{x3}
\end{eqnarray}

A slightly different way of arriving at this result is as follows.
We start with the conditional probability process $\pi_t(x)$.
Then, using the same notation as above, for the dynamics of
$\pi_t(x)$ we obtain
\begin{eqnarray}
\rd\pi_t(x)=\frac{\sigma T}{T-t}(x-D_{tT})\pi_t(x)\,\rd W_t.
\end{eqnarray}
Since the asset price is given by
\begin{eqnarray}
S_t={ \bf 1}_{\{t<T\}}P_{tT}\int_0^{\infty} x \pi_t(x)\,\rd x,
\end{eqnarray}
we are thus able to infer the dynamics of the price $\{S_t\}$ from
the dynamics of the conditional probability $\{\pi_t(x)\}$, once
we take into account the formula for the conditional variance.

As we have demonstrated earlier, in the context of a discrete
payment, the process $\{W_t\}$ defined in (\ref{eq:2-16}) is an
$\{{\mathcal F}_t\}$-Brownian motion. Hence \textit{from the point
of view of the market it is the process $\{W_t\}$ that drives the
asset price dynamics}. In this way our framework resolves the
somewhat paradoxical point of view usually adopted in financial
modelling in which $\{W_t\}$ is regarded as ``noise'', and yet
also generates the market information flow. And thus, instead of
hypothesising the existence of a driving process for the dynamics
of the markets, we are able from the information-based perspective
to deduce the existence of such a process.

The information-flow parameter $\sigma$ determines the overall
magnitude of the volatility. In fact, as we have remarked earlier,
the parameter $\sigma$ plays a role that is in many respects
analogous to the similarly-labelled parameter in the Black-Scholes
theory. Thus, we can say that the rate at which information is
revealed in the market determines the overall magnitude of the
market volatility. In other words, everything else being the same,
if we increase the information flow rate, then the market
volatility will increase as well. It is ironic that, according to
this point of view, those mechanisms that one might have thought
were designed to make markets more efficient---e.g., globalisation
of the financial markets, reduction of trade barriers, improved
communications, a more robust regulatory environment, and so
on---can have the effect of increasing market volatility, and
hence market risk, rather than reducing it.
\section{European-style options on a single-dividend paying asset}\label{sec5.6}
Before we turn to the consideration of more general cash flows and
more general market information structures, let us consider the
problem of pricing a derivative on an asset for which the price
process is governed by the dynamics (\ref{eq:2-18}). We shall look
at the valuation problem for a European-style call option on such
an asset, with strike price $K$, and exercisable at a fixed
maturity date $t$. The option is written on an asset that pays a
single dividend $D_T$ at time $T>t$. The value of the option at
time $0$ is clearly
\begin{eqnarray}
C_0=P_{0t}{\mathbb E}^{{\mathbb Q}}\left[(S_t-K)^+\right].
\label{Eopt}
\end{eqnarray}
Inserting the information-based expression for the price $S_t$
derived in the previous section into this formula, we obtain
\begin{eqnarray}
C_0=P_{0t}\,{\mathbb E}^{{\mathbb Q}} \left[ \left(
P_{tT}\int^{\infty}_0 x\,\pi_t(x)\rd x-K \right)^+\right].
\label{eq:3-2}
\end{eqnarray}
For convenience we write the conditional probability $\pi_t(x)$ in
the form
\begin{eqnarray}
\pi_t(x)=\frac{p_t(x)}{\int^{\infty}_0\,p_t(x)\rd x},
\label{eq:3-3}
\end{eqnarray}
where the ``unnormalised'' density process $\{p_t(x)\}$ is defined
by
\begin{eqnarray}
p_t(x)=p(x)\exp\left[\frac{T}{T-t}\left(\sigma x \xi_t
-\tfrac{1}{2}\sigma^2 x^2 t\right)\right]. \label{eq:3-4}
\end{eqnarray}
Substituting (\ref{eq:3-4}) into (\ref{eq:3-2}) we find that the
initial value of the option is given by
\begin{eqnarray}\label{eq:3-3a}
C_0=P_{0t}{\mathbb E}^{{\mathbb Q}}\left[ \frac{1}{\Phi_t}
\left(\int^{\infty}_0\left(P_{tT}x-K\right)p_t(x)\rd
x\right)^+\right],
\end{eqnarray}
where
\begin{eqnarray}
\Phi_t=\int^{\infty}_0 p_t(x)\rd x.
\end{eqnarray}
The random variable $\Phi_t$ can be used to introduce a measure
${\mathbb B}_T$ applicable over the time horizon $[0,t]$, which as
before we call the ``bridge measure''. The call option price can
thus be written:
\begin{eqnarray}
C_0=P_{0t}{\mathbb E}^{{\mathbb B}_T} \left[\left(
\int^{\infty}_0\left(P_{tT}x-K\right)p_t(x)\rd x\right)^+\right].
\end{eqnarray}
The special feature of the bridge measure is that the random
variable $\xi_t$ is Gaussian under ${\mathbb B}_T$. In particular,
under the measure ${\mathbb B}_T$ we find that $\{\xi_t\}$ has
mean $0$ and variance $t(T-t)/T$. Since $p_t(x)$ can be expressed
as a function of $\xi_t$, when we carry out the expectation above
we are led to a tractable formula for $C_0$.

To obtain the value of the option we define a constant $\xi^*$
(the critical value) by the following condition:
\begin{eqnarray}
\int^{\infty}_0\left(P_{tT}x-K\right)p(x)\exp\left[\frac{T}{T-t}
\left(\sigma x \xi^*-\tfrac{1}{2}\sigma^2 x^2 t\right)\right]\rd
x=0.
\end{eqnarray}
Then the option price is given by:
\begin{equation}\label{eq:ec1}
C_0=P_{0T}\int^{\infty}_0 x\,p(x)\,N\Big(-z^*+\sigma
x\sqrt{\tau}\Big)\rd x-P_{0t}K\int^{\infty}_0 p(x) \,N
\Big(-z^*+\sigma x\sqrt{\tau}\Big)\rd x,
\end{equation}
where
\begin{eqnarray}
\tau=\frac{tT}{T-t},\qquad z^*=\xi^*\sqrt{\frac{T}{t(T-t)}}\, ,
\end{eqnarray}
and $N(x)$ denotes the standard normal distribution function. We
see that a tractable expression is obtained, and that it is of the
Black-Scholes type. The option pricing problem, even for general
$p(x)$, reduces to an elementary numerical problem. It is
interesting to note that although the probability distribution for
the price $S_t$ at time $t$ is not of a ``standard'' type,
nevertheless the option valuation problem remains a solvable one.
\section{Dividend structures: specific examples}\label{sec5.7}
In this section we consider the dynamics of assets with various
specific continuous dividend structures. First we look at a simple
asset for which the cash flow is exponentially distributed. The
\textit{a priori} probability density for $D_T$ is thus of the
form
\begin{eqnarray}
p(x)=\frac{1}{\delta}\,\exp\left(-\frac{x}{\delta}\right),
\end{eqnarray}
where $\delta$ is a constant. The idea of an exponentially
distributed payout is of course somewhat artificial; nevertheless
we can regard this as a useful model for the situation where
little is known about the probability distribution of the
dividend, apart from its mean. Then from formula (\ref{eq:2-11})
we find that the corresponding asset price is:
\begin{eqnarray}
S_t = {\bf 1}_{\{t<T\}} P_{tT}\frac{\int^{\infty}_0
x\exp(-x/\delta)\exp\left[\frac{T}{T-t}(\sigma x\xi_t-
\tfrac{1}{2}\sigma^2 x^2 t)\right]\rd
x}{\int^{\infty}_0\exp(-x/\delta)\exp\left[\frac{T}{T-t}(\sigma
x\xi_t-\tfrac{1}{2}\sigma^2 x^2 t)\right]\rd x}.
\end{eqnarray}

We note that $S_0=P_{0T}\delta$, so we can calibrate the choice of
$\delta$ by use of the initial price. The integrals in the
numerator and denominator in the expression above can be worked
out explicitly. Hence, we obtain a closed-form expression for the
price in the case of an asset with an exponentially-distributed
terminal cash flow. This is given by:
\begin{eqnarray}
S_t={\bf 1}_{\{t<T\}} P_{tT}\left[\frac{\exp
\left(-\tfrac{1}{2}B_t^2/A_t\right)} {\sqrt{2\pi A_t}\
N(B_t/\sqrt{A_t})}+\frac{B_t}{A_t}\right],
\end{eqnarray}
where
\begin{equation}
A_t=\sigma^2\frac{tT}{(T-t)},
\end{equation}
and
\begin{equation}
B_t= \frac{\sigma T}{(T-t)}\xi_t-\delta^{-1}.
\end{equation}

Next we consider the case of an asset for which the single
dividend paid at time $T$ has a gamma distribution. More
specifically, we assume the probability density is of the form
\begin{eqnarray}
p(x)=\frac{\delta^n}{(n-1)!}\,x^{n-1}\exp(-\delta x),
\end{eqnarray}
where $\delta$ is a positive real number and $n$ is a positive
integer. This choice for the probability density also leads to a
closed-form expression. We find that
\begin{eqnarray}
S_t={\bf 1}_{\{t<T\}}P_{tT}\frac{\sum\limits^n_{k=0} {n\choose k}
A_t^{\frac{1}{2}k-n}B_t^{n-k}F_k(-B_t/\sqrt{A_t})}
{\sum\limits^{n-1}_{k=0}{n-1 \choose
k}A_t^{\frac{1}{2}k-n+1}B_t^{n-k-1} F_k(-B_t/\sqrt{A_t})},
\end{eqnarray}
where $A_t$ and $B_t$ are as above, and
\begin{eqnarray}
F_k(x)=\int^{\infty}_x\,z^k\exp\left(-\tfrac{1}{2}z^2\right)dz.
\end{eqnarray}
A recursion formula can be worked out for the function $F_k(x)$.
This is given by
\begin{equation}\label{}
(k+1)F_k(x)=F_{k+2}(x)-x^{k+1}\exp\lb(-\tfrac{1}{2}x^2\rb),
\end{equation}
from which it follows that
\begin{eqnarray}
F_0(x) &=& \sqrt{2\pi} N(-x),\nn\\
F_1(x) &=& \re^{-\frac{1}{2}x^2},\nn\\
F_2(x) &=& x \re^{-\frac{1}{2}x^2} + \sqrt{2\pi} N(-x),\nn\\
F_3(x) &=& (x^2+2) \re^{-\frac{1}{2}x^2},
\end{eqnarray}
and so on. In general, the polynomial parts of
$\{F_k(x)\}_{k=0,1,2,\ldots}$ are related to the Legendre
polynomials.

\chapter{$X$-factor analysis and applications}\label{chap6}
%
%
\section{Multiple cash flows}\label{sec6.1}
In this chapter we generalise the preceding material to the
situation where the asset pays multiple dividends. This allows us
to consider a wider range of financial instruments. Let us write
$D_{T_k}$ $(k=1,\ldots,n)$ for a set of random cash flows paid at
the pre-designated dates $T_k$  $(k=1,\ldots,n)$. Possession of
the asset at time $t$ entitles the bearer to the cash flows
occurring at times $T_k>t$. For simplicity we assume $n$ is
finite.

For each value of $k$ we introduce a set of independent random
variables $X^{\alpha}_{T_k}$ $(\alpha=1,\ldots,m_k)$, which again
we call market factors or $X$-factors. For each value of $\alpha$
we assume that the factor $X^{\alpha}_{T_k}$ is ${\mathcal
F}_{T_k}$-measurable, where $\{{\mathcal F}_t\}$ is the market
filtration.

Intuitively speaking, for each value of $k$ the market factors
$\{X^{\alpha}_{T_j}\}_{j\le k}$ represent the independent elements
that determine the cash flow occurring at time $T_k$. Thus for
each value of $k$ the cash flow $D_{T_k}$ is assumed to have the
following structure:
\begin{equation}
D_{T_k}=\Delta_{T_k}(X^{\alpha}_{T_1},
X^{\alpha}_{T_2},...,X^{\alpha}_{T_k}),
\end{equation}
where $\Delta_{T_k}(X^{\alpha}_{T_1},
X^{\alpha}_{T_2},...,X^{\alpha}_{T_k})$ is a function of
$\sum_{j=1}^k m_j$ variables. For each cash flow it is, so to
speak, the job of the financial analyst (or actuary) to determine
the relevant independent market factors, and the form of the
cash-flow function $\Delta_{T_k}$ for each cash flow. With each
market factor $X^{\alpha}_{T_k}$ we associate an information
process $\{\xi^{\alpha}_{tT_k}\}_{0\leq t\leq T_k}$ of the form
\begin{eqnarray}
\xi^{\alpha}_{tT_k}=\sigma^{\alpha}_{T_k}X^{\alpha}_{T_k}t+
\beta^{\alpha}_{tT_k}.
\end{eqnarray}
Here $\sigma^{\alpha}_{T_k}$ is an information flux parameter, and
$\{\beta^{\alpha}_{tT_k}\}$ is a standard Brownian bridge process
over the interval $[0,T_k]$. We assume that the $X$-factors and
the Brownian bridge processes are all independent. The parameter
$\sigma^{\alpha}_{T_k}$ determines the rate at which information
about the value of the market factor $X^{\alpha}_{T_k}$ is
revealed. The Brownian bridge $\beta^{\alpha}_{tT_k}$ represents
the associated noise. We assume that the market filtration
$\{{\mathcal F}_t\}$ is generated by the totality of the
independent information processes $\{\xi^{\alpha}_{tT_k}\}_{0\leq
t\leq T_k}$ for $k=1,2,\ldots,n$ and $\alpha=1,2,\ldots,m_k$.
Hence, the price process of the asset is given by
\begin{eqnarray}
S_t=\sum^n_{k=1}{\bf 1}_{\{t<T_k\} }P_{tT_k} {\mathbb E}^{{\mathbb
Q}}\left[D_{T_k}\bigg\vert{\mathcal F}_t\right]. \label{eq:42-1}
\end{eqnarray}
Again, here $\Q$ represents the risk-neutral measure, and the
default-free interest rate term structure is assumed to be
deterministic.
%
%
%
\section{Simple model for dividend growth}\label{sec6.2}
As an elementary example of a multi-dividend structure, we shall
look at a simple growth model for dividends in the equity markets.
We consider an asset that pays a sequence of dividends $D_{T_k}$,
where each dividend date has an associated $X$-factor. Let
$\{X_{T_k}\}_{k=1,\ldots,n}$ be a set of independent,
identically-distributed $X$-factors, each with mean $1+g$. The
dividend structure is assumed to be of the form
\begin{eqnarray}
D_{T_k}=D_0\prod^k_{j=1}X_{T_j},
\end{eqnarray}
where $D_0$ is a constant. The parameter $g$ can be interpreted as
the dividend growth factor, and $D_0$ can be understood as
representing the most recent dividend before time zero. For the
price process of the asset we have:
\begin{eqnarray}
S_t=D_0\sum^n_{k=1}{\bf 1}_{\{t<T_k\}}P_{tT_k}{\mathbb
E}^{{\mathbb Q}}\left[\prod^k_{j=1}X_{T_j}\bigg\vert{\mathcal
F}_t\right].
\end{eqnarray}
Since the $X$-factors are independent of one another, the
conditional expectation of the product appearing in this
expression factorises into a product of conditional expectations,
and each such conditional expectation can be written in the form
of an expression of the type we have already considered. As a
consequence we are led to a completely tractable family of
dividend growth models.
\section{A natural class of stochastic volatility models}\label{sec6.3}
Based on the general model introduced in the previous section, we
are now in a position to make an observation concerning the nature
of stochastic volatility. In particular, we shall show how
stochastic volatility arises in the information-based
framework. This is achieved without the need for any \textit{ad
hoc} assumptions concerning the dynamics of the stochastic
volatility. In fact, a very specific dynamical model for
stochastic volatility is obtained---thus leading in principal to a
possible means by which the theory proposed here might be tested.

We shall work out the volatility associated with the dynamics of
the general asset price process $\{S_t\}$ given by equation
(\ref{eq:42-1}). The result is given in Proposition \ref{propMDA}
below.

First, as an example, we consider the dynamics of an asset that
pays a single dividend $D_T$ at time $T$. We assume that the
dividend depends on a set of market factors $\{X^{\alpha}_T
\}_{\alpha=1,\ldots,m}$. For $t<T$ we then have:
\begin{eqnarray}\label{GenAPP}
S_t&=&P_{tT}{\mathbb E}^{{\mathbb Q}}
\left[\left.\Delta_{T}\left(X^1_{T},\ldots,X^m_{T}\right)
\right|\xi^1_{tT},\ldots,\xi^m_{tT}\right] \nonumber \\
&=& P_{tT}\int\cdots\int\Delta_{T}(x^1,\ldots,x^m)\, \pi^1_{tT}
(x_1)\cdots\pi^{m}_{tT}(x_{m})\,\rd x_1\cdots \rd x_{m}.
\end{eqnarray}
Here the various conditional probability density functions
$\pi^{\alpha}_{tT}(x)$ for $\alpha=1,\ldots,m$ are
\begin{eqnarray}
\pi^{\alpha}_{tT}(x)=\frac{p^{\alpha}(x)\exp\left[\frac{T}{T-t}
\left(\sigma^{\alpha}\,x\,\xi^{\alpha}_{tT}-\tfrac{1}{2}
(\sigma^{\alpha})^2\,x^2 t\right)\right]}
{\int^{\infty}_{0}\,p^{\alpha}(x)
\exp\left[\frac{T}{T-t}\left(\sigma^{\alpha}\,x\,
\xi^{\alpha}_{tT}-\tfrac{1}{2}(\sigma^{\alpha})^2\,x^2
t\right)\right]\rd x},
\end{eqnarray}
where $p^\alpha(x)$ denotes the \textit{a priori} probability
density function for the market factor $X_T^\alpha$. The drift of
$\{S_t\}_{0\leq t<T}$ is given by the short rate of interest. This
is because ${\mathbb Q}$ is the risk-neutral measure, and no
dividend is paid before $T$.

Thus, we are left with the problem of determining the volatility
of $\{S_t\}$. We find that for $t<T$ the dynamical equation of
$\{S_t\}$ assumes the following form:
\begin{eqnarray}
\rd S_t=r_t S_t\rd t+\sum^m_{\alpha=1}\Gamma^{\alpha}_{tT}\rd
W^{\alpha}_t. \label{eq:50}
\end{eqnarray}
Here the volatility term associated with factor number $\alpha$ is
given by
\begin{eqnarray}
\Gamma^{\alpha}_{tT}=\sigma^{\alpha} \frac{T}{T-t}
P_{tT}\,\textrm{Cov}\left[\left.\Delta_{T}\left(
X^{1}_{T},\ldots,X^{m}_{T}\right),X^{\alpha}_{T} \right|{\mathcal
F}_t\right], \label{vol}
\end{eqnarray}
and $\{W_t^\alpha\}$ denotes the Brownian motion associated with
the information process $\{\xi_t^\alpha\}$, as defined in
(\ref{eq:2-16}). The absolute volatility of $\{S_t\}$ is evidently
of the form
\begin{eqnarray}
\Gamma_t=\left(\textstyle{\sum\limits_{\alpha=1}^m}
\left(\Gamma^{\alpha}_{tT} \right)^2\right)^{1/2}.
\end{eqnarray}
For the dynamics of a multi-factor single-dividend-paying asset we
can thus write
\begin{eqnarray}
\rd S_t=r_t S_t\rd t+\Gamma_t \rd Z_t,
\end{eqnarray}
where the $\{{\mathcal F}_t\}$-Brownian motion $\{Z_t\}$ that
drives the asset-price process is defined by
\begin{eqnarray}
Z_t=\int_0^t\frac{1}{\Gamma_s}\sum^m_{\alpha=1}
\Gamma^{\alpha}_{sT}\,\rd W^{\alpha}_{s}.
\end{eqnarray}

The key point is that in the case of a multi-factor
model we obtain an unhedgeable stochastic volatility. That is,
although the asset price is in effect driven by a single Brownian
motion, its volatility depends on a multiplicity of
Brownian motions. This means that in general an option position
cannot be hedged with a position in the underlying asset. The
components of the volatility vector are given by the covariances
of the terminal cash flow with the independent market factors.
Unhedgeable stochastic volatility emerges from the multiplicity of
uncertain elements in the market that affect the value of the
future cash flow. As a consequence we see that \emph{in this
framework we obtain a possible explanation for the origin of
stochastic volatility}.

This result can be contrasted with, say, the Heston model (Heston
1993), which despite its popularity suffers somewhat from the fact
that it is essentially \textit{ad hoc} in nature. Much the same
has to be said for the various generalisations of the Heston model
that have been so widely used in commercial applications. The
approach to stochastic volatility proposed in this thesis is thus
of a new character.

Of course, stochastic volatility is the ``rule" rather than the
exception in asset price modelling---that is to say, the
``generic" model will have stochastic volatility; the problem is,
to select on a rational basis a natural class of stochastic
volatility models from the myriad of possible such models. In the
present analysis, what we mean by ``rational basis" is that the
model is deduced from a set of specific, simple assumptions
concerning the underlying cash flows and market filtration. It is
an open question whether some of the well-known stochastic
volatility models can be re-derived from an information-based
perspective.

Expression (\ref{eq:50}) generalises naturally to the case in
which the asset pays a set of dividends $D_{T_k}$ $(k=1,
\ldots,n)$, and for each $k$ the dividend depends on the
$X$-factors $\{\{ X^\alpha_{T_j}\}^{\alpha=1,\ldots,m_j}_{j=1,
\ldots,k}\}$. The result can be summarised as below.

\begin{prop}\label{propMDA}
The price process of a multi-dividend asset has the following
dynamics:
\begin{eqnarray}
\rd S_t &=& r_t S_t\rd t\nn\\
        &+& \sum^n_{k=1}{\bf 1}_{\{t<T_k\}}P_{tT_k}\sum^k_{j=1}\sum^{m_j}_{\alpha=1}\frac{\sigma^{\alpha}_j
        T_j}{T_j-t}\,{\rm Cov}_t\left[\Delta_{T_k},X^{\alpha}_{T_j}\right]\rd W^{\alpha j}\nn\\
        &+&\sum^n_{k=1}\Delta_{T_k}\rd{\bf 1}_{\{t<T_k\}},
\end{eqnarray}
where $\Delta_{T_k}=\Delta_{T_k}(X^{\alpha}_{T_1},
X^{\alpha}_{T_2}, \cdots, X^{\alpha}_{T_k})$ is the dividend at
time $T_k$ $(k=1,2,\ldots,n)$.
\end{prop}

We conclude that the multi-factor, multi-dividend situation is
also fully tractable. A straightforward extension of Proposition
\ref{propMDA} then allows us to formulate the joint price dynamics
of a system of assets, the associated dividend flows of which may
depend on common market factors. As a consequence, it follows that
a rather specific model for stochastic volatility and correlation
emerges for such a system of assets, and it is one of the
principal conclusions of this work that such a model, which is
entirely natural in character, can indeed be formulated.

The information-based $X$-factor approach presented here thus
offers new insights into the nature of volatility and correlation,
and as such may find applications in a number of different areas
of financial risk analysis. We have in mind, in particular,
applications to equity portfolios, credit portfolios, and
insurance, all of which exhibit important intertemporal market
correlation effects. We also have in mind the problem of firm-wide
risk management and optimal capital allocation for banking
institutions. A further application of the $X$-factor method may
arise in connection with the modelling of asymmetric information
flows and insider trading, i.e. in stratified markets, where some
participants have better access to information than others, but
all agents act optimally  (cf. F\"ollmer {\it et al}. 1999).
\section{Black-Scholes model from an information-based
perspective}\label{sec6.3a} In the section above we have derived
the dynamics of the price process of a multi-dividend-paying asset
in the case where the dividends depend on a set of market factors,
and we have seen how a model for stochastic volatility emerges in
this context. An interesting question that can be asked now is
whether it is possible to recover the standard Black-Scholes
geometric Brownian motion asset-price model by a particular choice
of the dividend structure and market factors. It is arguable that
any asset pricing framework with a claim of generality should
include the Black-Scholes asset price model as a special case. The
set-up we consider is the following.

We consider a limited-liability asset that pays no interim
dividends, and that at time $T$ is sold off for the value $S_T$.
Thus in the present example $S_T$ plays the role of the ``single"
cash flow $\Delta_T$. Our goal is to find the price process
$\{S_t\}_{0\le t\le T}$ of such an asset. In particular, we look
at the case when $S_T$ is log-normally distributed and is of the
form
\begin{equation}
S_T=S_0\exp\left(rT-\tfrac{1}{2}\nu^2 T+\nu\sqrt{T}X_T\right),
\end{equation}
where $S_0$, $r$, and $\nu$ are given constants and $X_T$ is a
standard normally distributed random variable. The corresponding
information process is given by
\begin{equation}
\xi_t=\sigma X_T t+\beta_{tT},
\end{equation}
and the price process $\{S_t\}_{0\le t\le T}$ is then obtained by
using (\ref{GenAPP}). The result is:
\begin{equation}\label{GenBS}
S_t={ \bf 1}_{\{t<T\}}P_{tT}S_0\exp\left[rT-\frac{1}{2}\nu^2
T+\frac{1}{2}\frac{\nu\sqrt{T}}{\sigma^2\tau+1}+\frac{\nu\sqrt{T}\sigma\tau}{t(\sigma^2\tau+1)}\,\xi_t\right],
\end{equation}
where we set $\tau=tT/(T-t)$.

The dynamics of a single-dividend paying asset, in the case that
the dividend is a function of a single random variable, are given
by the following stochastic differential equation, which is a special case of Proposition \ref{propMDA}:
\begin{equation}
\rd S_t=rS_t\rd t+\frac{\sigma
T}{T-t}\textrm{Cov}_t\left[S_T,X_T\right]\rd W_t.
\end{equation}
The conditional covariance $\textrm{Cov}_t\left[S_T,X_T\right]$
between the random variables $S_T$ and $X_T$ can be written as
follows:
\begin{equation}\label{ConCov}
\textrm{Cov}_t\left[S_T,X_T\right]=\E_t\left[S_T
X_T\right]-\E_t\left[S_T\right]\E_t\left[X_T\right].
\end{equation}
The conditional expectation, with respect to $\xi_t$, of a
function $f(X_T)$ is
\begin{equation}
\E_t\left[f\left(X_T\right)\right]=\frac{\int^{\infty}_{-\infty}\,f(x)p(x)\exp\left[\frac{T}{T-t}(\sigma
x\xi_t- \tfrac{1}{2}\sigma^2 x^2 t)\right]\rd
x}{\int^{\infty}_{-\infty}\,p(x)\exp\left[\frac{T}{T-t}(\sigma
x\xi_t- \tfrac{1}{2}\sigma^2 x^2 t)\right]\rd x},
\end{equation}
where $p(x)$ is the {\it a priori} probability density function
associated with the random variable $X_T$. Since we assume that
$X_T$ is standard normally distributed we have,
\begin{equation}
p(x)=\frac{1}{\sqrt{2\pi}}\exp\left(-\tfrac{1}{2}x^2\right).
\end{equation}
The computation of the conditional covariance requires
the results of four integrals of the form
\begin{equation}
\int^{\infty}_{-\infty}\,f(x)p(x)\exp\left[\frac{T}{T-t}(\sigma x\xi_t- \tfrac{1}{2}\sigma^2 x^2 t)\right]\rd x,
\end{equation}
with $f(x)=1$, $f(x)=x$, $f(x)=S(x)$, and $f(x)=xS(x)$, where
\begin{equation}
S(x)=S_0\exp\left(rT+\nu\sqrt{T}x-\frac{1}{2}\nu^2 T\right).
\end{equation}
To proceed it will be useful to have at our disposal the following two well-known Gaussian integrals, with parameters $a\in\mathbb{R}$ and $b\in\mathbb{R}^+$. These are:
\begin{align}\label{GenInt1}
&\frac{1}{\sqrt{2\pi}}\int^{\infty}_{-\infty}\,\exp\left(-\tfrac{1}{2}\,x^2\right)\exp\left(ax-bx^2\right)\rd
x=\frac{1}{\sqrt{2b+1}}\exp\left(\frac{1}{2}\frac{a^2}{2b+1}\right),\\
\phantom{x}\nn\\
&\frac{1}{\sqrt{2\pi}}\int^{\infty}_{-\infty}\,\exp\left(-\tfrac{1}{2}\,x^2\right)\,x\,\exp\left(ax-bx^2\right)\rd
x=\frac{a}{(2b+1)^{3/2}}\,\exp\left(\frac{1}{2}\frac{a^2}{2b+1}\right)\label{GenInt2}.
\end{align}
Armed with these results we can now proceed to calculate the
four integrals involved in the computation of the conditional
covariance (\ref{ConCov}). The first of these integrals, for which $f(x)=1$, gives:
\begin{align}\label{I1}
&\frac{1}{\sqrt{2\pi}}\int^{\infty}_{-\infty}\,\exp\left(-\tfrac{1}{2}\,x^2\right)\exp\left[\frac{T}{T-t}(\sigma
x\xi_t- \tfrac{1}{2}\sigma^2 x^2 t)\right]\rd
x\nn\\
&\hspace{7cm}=\frac{1}{\sqrt{\sigma^2\tau+1}}\exp\left(\frac{1}{2}\frac{\sigma^2\tau^2\xi_t^2}{t^2\left(\sigma^2\tau+1\right)}\right),
\end{align}
where for the parameters $a$ and $b$ in (\ref{GenInt1}) we have
\begin{equation}
a=\frac{\sigma T}{T-t}\xi_t,\qquad b=\tfrac{1}{2}\,\sigma^2\tau.
\end{equation}
The second integral, with $f(x)=x$, gives:
\begin{align}\label{I2}
&\frac{1}{\sqrt{2\pi}}\int^{\infty}_{-\infty}\,x\,\exp\left(-\tfrac{1}{2}\,x^2\right)\exp\left[\frac{T}{T-t}(\sigma
x\xi_t- \tfrac{1}{2}\sigma^2 x^2 t)\right]\rd
x\nn\\
&\hspace{7cm}=\frac{\sigma\tau\xi_t}{t(\sigma^2\tau+1)^{3/2}}\exp\left(\frac{1}{2}\frac{\sigma^2\tau^2\xi_t^2}{t^2\left(\sigma^2\tau+1\right)}\right),
\end{align}
where here we use the result (\ref{GenInt2}) with
\begin{equation}
a=\frac{\sigma T}{T-t}\xi_t,\qquad b=\tfrac{1}{2}\,\sigma^2\tau.
\end{equation}
The third integral, with $f(x)=S(x)$, gives:
\begin{align}\label{I3}
&\frac{1}{\sqrt{2\pi}}\int^{\infty}_{-\infty}\,\exp\left(-\tfrac{1}{2}\,x^2\right)S_0\exp\left(rT+\nu\sqrt{T}x-\frac{1}{2}\nu^2 T\right)\exp\left[\frac{T}{T-t}(\sigma
x\xi_t- \tfrac{1}{2}\sigma^2 x^2 t)\right]\rd
x\nn\\
&=S_0\exp\left(rT-\frac{1}{2}\nu^2 T\right)\frac{1}{\sqrt{\sigma^2\tau+1}}\exp\left[\frac{\left(\nu\sqrt{T}+\frac{\sigma\tau}{t}\xi_t\right)^2}{2(\sigma^2\tau+1)}\right],
\end{align}
where in (\ref{GenInt1}) we insert
\begin{equation}
a=\nu\sqrt{T}+\frac{\sigma T}{T-t}\xi_t,\qquad b=\tfrac{1}{2}\,\sigma^2\tau.
\end{equation}
Finally we calculate that the fourth integral, with $f(x)=xS(x)$, gives:
\begin{align}\label{I4}
&\frac{1}{\sqrt{2\pi}}\int^{\infty}_{-\infty}\,\exp\left(-\tfrac{1}{2}\,x^2\right)S_0\exp\left(rT+\nu\sqrt{T}x-\frac{1}{2}\nu^2 T\right)\,x\,\exp\left[\frac{T}{T-t}(\sigma
x\xi_t- \tfrac{1}{2}\sigma^2 x^2 t)\right]\rd
x\nn\\
&=S_0\exp\left(rT-\frac{1}{2}\nu^2 T\right)\frac{\nu\sqrt{T}+\frac{\sigma T}{T-t}\xi_t}{(\sigma^2\tau+1)^{3/2}}\,\exp\left[\frac{\left(\nu\sqrt{T}+\frac{\sigma\tau}{t}\xi_t\right)^2}{2(\sigma^2\tau+1)}\right],
\end{align}
where in this case we have
\begin{equation}
a=\nu\sqrt{T}+\frac{\sigma T}{T-t}\xi_t,\qquad b=\tfrac{1}{2}\,\sigma^2\tau.
\end{equation}
The results of the integrals above enable us to calculate the conditional expectations composing the conditional covariance (\ref{ConCov}). In particular, we have:
\begin{eqnarray}\label{EXS}
\E_t\left[X_T S_T\right]=S_0\exp\left(rT-\frac{1}{2}\nu^2 T\right)\frac{\nu\sqrt{T}+\frac{\sigma T}{T-t}\xi_t}{\sigma^2\tau+1}
\exp\left[\frac{1}{2}\frac{\nu\sqrt{T}}{\sigma^2\tau+1}+\frac{\nu\sqrt{T}\sigma\tau}{t(\sigma^2\tau+1)}\xi_T\right].\nn\\
\phantom{x}
\end{eqnarray}
The second term in the conditional covariance is given by
\begin{eqnarray}\label{ESEX}
\E_t\left[S_T\right]\E_t\left[X_T\right]=S_0\exp\left(rT-\frac{1}{2}\nu^2 T\right)\frac{\sigma\tau\xi_t}{t(\sigma^2\tau+1)}
\exp\left[\frac{1}{2}\frac{\nu\sqrt{T}}{\sigma^2\tau+1}+\frac{\nu\sqrt{T}\sigma\tau}{t(\sigma^2\tau+1)}\xi_T\right].\nn\\
\phantom{x}
\end{eqnarray}
We now subtract the term (\ref{ESEX}) from the first term (\ref{EXS}) and after some cancellations we obtain the following formula for the conditional covariance:
\begin{equation}
P_{tT}\textrm{Cov}_t\left[X_T,S_T\right]=\frac{\nu\sigma T^{3/2}}{T+(\sigma^2 T-1)t}\,S_t,
\end{equation}
where the price $S_t$ of the asset at time $t$ is given by equation (\ref{GenBS}).
This expression for $S_t$ is consistent with the formula
\begin{equation}\label{GenBSasset}
S_t={ \bf 1}_{\{t<T\}}P_{tT}\E_t\left[S_T\,\vert\,\xi_t\right].
\end{equation}
In particular, the conditional expectation in (\ref{GenBSasset}) can be computed by use of the integrals (\ref{I1}) and (\ref{I3}). Thus, for the dynamics of the asset price (\ref{GenBSasset}) we have the following relation:
\begin{equation}
\frac{\rd S_t}{S_t}=r\rd t+\frac{\nu\sigma T^{3/2}}{T+(\sigma^2 T-1)t}\,\rd W_t.
\end{equation}
We see therefore that the volatility is a deterministic function. But . . . if! . . .
\begin{equation}\label{butif}
\sigma^2 T=1,
\end{equation}
then the volatility is {\it constant}, and for the asset price dynamics we obtain
\begin{equation}
\frac{\rd S_t}{S_t}=r\rd t+\nu\,\rd W_t.
\end{equation}
In other words, we have a geometric Brownian motion.

An alternative way of deriving this result is as follows. We begin with the expression (\ref{GenBS}) and impose the condition (\ref{butif}). It may not be immediately evident, but one can easily verify in this case that equation (\ref{GenBS}) reduces to
\begin{equation}
S_t=S_0\exp\left(rt+\nu\xi_t-\tfrac{1}{2}\nu^2 t\right).
\end{equation}
Now, in the situation where $X_T$ is a standard Gaussian random
variable, and where $\sigma^2 T=1$, the information process
$\{\xi_t\}$ takes the form
\begin{equation}\label{varinfo}
\xi_t=X_T\frac{t}{\sqrt{T}}+\beta_{tT},
\end{equation}
and, in particular, is a Gaussian process. A short calculation, making use of the fact that $X_T$ and $\{\beta_{tT}\}$ are independent, shows that (\ref{varinfo}) implies that
\begin{equation}
\E\left[\xi_s\xi_t\right]=s
\end{equation}
for $s\le t$. This shows that $\{\xi_t\}_{0\le t\le T}$ is a standard Brownian motion. We note further that $\xi_T=X_T\sqrt{T}$, and thus that
\begin{equation}
\xi_t-\frac{t}{T}\xi_T=\beta_{tT}.
\end{equation}
But this demonstrates that $\{\beta_{tT}\}$ is the natural Brownian
bridge associated with $\{\xi_t\}_{0\le t\le T}$. Thus we have
shown that in the case of a Gaussian $X$-factor and an information
flow rate $\sigma=1/\sqrt{T}$, the information process coincides
with the innovation process, and the noise $\{\beta_{tT}\}$ is the
associated Brownian bridge.
\section{Defaultable $n$-coupon bonds with multiple recovery levels}\label{sec6.4}
Now we consider the case of a defaultable coupon bond where
default can occur at any of the $n$ coupon payment dates. Let us
introduce a sequence of coupon dates $T_k$, $k=1,2,\ldots,n$. At
each date $T_k$ a cash-flow $H_{T_k}$ occurs. We introduce a set
of independent binary random variables $X_{T_j}$, $j=1,\ldots,k$
(the $X$-factors), where they take either the value 0 (default) or
1 (no default) with {\it a priori} probability
$p_{X_j}=\Q(X_{T_j}=0)$ and $1-p_{X_j}=\Q(X_{T_j}=1)$ respectively.\\

\begin{prop}
Let $T_k$, $k=1,2,\ldots,n$, be the pre-specified payment dates.
The coupon is denoted by ${\bf c}$ and the principal by ${\bf p}$.
In case of default at the date $T_k$ we have the recovery payment
$R_k({\bf c}+{\bf p})$ instead, where $R_k$ is a percentage of the
owed coupon and principal payment. At each date $T_k$ the
cash-flow structure is given by:
\\
\newline
For $k=1,2,\ldots,n-1$
\begin{equation}\label{form1a}
    H_{T_k}={\bf c}\prod^k_{j=1}\,X_{T_j}+R_k({\bf c}+{\bf p})\prod^{k-1}_{j=1}X_{T_j}(1-X_{T_k}),
\end{equation}
and for $k=n$
\begin{equation}\label{form1}
    H_{T_n}=({\bf c}+{\bf p})\prod^n_{j=1}\,X_{T_j}+R_n({\bf c}+{\bf p})\prod^{n-1}_{j=1}X_{T_j}(1-X_{T_n}).
\end{equation}
\end{prop}

Now we introduce a set of market information processes, given by
\begin{equation}\label{form2}
    \xi_{tT_j}=\sigma_{j}X_{T_j}t+\beta_{tT_j},
\end{equation}
and we assume that the market filtration $\{\F^{\xi}_t\}$ is
generated collectively by the market information processes.

\begin{prop}
Let the default-free discount bond system $\{P_{tT}\}$ be
deterministic. Then the information-based price process of a
binary defaultable coupon bond is given by
\begin{equation}\label{form4}
    S_{t}=\sum^{n-1}_{k=1}{\bf{1}}_{\{t<T_k\}}P_{tT_k}\E^{\Q}\lb[H_{T_k}\bigg\vert\F^{\boldsymbol\xi}_t\rb]
    +P_{tT_n}{\bf{1}}_{\{t<T_n\}}\E^{\Q}\lb[H_{T_n}\bigg\vert\F^{\boldsymbol\xi}_t\rb],
\end{equation}
where $H_{T_k}$ and $H_{T_n}$ are defined by equations
{\rm(\ref{form1a})} and {\rm(\ref{form1})}.
\end{prop}

Example. The price process of a binary defaultable bond paying a
coupon ${\bf c}$ at the dates $T_1$ and $T_2$, as well as a third
coupon ${\bf c}$ plus the principal ${\bf p}$ at the date $T_3$,
is given for $t<T_1$ by:
\begin{equation}\label{form5}
    S_t=P_{tT_1}\E^{\Q}\lb[H_{T_1}\bigg\vert\F^{\boldsymbol\xi}_t\rb]+
    P_{tT_2}\E^{\Q}\lb[H_{T_2}\bigg\vert\F^{\boldsymbol\xi}_t\rb]+
    P_{tT_3}\E^{\Q}\lb[H_{T_3}\bigg\vert\F^{\boldsymbol\xi}_t\rb],
\end{equation}
where
\begin{eqnarray}
  H_{T_1} &=& {\bf c}X_{T_1}+R_1({\bf c}+{\bf p})(1-X_{T_1}), \\
  H_{T_2} &=& {\bf c}X_{T_1}X_{T_2}+R_2({\bf c}+{\bf p})X_{T_1}(1-X_{T_2}), \\
  H_{T_3} &=& ({\bf c}+{\bf p})X_{T_1}X_{T_2}X_{T_3}+R_3({\bf c}+{\bf
  p})X_{T_1}X_{T_2}(1-X_{T_3}).
\end{eqnarray}
\section{Correlated cash flows}\label{sec6.5}
The multiple-dividend asset pricing model introduced in this
chapter can be extended in a natural way to the situation where
two or more assets are being priced. In this case we consider a
collection of $N$ assets with price processes
$\{S^{(i)}_t\}_{i=1,2,\ldots,N}$. With asset number $(i)$ we
associate the cash flows $\{D^{(i)}_{T_k}\}$ paid at the dates
$\{T_k\}_{k=1,2,\ldots,n}$. We note that the dates
$\{T_k\}_{k=1,2,\ldots,n}$ are not tied to any specific asset, but
rather represent the totality of possible cash-flow dates of any
of the given assets. If a particular asset has no cash flow on one
of the given dates, then it is simply assigned a zero cash-flow
for that date. From this point, the theory proceeds in the single
asset case. That is to say, with each value of $k$ we associate a
set of $X$-factors $X^{\alpha}_{T_k}$ $(\alpha=1,2,\ldots,m_k)$,
and a corresponding system of market information processes,
$\{\xi^{\alpha}_{tT_k}\}$. The $X$-factors and the information
processes are not tied to any particular asset. The cash flow
$D^{(i)}_{T_k}$ occurring at time $T_k$ for asset $(i)$ is assumed
to be given by a cash flow function of the form
\begin{equation}
D^{(i)}_{T_k}=\Delta^{(i)}_{T_k}(X^{\alpha}_{T_1},
X^{\alpha}_{T_2},...,X^{\alpha}_{T_k}).
\end{equation}
In other words, for each asset each cash flow can depend on all of
the $X$-factors that have been ``activated'' so far.

In particular, it is possible for two or more assets to ``share''
an $X$-factor associated with one or more of the cash flows of
each of the assets. This in turn implies that the various assets
will have at least one Brownian motion in common in the dynamics
of their price processes. As a consequence we obtain a natural
model for the existence of correlation structures in the prices of
these assets. The intuition is that as new information comes in
(whether ``true'' or ``bogus'') there will be several different
assets affected by the news, and as a consequence there will be a
correlated movement in their prices. Thus for the general
multi-asset model we have the following price process system:
\begin{equation}
S^{(i)}_t=\sum^n_{k=1}{\bf 1}_{\{t<T_k\}} P_{tT_k} {\mathbb
E}^{{\mathbb Q}} \left[D^{(i)}_{T_k}\left|{\mathcal F}_t
\right.\right].
\end{equation}

As an illustration we imagine the following situation: A big
factory has an outstanding debt that needs to be honoured at time
$T_1$. Across the street there is a restaurant that has also an
outstanding loan that has to be paid back at time $T_2$, where we
assume that $T_1<T_2$.

Let us suppose that the main source of income of the restaurant
comes from the workers of the factory, who regularly have their
lunch at the restaurant. However, the factory has been
encompassing a long period of decreasing profits and as a result
fails to pay its debt. The factory decides to send home almost all
of its workers thus putting, as a side effect, the restaurant in a
financially distressed state. As a consequence, despite its
overall good management, the restaurant fails to repay the loan.

The second scenario is that the factory is financially robust and
honours its debt, but the restaurant has bad management. In this
case the workers continue having their lunch at the restaurant,
but the restaurant is unable to repay its debt.

The third scenario is the worst-case picture, in which the factory
fails to repay its debt and then the restaurant fails to repay its
debt as a result both of its bad management and the factory's
decision to reduce the number of its workers.

These various situations can be modelled as follows in the
$X$-factor theory: The two debts can be viewed as a pair two
zero-coupon bonds. The first discount bond is defined by a cash
flow $H_{T_1}$ at time $T_1$. The second discount bond associated
with the restaurant's debt is defined by a cash flow $H_{T_2}$ at
time $T_2$. We introduce two independent market factors $X_{T_1}$
and $X_{T_2}$ that can take the values zero and one. The three
default scenarios described above can be reproduced by defining
the cash flows $H_{T_1}$ and $H_{T_2}$ in such a way that the
dependence between the two cash flows is captured. This is
equivalent to describing the economic microstructure intertwining
the factory and the restaurant, and is thus a natural way of
constructing the dependence between the businesses by analysing
how the relevant cash flows are linked to each other. For the
particular example presented above the cash flow structure is
given by:
\begin{eqnarray}
H_{T_1}&=&{\bf n}_1 X_{T_1}+R_1 {\bf n}_1(1-X_{T_1})\\
H_{T_2}&=&{\bf n}_2 X_{T_1}X_{T_2}+R^a_{2}{\bf
n}_2(1-X_{T_1})X_{T_2}\nonumber\\
&+&R^b_{2}{\bf n}_2 X_{T_1}(1-X_{T_2})+R^c_2{\bf
n}_2(1-X_{T_1})(1-X_{T_2}).
\end{eqnarray}
Here ${\bf n}_1$ and ${\bf n}_2$ denote the amounts of the two
outstanding debts (or, equivalently, the bond principals). We also
introduce recovery rates $R_1$, $R^a_2$, $R^b_2$, and $R^c_2$ for
the cases when the factory and/or the restaurant are not able to
pay back the loans. The recovery rates take into account the
salvage values that can be extracted in the various scenarios. It
is natural, for example to suppose that $R^b_2>R^a_2>R^c_2$. In
this example, the price of the factory bond will be driven by a
single Brownian motion, whereas the restaurant will have a pair of
Brownian drivers. Since one of these coincides with the driver of
the factory bond, the dynamics of the two bonds will be
correlated; analytic formulae can be derived for the resulting
volatilities and correlations.
\section{From $Z$-factors to $X$-factors}\label{sec6.6}
So far we have always assumed that the relevant market factors
($X$-factors) are independent random variables. In reality, what
one often loosely refers to as ``market factors" tend to be
dependent quantities and it may be a difficult task to find a
suitable set of independent $X$-factors. In this section we
illustrate a method to express a set of dependent factors, which
we shall call $Z$-factors, in terms of a set of $X$-factors. In
particular, the scheme described below allows one to reduce a set
of binary $Z$-factors, i.e.~random variables that can take two
possible values, into a set of binary $X$-factors.

Let us introduce a set of $n$ dependent binary $Z$-factors denoted
$\{Z_j\}_{j=1,\ldots,n}$ for which reduction to a set
$\{X_k\}_{k=1,\ldots,2^n-1}$ of $2^n-1$ independent binary
$X$-factors is required. We shall establish a system of reduction
equations of the form $Z_j=Z_j(X_1,\ldots,X_{2^j-1})$ for
$j=1,\ldots n$. For example, in the case $n=2$ the set
$\{Z_1,Z_2\}$ admits a reduction of the form
\begin{eqnarray}
Z_1&=&Z_1(X_1)\\
Z_2&=&Z_2(X_1, X_2, X_3).
\end{eqnarray}
in terms of a set of three independent $X$-factors.

The idea is to construct an algorithm for the reduction scheme for
any number of dependent binary $Z$-factors. For the two possible
values of the binary random variable $Z_j$ let us write
$\{z_j,\bar{z}_j\}$. The independent $X$-factors will be assumed
to take values in $\{0,1\}$. Since we deal with binary random
variables, one can guess that behind the reduction method a binary
tree structure must play a fundamental role.

Indeed, this is the case, and the reduction for each $Z$-factor
builds upon an embedding scheme. For instance the reduction
equation for three dependent $Z$-factors $\{Z_1,Z_2,Z_3\}$ embeds
the reduction system for a set of two $Z$-factors, which in turn
is based on the reduction for one dependent factor. Before
constructing the reduction algorithm we shall as an example write
down the reduction equations for a set of five $Z$-factors
$\{Z_1,Z_2,\ldots,Z_5\}$. This gives us an opportunity to
recognise the above-mentioned embedding property of the reduction
scheme, and also the pattern of the same which is needed in order
to produce the general reduction algorithm. The reduction
equations for the set $\{Z_1,Z_2,\ldots,Z_5\}$ read as follows.
For each market factor $X$ let us define $\bar{X}=1-X$, and call
$\bar{X}$ the co-factor of $X$. Then the reduction of the
dependent variables $\{Z_1,\ldots,Z_5\}$ in terms of the
independent variables $\{X_1,\ldots,X_{31}\}$ is given explicitly
by the following scheme:
\begin{eqnarray}\label{REZ1}
Z_1=z_1 X_1+\bar{z}_1\bar{X}_1
\end{eqnarray}
\begin{eqnarray}\label{REZ2}
Z_2=X_1(z_2 X_2+\bar{z}_2\bar{X}_2)+\bar{X}_1(z_2
X_3+\bar{z}_2\bar{X}_3)
\end{eqnarray}
\begin{eqnarray}\label{REZ3}
  Z_3 &=& X_1 X_2(z_3 X_4+\bar{z}_3\bar{X}_4)+X_1\bar{X}_2(z_3 X_5+\bar{z}_3\bar{X}_5)  \nn\\
      &+& \bar{X}_1 X_3(z_3 X_6+\bar{z}_3\bar{X}_6)+\bar{X}_1\bar{X}_3(z_3X_7+\bar{z}_3\bar{X}_7)
\end{eqnarray}
\begin{eqnarray}\label{REZ4}
  Z_4 &=& X_1 X_2 X_4(z_4 X_8+\bar{z}_4\bar{X}_8)+X_1 X_2 \bar{X}_4(z_4 X_9+\bar{z}_4\bar{X}_9)  \nn\\
      &+& X_1 \bar{X}_2 X_5(z_4 X_{10}+\bar{z}_4\bar{X}_{10})+X_1 \bar{X}_2 \bar{X}_5(z_4 X_{11}+\bar{z}_4\bar{X}_{11}) \nn\\
      &+& \bar{X}_1 X_3 X_6(z_4 X_{12}+\bar{z}_4\bar{X}_{12})+\bar{X}_1 X_3 \bar{X}_6(z_4 X_{13}+\bar{z}_4\bar{X}_{13}) \nn\\
      &+& \bar{X}_1\bar{X}_3 X_7(z_4 X_{14}+\bar{z}_4\bar{X}_{14})+\bar{X}_1\bar{X}_3 \bar{X}_7(z_4 X_{15}+\bar{z}_4\bar{X}_{15})
\end{eqnarray}
\begin{eqnarray}\label{REZ5}
  Z_5 &=& X_1 X_2 X_4X_8(z_5 X_{16}+\bar{z}_5\bar{X}_{16})+X_1 X_2 X_4\bar{X_8}(z_5 X_{17}+\bar{z}_5\bar{X}_{17})\nonumber\\
      &+& X_1 X_2 \bar{X}_4 X_9(z_5 X_{18}+\bar{z}_5\bar{X}_{18})+X_1 X_2 \bar{X}_4\bar{X}_9(z_5 X_{19}+\bar{z}_5\bar{X}_{19})\nonumber\\
      &+& X_1 \bar{X}_2 X_5X_{10}(z_5 X_{20}+\bar{z}_5\bar{X}_{20})+X_1 \bar{X}_2 X_5\bar{X}_{10}(z_5 X_{21}+\bar{z}_5\bar{X}_{21})\nonumber\\
      &+& X_1 \bar{X}_2\bar{X}_5 X_{11}(z_5 X_{22}+\bar{z}_5\bar{X}_{22})+X_1 \bar{X}_2\bar{X}_5\bar{X}_{11}(z_5 X_{23}+\bar{z}_5\bar{X}_{23})\nonumber\\
      &+& \bar{X}_1 X_3 X_6 X_{12}(z_5 X_{24}+\bar{z}_5\bar{X}_{24})+\bar{X}_1 X_3 X_6\bar{X}_{12}(z_5 X_{25}+\bar{z}_5\bar{X}_{25})\nonumber\\
      &+& \bar{X}_1 X_3\bar{X}_6 X_{13}(z_5 X_{26}+\bar{z}_5\bar{X}_{26})+\bar{X}_1 X_3\bar{X}_6\bar{X}_{13}(z_5 X_{27}+\bar{z}_5\bar{X}_{27})\nonumber\\
      &+& \bar{X}_1\bar{X}_3 X_7 X_{14}(z_5 X_{28}+\bar{z}_5\bar{X}_{28})+\bar{X}_1\bar{X}_3 X_7\bar{X}_{14}(z_5 X_{29}+\bar{z}_5\bar{X}_{29})\nonumber\\
      &+& \bar{X}_1\bar{X}_3\bar{X}_7 X_{15}(z_5 X_{30}+\bar{z}_5\bar{X}_{30})+\bar{X}_1\bar{X}_3\bar{X}_7\bar{X}_{15}(z_5 X_{31}+\bar{z}_5\bar{X}_{31})
\end{eqnarray}

Before we discuss the issue of how to produce an algorithm that
gives the reduction system for any number of dependent market
factors, we address the question of how the {\it a priori}
probability distributions of the independent $X$-factors can be
expressed in terms of the {\it a priori} joint probability
distribution of the dependent $Z$-factors. We will shortly see
that the derivation of the {\it a priori} probability distribution
of the $X$-factors in terms of the {\it a priori} joint
probability distributions of the $Z$-factors is closely related to
the reduction system of the considered set of dependent random
variables.

As an example we investigate a set of three $Z$-factors
$\{Z_1,Z_2,Z_3\}$, for which the corresponding reduction in terms
of $X$-factors is given by equations (\ref{REZ1}), (\ref{REZ2}),
and (\ref{REZ3}) above, and with which the following {\it a
priori} joint probability distribution are associated:
\begin{align}
&q_{z_1 z_2 z_3}:=\Q[Z_1=z_1,Z_2=z_2,Z_3=z_3]=p_{X_1}p_{X_2}p_{X_4} \\
&q_{z_1 z_2\bar{z}_3}:=\Q[Z_1=z_1,Z_2=z_2,Z_3=\bar{z}_3]=p_{X_1}p_{X_2}(1-p_{X_4}) \\
&q_{z_1\bar{z}_2 z_3}:=\Q[Z_1=z_1,Z_2=\bar{z}_2,Z_3=z_3]=p_{X_1}(1-p_{X_2})p_{X_5} \\
&q_{z_1\bar{z}_2\bar{z}_3}:=\Q[Z_1=z_1,Z_2=\bar{z}_2,Z_3=\bar{z}_3]=p_{X_1}(1-p_{X_2})(1-p_{X_5})
\end{align}
\begin{align}
&q_{\bar{z}_1 z_2 z_3}:=\Q[Z_1=\bar{z}_1,Z_2=z_2,Z_3=z_3]=(1-p_{X_1})p_{X_3}p_{X_6} \\
&q_{\bar{z}_1 z_2 \bar{z}_3}:=\Q[Z_1=\bar{z}_1,Z_2=z_2,Z_3=\bar{z}_3]=(1-p_{X_1})p_{X_3}(1-p_{X_6}) \\
&q_{\bar{z}_1\bar{z}_2 z_3}:=\Q[Z_1=\bar{z}_1,Z_2=\bar{z}_2,Z_3=z_3]=(1-p_{X_1})(1-p_{X_3})p_{X_7} \\
&q_{\bar{z}_1\bar{z}_2\bar{z}_3}:=\Q[Z_1=\bar{z}_1,Z_2=\bar{z}_2,Z_3=\bar{z}_3]=(1-p_{X_1})(1-p_{X_3})(1-p_{X_7})
\end{align}
Recalling that $\bar{X}_k=1-X_k$ and hence
$p_{\bar{X}_k}=1-p_{X_k}$, we recognise that each joint
probability distribution is associated with the corresponding term
in the reduction equation for the set $\{Z_1,Z_2,Z_3\}$. For
instance $q_{z_1 z_2 z_3}$ is the probability distribution
connected with the first term in (\ref{REZ3}), i.e.~$X_1 X_2 X_4
z_3$.

The relations above can be inverted to give the corresponding
univariate probability distributions for the independent
$X$-factors in terms of the joint distributions of the
$Z$-factors:
\begin{equation}
p_{X_1}=q_{z_1 z_2 z_3}+q_{z_1 z_2\bar{z}_3}+q_{z_1\bar{z}_2
z_3}+q_{z_1\bar{z}_2 \bar{z}_3}\nn
\end{equation}
\begin{align}
&p_{X_2}=\frac{q_{z_1 z_2 z_3}+q_{z_1 z_2\bar{z}_3}}{q_{z_1 z_2
z_3}+q_{z_1 z_2\bar{z}_3}+q_{z_1\bar{z}_2 z_3}+q_{z_1\bar{z}_2
\bar{z}_2}} &p_{X_3}=\frac{q_{\bar{z}_1 z_2 z_3}+q_{\bar{z}_1 z_2
\bar{z}_3}}{q_{\bar{z}_1 z_2 z_3}+q_{\bar{z}_1 z_2
\bar{z}_3}+q_{\bar{z}_1\bar{z}_2
z_3}+q_{\bar{z}_1\bar{z}_2\bar{z}_3}}\nn\\
\nn\\
&p_{X_4}=\frac{q_{z_1 z_2 z_3}}{q_{z_1 z_2 z_3}+q_{z_1
z_2\bar{z}_3}} &p_{X_5}=\frac{q_{z_1\bar{z}_2
z_3}}{q_{z_1\bar{z}_2 z_3}+q_{z_1\bar{z}_2 \bar{z}_3}}\nn
\end{align}
\begin{align}
&p_{X_6}=\frac{q_{\bar{z}_1 z_1 z_2}}{q_{\bar{z}_1 z_2
z_3}+q_{\bar{z}_1 z_2 \bar{z}_3}}
&p_{X_7}=\frac{q_{\bar{z}_1\bar{z}_2 z_3}}{q_{\bar{z}_1\bar{z}_2
z_3}+q_{\bar{z}_1\bar{z}_2\bar{z}_3}}.\nn\\
\phantom{}
\end{align}
It is a short calculation to verify that the resulting system of
univariate probabilities is consistent in the sense that these
probabilities all lie in the range $[0,1]$.
\section{Reduction algorithm}\label{sec6.7}
In this section we show how to produce the reduction system for a
specific set of dependent random variables once a fixed $j$ has
been chosen; that is to say we show how to determine the function
\begin{equation}
Z_j=Z_j(X_1,X_2,\ldots,X_{2^{j}-1}).
\end{equation}
In particular, we aim at a reduction algorithm that allows us to
find the reduction equation for $Z_j$ directly without needing to
produce all the reduction equations for all other $Z$-factors
$\{Z_1,\ldots,Z_{j-1}\}$ implicitly involved due to the embedding
property.

To give an example, we would like for the reduction algorithm to
give us the reduction system (\ref{REZ4}) without us needing to
write down explicitly the equations (\ref{REZ1}), (\ref{REZ2}),
and (\ref{REZ3}) for $Z_1$, $Z_2$, and $Z_3$ respectively.

A possible application of a reduction scheme could be the
situation where one has economic quantities which are clearly
dependent on each other and can be represented by a system of
dependent binary random variables. In this case one may be
interested in the dependence structure linking the various
economic factors, and in being able to describe the dependence in
terms of a set of independent $X$-factors. These would be viewed
as the ``fundamental" quantities in the economy producing the
dependence among the economic quantities modelled by the set of
$Z$-factors.

Let us assume a set of $n$ $Z$-factors $\{Z_j\}_{j=1,\ldots,n}$.
We now write down the reduction algorithm for $Z_j$ for a number $j\in\{1,\ldots,n\}$
in a series of steps.\\

{\bf Step 1}. We observe first that the reduction equation for
$Z_j$ contains $2^{j-1}$ so-called $X$-terms, where an $X$-term is
a summand of the form
\begin{equation}
X\cdots X(z_j X+\bar{z}_j\bar{X}).
\end{equation}
For example, $Z_3$ is composed by four $X$-terms, as can be
verified in formula (\ref{REZ3}).\\

{\bf Step 2}. In order to describe the building blocks of the
reduction procedure we introduce some further terminology. An
$X$-term, for fixed $k<j$, of the form
\begin{equation}
X\cdots X_k(\bar{z}_j X_{2k}+\bar{z}_j\bar{X}_{2k})
\end{equation}
is called branch number $k$, and an $X$-term of the form
\begin{equation}
X\cdots \bar{X}_k(\bar{z}_j X_{2k+1}+\bar{z}_j\bar{X}_{2k+1})
\end{equation}
is called co-branch number $k$. For example, the $X$-term $X_1
X_2(z_3 X_4+\bar{z}_3\bar{X}_4)$ in (\ref{REZ3}) is branch number
two, and the $X$-term $\bar{X}_1\bar{X}_3(z_3
X_7+\bar{z}_3\bar{X}_7)$ is the co-branch number three. We also
introduce the concept of a node in the reduction system. The
branch $k$ leads to the ``node" $2k$, i.e.~the term given by:
\begin{equation}
X\cdots X_k\underbrace{(z_j
X_{2k}+\bar{z}_j\bar{X}_{2k})}_{\textrm{node number $2k$}}.
\end{equation}
The co-branch $k$ leads, instead, to the node $2k+1$, that is:
\begin{equation}
X\cdots \bar{X}_k\underbrace{(z_j
X_{2k+1}+\bar{z}_j\bar{X}_{2k+1})}_{\textrm{node number $2k+1$}}.
\end{equation}
For instance, co-branch number three in (\ref{REZ3}),
$\bar{X}_1\bar{X}_3(z_3 X_7+\bar{z}_3\bar{X}_7)$, leads to
node number seven.\\

{\bf Step 3}. The first branch of the reduction system for $Z_j$
is given by:
\begin{equation}
Z_j=X_1\cdots X_{2^{j-2}}(z_j
X_{2^{j-1}}+\bar{z}_j\bar{X}_{2^{j-1}})+\ldots
\end{equation}
To the first branch we then add the corresponding co-branch, i.e.
\begin{equation}
Z_j=X_1\cdots X_{2^{j-2}}(z_j
X_{2^{j-1}}+\bar{z}_j\bar{X}_{2^{j-1}})+X_1\cdots
\bar{X}_{2^{j-2}}(z_j
X_{2^{j-1}+1}+\bar{z}_j\bar{X}_{2^{j-1}+1})\ldots
\end{equation}
and declare branch number $2^{j-2}$ complete. Here we adopt the
convention, for convenience, that $2^{j-2}=0$ for all $j<2$. We set $X_0=1$.\\

{\bf Step 4}. Before we continue with the remaining terms in the
reduction equation for $Z_j$, we need to establish two rules,
which we call ``connection rules". These rules tell us how to
``hop" from one branch or co-branch to the next.

(i) If the preceding independent random variable is an $X$-factor,
then the next market factor is obtained by doubling the index of
the predecessor, that is to say:
\begin{equation}
X\cdots X_k\longrightarrow X\cdots X_k\cdot\left\{
\begin{array}{l}
X_{2k}\\
\phantom{}\\
\bar{X}_{2k}\ .
\end{array}
\right.
\end{equation}

Example: In the reduction equation (\ref{REZ4}) we
have in the first term the expression $X_1 X_2 X_4(z_4
X_8+\bar{z}_4\bar{X}_8)$. We see that all $X$-factor indexes are
doubles of the predecessors.

(ii) If the preceding factor is an $X$-co-factor, i.e.~of the form
$\bar{X}_k=1-X_k$, then the next factor is obtained by doubling
the index of the predecessor and adding one; that is to say:
\begin{equation}
X\cdots \bar{X}_k\longrightarrow X\cdots \bar{X}_k\cdot\left\{
\begin{array}{l}
X_{2k+1}\\
\phantom{}\\
\bar{X}_{2k+1}\ .
\end{array}
\right.
\end{equation}
\\
Example: In the reduction equation (\ref{REZ4}) we have,
respectively, in the third and in the fourth terms, $X_1 \bar{X}_2
X_5(z_4 X_{10}+\bar{z}_4\bar{X}_{10})$ and $X_1 \bar{X}_2
\bar{X}_5(z_4 X_{11}+\bar{z}_4\bar{X}_{11})$, where the
$X$-co-factor $\bar{X}_2$ leads, respectively, to branch number 5
and co-branch number 5.

Now we return to the task of reducing the dependent factor $Z_j$
into its independent constituents. We started with the first
$X$-term (or branch, in this case) given by
\begin{equation}
X_1\cdots X_{2^{j-2}}(z_j X_{2^{j-1}}+\bar{z}_j\bar{X}_{2^{j-1}}),
\end{equation}
and added its complementary co-branch
\begin{equation}
X_1\cdots \bar{X}_{2^{j-2}}(z_j
X_{2^{j-1}+1}+\bar{z}_j\bar{X}_{2^{j-1}+1}).
\end{equation}

Once we have added to a particular branch its complementary
co-branch, we then say that the specific branch/co-branch pair is
complete. We then work back through all preceding branches using
the connection rules and completing the various branches. For the
case of $Z_j$ this is carried out as follows: To the first branch
and its complementary co-branch
\begin{equation}
X_1\cdots X_{2^{j-2}}(z_j
X_{2^{j-1}}+\bar{z}_j\bar{X}_{2^{j-1}})+X_1\cdots
\bar{X}_{2^{j-2}}(z_j X_{2^{j-1}+1}+\bar{z}_j\bar{X}_{2^{j-1}+1})
\end{equation}
we add the term
\begin{eqnarray}
&\phantom{+}&X_1\cdots \bar{X}_{2^{j-3}}X_{2^{j-2}+1}(z_j
X_{2(2^{j-2}+1)}+\bar{z}_j\bar{X}_{2(2^{j-2}+1)})\nn\\
&+&X_1\cdots \bar{X}_{2^{j-3}}\bar{X}_{2^{j-2}+1}(z_j
X_{2(2^{j-2}+1)+1}+\bar{z}_j\bar{X}_{2(2^{j-2}+1)+1}).
\end{eqnarray}
Hence branch number $2^{j-3}$ is complete as well. The next step
is to add branch number $2^{j-4}$ and then to complete it with its
complementary co-branch number $2^{j-4}$. This procedure stops
once all branches are added up and completed, where the last pair
is composed of the two final $X$-terms, namely the final branch
and final co-branch:
\begin{eqnarray}
&\phantom{+}&\bar{X}_1\bar{X}_3\bar{X}_7\cdots X_{2^{j-1}-1}(z_j
X_{2(2^{j-1}-1)}+\bar{z}_j\bar{X}_{2(2^{j-1}-1)})\nn\\
&+&\bar{X}_1\bar{X}_3\bar{X}_7\cdots \bar{X}_{2^{j-1}-1}(z_j
X_{2(2^{j-1}-1)+1}+\bar{z}_j\bar{X}_{2(2^{j-1}-1)+1}).
\end{eqnarray}

At this point the reduction procedure is complete and we obtain
the full reduction of the $Z$-factor $Z_j$. Recapping, the
reduction for $Z_j$ is given by:
\begin{eqnarray}
Z_j&=&X_1\cdots \bar{X}_{2^{j-2}}(z_j
X_{2^{j-1}+1}+\bar{z}_j\bar{X}_{2^{j-1}+1})\nn\\
&+&X_1\cdots \bar{X}_{2^{j-3}}X_{2^{j-2}+1}(z_j
X_{2(2^{j-2}+1)}+\bar{z}_j\bar{X}_{2(2^{j-2}+1)})\nn\\
&\phantom{x}&\qquad+\ X_1\cdots
\bar{X}_{2^{j-3}}\bar{X}_{2^{j-2}+1}(z_j
X_{2(2^{j-2}+1)+1}+\bar{z}_j\bar{X}_{2(2^{j-2}+1)+1})\nn\\
&+&\ldots\nn\\
&\vdots&\nonumber\\
&+&\bar{X}_1\bar{X}_3\bar{X}_7\cdots X_{2^{j-1}-1}(z_j
X_{2(2^{j-1}-1)}+\bar{z}_j\bar{X}_{2(2^{j-1}-1)})\nn\\
&\phantom{x}&\qquad+\ \bar{X}_1\bar{X}_3\bar{X}_7\cdots
\bar{X}_{2^{j-1}-1}(z_j
X_{2(2^{j-1}-1)+1}+\bar{z}_j\bar{X}_{2(2^{j-1}-1)+1}).
\end{eqnarray}
This completes the derivation of the reduction equation for the
dependent $Z$-factor $Z_j$, where $j\in\{1,2,\ldots,n\}$, and also
the derivation of a general reduction scheme for the
disentanglement of a set of dependent binary random variables into
to a set of binary $X$-factors.

We have confined the discussion to the case of a set of dependent
binary $Z$-factors, and we have shown how such a set can be
``reduced" to a corresponding set of independent $X$-factors. In
practice, of course, we would like to have a similar collection of
results for wider classes of dependent random variables. For the
moment, the binary case offers a useful heuristic motivation for
the notion that in a financial context we may assume the
existence, as a basis for our modelling framework, of a set of
underlying independent $X$-factors upon which the observed
$Z$-factors depend.
\section{Information-based Arrow-Debreu securities and option pricing}\label{sec6.8}
In Section \ref{sec:3.5} we presented the concept of the
information-based Arrow-Debreu technique applied to an asset with
cash flows modelled by discrete random variables. In this section
we make a further step forward and present the technique for the
case that cash flows are expressed in terms of continuous random
variables. We shall construct the information-based price of an
Arrow-Debreu security having in mind, as an example, a
single-dividend-paying asset defined in terms of a single cash
flow $D_T$ occurring at time $T$. The price process of such an
asset is given in the information-based approach by:
\begin{eqnarray}
S_t = { \bf 1}_{\{t<T\}} P_{tT}\frac{\int^{\infty}_0 z
p(z)\exp\left[ \frac{T}{T-t}(\sigma z\xi_t-\tfrac{1}{2}\sigma^2
z^2 t)\right]\rd z}{\int^{\infty}_0
p(z)\exp\left[\frac{T}{T-t}(\sigma z\xi_t-\tfrac{1}{2}\sigma^2 z^2
t)\right]\rd z}, \label{infobasedasset}
\end{eqnarray}
where $p(z)$ is the {\it a priori} probability density for a
continuous random variable $D_T$ that takes values in the range
$z\in[0,\infty)$. We remark, incidentally, that in the case of an
$X$-factor with a more general distribution, not necessarily
continuous, then appropriate formulae can be obtained by replacing
$p(z)\rd z$ with $\mu(\rd z)$ in the formula above, and elsewhere.
For simplicity we will refer here to the continuous case. The
information process $\{\xi_t\}$ associated with the cash flow
$D_T$ is given by
\begin{eqnarray}\label{DTinfoproc}
\xi_t=\sigma D_{T} t+\beta_{tT}.
\end{eqnarray}
We observe that, since the maturity date $T$ is a fixed
pre-specified date, the price $S_t$ at time $t$ is given by a
function of the value of the information process at time $t$. That
is to say $S_t=S(\xi_t,t)$. Since the asset price $S_t$ is a
positive strictly-increasing function of $\{\xi_t\}$, it is
possible to invert the function $S(t,\xi_t)$ such that the
information process at time $t$ can be given in a unique way in
terms of the asset price $S_t$ at the time $t$. In the special
case that the cash flow $D_T$ is given in terms of a binary random
variable (reverting briefly to the discrete case), it is possible
to express the information process $\{\xi_t\}$ in the following
form:
\begin{equation}\label{infoprocbinaryasset}
\xi_t=\frac{T-t}{\sigma(h_1-h_0)T}\ln\left[\frac{p_0(P_{tT}d_0-S_t)}{p_1(S_t-P_{tT}d_1)}\right]
+\tfrac{1}{2}\sigma(h_1+h_0)\frac{t(T-t)}{T},
\end{equation}
where $p_0$ and $p_1$ are the probabilities that $D_T$ takes the
values $h_0$ and $h_1$. This shows that the price of an
Arrow-Debreu security can in principal be calibrated by use of
other assets with which the information process is associated.

Following the notation of Section \ref{sec:3.5}, we now define the
payoff of an Arrow-Debreu security where the underlying is given
by the information process $\{\xi_t\}$. Thus for the payoff we
have:
\begin{equation}\label{ADpayoffdelta}
f(\xi_t)=\delta(\xi_t-x),
\end{equation}
where $\delta(\xi_t)$ is the delta function. As before, the value
of the Arrow-Debreu security is given by the discounted
risk-neutral expectation of the payoff above:
\begin{equation}\label{ADpricedelta}
A_{0t}(x)=P_{0t}\E^{Q}[\delta(\xi_t-x)].
\end{equation}
The only difference between the expressions (\ref{ADpricedelta})
and (\ref{eisprice}), is that the random variable $D_T$ with which
the information process $\{\xi_t\}$ in \ref{ADpricedelta} is
associated is assumed to be continuous.

Following through the same steps as in Section (\ref{sec:3.5}), we
now calculate explicitly the Arrow-Debreu price $A_{0t}$. For the
delta function we use the representation
\begin{equation}\label{deltarepresentation}
\delta(\xi_t-x)=\frac{1}{2\pi}\int^{\infty}_{-\infty}\re^{\im(\xi_t-x)\kappa}\,\rd\kappa.
\end{equation}
Inserting (\ref{deltarepresentation}) into the expectation in
equation (\ref{ADpricedelta}), we now calculate as follows. We
shall assume that $D_T$ has properties sufficient to ensure that
\begin{equation}\label{eq7}
\E^{Q}[\delta(\xi_t-x)]=\frac{1}{2\pi}\int^{\infty}_{-\infty}\re^{-\im
x\kappa}\,\E^{\Q}[\re^{\im\kappa\xi_t}]\,\rd\kappa.
\end{equation}
We insert the definition of the information process
(\ref{DTinfoproc}) into the expectation under the integral, and
recall that $D_T$ and $\{\beta_{tT}\}$ are independent, and the
fact that $\{\beta_{tT}\}$ is normally distributed with mean zero
and variance $t(T-t)/T$. This leads us to the relation
\begin{equation}\label{eq8}
    \E^{\Q}\left[\e^{\im\kappa\xi_t}\right]=\int^{\infty}_0\,p(z)
    \e^{\im\kappa\sigma zt-\frac{1}{2}\kappa^2\frac{t(T-t)}{T}}\,\rd z.
\end{equation}
where $p(z)$ is the {\it a priori} density for $D_T$. Hence, the
expectation in (\ref{ADpricedelta}) turns out to be
\begin{equation}\label{eq9}
    \E^{Q}[\delta(\xi_t-x)]=\frac{1}{2\pi}\int^{\infty}_{-\infty}\re^{-\im
    x\kappa}\int^{\infty}_{0}\,p(z)\,\re^{\im\kappa\sigma
    z t-\frac{1}{2}\kappa^2\frac{t(T-t)}{T}}\,\rd z\,\rd\kappa.
\end{equation}
If we swap the integration order and re-arrange slightly the terms
in the exponent we obtain
\begin{equation}\label{eq10}
    \E^{Q}[\delta(\xi_t-x)]=\frac{1}{2\pi}\int^{\infty}_{0}\,p(z)\int^{\infty}_{-\infty}\re^{\im(
    \sigma zt-x)\kappa-\frac{1}{2}\frac{t(T-t)}{T}\kappa^2}\,\rd\kappa\,\rd z.
\end{equation}
Working out the inner integral, we eventually obtain
\begin{equation}\label{eq13}
    A_{0t}(x):=P_{0t}\E^{\Q}[\delta(\xi_t-x)]=P_{0t}\int\limits^{\ \infty}_{0}p(z)\sqrt{\frac{T}{2\pi
    t(T-t)}}\exp\left[-\frac{1}{2}\frac{(\sigma zt-x)^2 T}{t(T-t)}\right]\rd
    z.
\end{equation}
We call $A_{0t}(x)$ the information-based price of an Arrow-Debreu
security for the case that the underlying information process is
associated with a continuous random variable. The price
$A_{0t}(x)$ is the present value of a security paying at time $t$
a delta distribution centered at $\xi_t=x$.

The calculation above serves the purpose of illustrating an
application of the information-based approach to the pricing of an
Arrow-Debreu security. However the derived expressions, especially
equation (\ref{eq13}), can be used to price a series of more complex
derivatives. In Section \ref{sec5.6}  the price of a European-style
call option written on a single-dividend-paying asset with
continuous payoff is calculated by use of a change-of-measure
technique. In what follows we show the derivation of the price of a
vanilla call within the information-based approach, avoiding
changing the measure to the bridge measure used in Sections
\ref{sec:3.2} and \ref{sec5.6}.

The price $C_0$ of a European-style call option written on a
single-dividend-paying asset with price process
(\ref{infobasedasset}) is given by
\begin{equation}\label{eq15}
    C_0=P_{0t}\E^{\Q}\left[(S_t-K)^+\right],
\end{equation}
where $K$ is the strike price, and the option matures at time
$t\le T$. Let us recall that $T$ is the time at which the single
dividend $D_T$ is paid. The idea now is to view the option payoff
as a ``continuum" of delta functions. In more detail, the value of
the payoff
\begin{equation}
C_t=(S(\xi_t,t)-K)^+
\end{equation}
can be regarded as being a continuous ``superposition" of delta
distributions, each of which depends on the value of the
information process at time $t$.

Treating the call option pricing formula no differently than the
formula for an Arrow-Debreu security (\ref{ADpricedelta}), we now
express the value of a call given by (\ref{eq15}) in terms of the
function
\begin{equation}
A(x)=\frac{A_{0t}(x)}{P_{0t}},
\end{equation}
which we call the (non-discounted) Arrow-Debreu density. The price
of a call in terms of the information-based Arrow-Debreu density
is thus:
\begin{equation}\label{eq16}
    C_0=P_{0t}\int^{\infty}_{-\infty}(S(x)-K)^+A(x)\rd x.
\end{equation}
Then we substitute $S(x)$ with the according expression in
(\ref{infobasedasset}) to obtain:
\begin{equation}\label{eq17}
    C_0=P_{0t}\int\limits^{\ \infty}_{-\infty}\left(P_{tT}\frac{\int^{\infty}_0 z
p(z)\exp\left[ \frac{T}{T-t}(\sigma zx-\tfrac{1}{2}\sigma^2 z^2
t)\right]\rd z}{\int^{\infty}_0 p(z)\exp\left[\frac{T}{T-t}(\sigma
zx-\tfrac{1}{2}\sigma^2 z^2 t)\right]\rd z}-K\right)^+A(x)\,\rd x.
\end{equation}
We observe that the Arrow-Debreu density $A(x)$ is a positive
function, which enables us to take it inside the maximum function.
That is to say:
\begin{equation}\label{eq18}
    C_0=P_{0t}\int\limits^{\ \infty}_{-\infty}\left(P_{tT}\frac{\int^{\infty}_0 z
p(z)\exp\left[ \frac{T}{T-t}(\sigma zx-\tfrac{1}{2}\sigma^2 z^2
t)\right]\rd z}{\int^{\infty}_0 p(z)\exp\left[\frac{T}{T-t}(\sigma
zx-\tfrac{1}{2}\sigma^2 z^2 t)\right]\rd
z}\,A(x)-K\,A(x)\right)^+\rd x.
\end{equation}
Next, we rewrite the density $A(x)$ in the form
\begin{equation}\label{eq19}
    A(x)=\sqrt{\frac{T}{2\pi
    t(T-t)}}\,\exp\left[-\frac{T}{2t(T-t)}x^2\right]\int\limits^{\ \infty}_0 p(z)
    \exp\left[\frac{T}{T-t}(\sigma zx-\tfrac{1}{2}\,\sigma^2 z^2 t)\right]\rd z.
\end{equation}
We see that the equation (\ref{eq18}) simplifies due the
cancellation of the denominator in the first term. Thus, we obtain
\begin{eqnarray}\label{eq20}
C_0=P_{0t}\sqrt{\frac{T}{2\pi t(T-t)}}\int\limits^{\ \infty}_{-\infty}\left(P_{tT}\exp\left[-\frac{T}{2t(T-t)}x^2\right]\int\limits^{\ \infty}_{0}p_t(z)z\rd z\right.\nn\\
\left.-K\exp\left[-\frac{T}{2t(T-t)}x^2\right]\int\limits^{\ \infty}_{0}p_t(z)\rd z\right)^+\rd x.
\end{eqnarray}
where $p_t(z)$ is the ``unnormalised" conditional probability
density, given by
\begin{equation}
p_t(z)=p(z)\exp\left[\frac{T}{T-t}\left(\sigma xz-\sigma^2 z^2 t\right)\right].
\end{equation}
We observe that the argument of the maximum function (\ref{eq20})
vanishes when $x$ takes the value $x^{\ast}$---the critical
value---that solves the following equation:
\begin{align}
&P_{tT}\int\limits^{\
\infty}_{0}p(z)z\exp\left[\frac{T}{2t(T-t)}(x^{\ast}-\sigma
zt)^2\right]\rd z-K\int\limits^{\
\infty}_{0}p(z)\exp\left[\frac{T}{2t(T-t)}(x^{\ast}-\sigma
zt)^2\right]\rd z\nn\\
&=0.
\end{align}
Now we define a random variable
\begin{equation}
\eta^{\ast}(z)=x^{\ast}-\sigma zt,
\end{equation}
and re-scale it by the variance of the standard Brownian bridge,
to give us a new variable
\begin{equation}
\nu^{\ast}(z)=\frac{\eta^{\ast}(z)}{\sqrt{\frac{t(T-t)}{T}}}.
\end{equation}
The random variable $\nu^{\ast}$ is normally distributed with zero
mean and unit variance. Then the equation (\ref{eq20}) reads
\begin{eqnarray}
C_0=P_{0t}\left[P_{tT}\int^{\infty}_{0}\,p(z)\,z\left(\frac{1}{\sqrt{2\pi}}\int^{\infty}_{\nu^{\ast}}\,\exp\left(-\frac{1}{2}\nu^2\right)\rd\nu\right)\rd z\right.\\
\left.-K\int^{\infty}_0\,p(z)\left(\frac{1}{\sqrt{2\pi}}\int^{\infty}_{\nu^{\ast}}\,\exp\left(-\frac{1}{2}\nu^2\right)\rd\nu\right)\rd z\right].
\end{eqnarray}
We now make use of the identity
\begin{equation}
\frac{1}{\sqrt{2\pi}}\int^{\infty}_{x}\exp\left(-\frac{1}{2}\eta^2\right)\rd
\eta=N(-x),
\end{equation}
where $N(x)$ is the standard normal distribution function. And
thus, we obtain the price of European-style call option on a
single-dividend-paying asset:
\begin{equation}
C_0=P_{0t}\left[P_{tT}\int^{\infty}_0 p(z)\,zN[-\nu^{\ast}(z)]\rd
z-K\int^{\infty}_0 p(z)\,N[-\nu^{\ast}(z)]\rd z\right].
\end{equation}
Here we note that the above expression coincides with the formula
(\ref{eq:ec1}) derived using the change-of-measure technique
presented in Section \ref{sec5.5}, once this has been adapted to
the case of an asset with a continuous payoff function.

\section{Intertemporal Arrow-Debreu densities}\label{sec6.9}
In this section we develop the Arrow-Debreu price for the case
where we consider a single information process at two distinct
fixed times, $t_1$ and $t_2$ where $t_1\le t_2$. Thus, we write
\begin{eqnarray}
\xi_{t_1}&=&\sigma H_T t_1+\beta_{t_1 T}\label{xi1}\\
\xi_{t_2}&=&\sigma H_T t_2+\beta_{t_2 T}\label{xi2}.
\end{eqnarray}
The random variable $H_T$ is assumed to be discrete. Then we
follow the same scheme as in Section \ref{sec4.2} to compute the
bivariate Arrow-Debreu density
\begin{equation}\label{bivariateAD}
A(x_1,x_2)=\E[\delta(\xi_{t_1}-x_1)\delta(\xi_{t_2}-x_2)].
\end{equation}
It should be evident from properties of the delta function that an
equivalent way of writing (\ref{bivariateAD}) is given by
\begin{equation}
A(x_1,x_2)=\E\left[\delta\left(\xi_{t_1}-\frac{t_1}{t_2}\xi_{t_2}+\frac{t_1}{t_2}x_2-x_1\right)\delta(\xi_{t_2}-x_2)\right].
\end{equation}
\begin{lem}\label{lemma6.9.1}
The random variable
\begin{equation}
\xi_{t_1}-\frac{t_1}{t_2}\xi_{t_2}
\end{equation}
is independent of $\xi_{t_2}$.
\end{lem}
\noindent{\bf Proof}. A short calculation shows that
\begin{equation}
\xi_{t_1}-\frac{t_1}{t_2}\xi_{t_2}=\beta_{t_1
T}-\frac{t_1}{t_2}\beta_{t_2 T}.
\end{equation}
Then we show that
\begin{equation}\label{e7.1}
\beta_{t_1 T}-\frac{t_1}{t_2}\beta_{t_2 T}
\end{equation}
is independent of $\xi_{t_2}$ due to the fact that
{\rm(\ref{e7.1})} is independent of $H_T$, and $\beta_{t_2 T}$.
Expression {\rm(\ref{e7.1})} is by definition independent of $H_T$
and independence of $\beta_{t_2 T}$ is shown by computing the
following covariance:
\begin{eqnarray}
\E\left[\left(\beta_{t_1 T}-\frac{t_1}{t_2}\beta{t_2
T}\right)\beta_{t_2 T}\right]&=&\E[\beta_{t_1 T}\beta_{t_2
T}]-\frac{t_1}{t_2}\E[\beta_{t_2 T}\beta_{t_2 T}]\nonumber\\
&=&\frac{t_1(T-t_2)}{T}-\frac{t_1 t_2(T-t_2)}{t_2 T}\nonumber\\
&=&0.
\end{eqnarray}
\hfill$\Box$

\noindent Making use of Lemma \ref{lemma6.9.1}, we thus obtain
\begin{equation}
A(x_1,x_2)=\E\left[\delta\left(\xi_{t_1}-\frac{t_1}{t_2}\xi_{t_2}+\frac{t_1}{t_2}x_2-x_1\right)\right]\E\left[\delta(\xi_{t_2}-x_2)\right].
\end{equation}
Now we use the Fourier representation of the delta function and
rewrite the Arrow-Debreu density in the form
\begin{eqnarray}
A(x_1,x_2)&=&\E\left[\frac{1}{2\pi}\int^{\infty}_{-\infty}\exp\left[\,\im(\xi_{t_2}-x_2)y_2\right]\rd
y_2\right]\nonumber\\
&\times&\E\left[\frac{1}{2\pi}\int^{\infty}_{-\infty}\exp\left[\im\left(\xi_{t_1}-\frac{t_1}{t_2}\xi_{t_2}+\frac{t_1}{t_2}x_2-x_1\right)y_1\right]\rd
y_1\right].
\end{eqnarray}
After swapping the integral with the expectation we then obtain
\begin{eqnarray}\label{e7.1a}
A(x_1,x_2)&=&\frac{1}{2\pi}\int^{\infty}_{-\infty}\rd
y_1\frac{1}{2\pi}\int^{\infty}_{-\infty}\rd y_2\exp\left[\im x_2
y_2-\im\left(x_1-\frac{t_1}{t_2}x_2\right)y_1\right]\nonumber\\
&\times&\E\left[\exp\left(\im
y_2\xi_{t_2}\right)\right]\E\left[\exp\left(\im
y_1\left(\xi_{t_1}-\frac{t_1}{t_2}\xi_{t_2}\right)\right)\right].
\end{eqnarray}
We now compute both expectations in the equation above, and
observe that
\begin{equation}
\E\left[\exp\left(\kappa_1\left(\xi_{t_1}-\frac{t_1}{t_2}\xi_{t_2}\right)\right)\right]=\E\left[\exp\left(\kappa_1
\left(\beta_{t_1 T}-\frac{t_1}{t_2}\beta_{t_2
T}\right)\right)\right],
\end{equation}
where $\kappa_1=\im y_1$.
\begin{lem}\label{lem772}
Let $\beta_{t_1 T}$ and $\beta_{t_2 T}$ be values of a Brownian
bridge process over the interval $[0,T]$, with $0\le t_1\le t_2\le
T$. Then the random variable
\begin{equation}\label{e7.1aa}
\beta_{t_1 T}-\frac{t_1}{t_2}\beta_{t_2 T}
\end{equation}
is normally distributed with mean zero and variance
\begin{equation}\label{e7.2}
\frac{t_1(t_2-t_1)}{t_2}.
\end{equation}
\end{lem}
\noindent{\bf Proof}. The mean of a linear combination of standard
Brownian bridge variables is zero. The variance of expression
(\ref{e7.1aa}) is computed as follows:
\begin{eqnarray}
\Var\left[\beta_{t_1 T}-\frac{t_1}{t_2}\beta_{t_2
T}\right]&=&\E\left[\left(\beta_{t_1 T}-\frac{t_1}{t_2}\beta_{t_2
T}\right)^2\right]\nonumber\\
&=&\E\left[\beta_{t_1
T}^2\right]-2\,\frac{t_1}{t_2}\,\E\left[\beta_{t_1 T}\beta_{t_2
T}\right]+\frac{t_1^2}{t_2^2}\,\E\left[\beta_{t_2
T}^2\right]\nonumber\\
&=&\frac{t_1(T-t_1)}{T}-2\,\frac{t_1^2(T-t_2)}{t_2
T}+\frac{t_1^2(T-t_2)}{t_2 T}\nonumber\\
&=&\frac{t_1(t_2-t_1)}{t_2}.
\end{eqnarray}
\hfill$\Box$

From Lemma \ref{lem772} we conclude that
\begin{equation}
\E\left[\exp\left(\kappa_1\left(\xi_{t_1}-\frac{t_1}{t_2}\xi_{t_2}\right)\right)\right]
=\exp\left[-\frac{1}{2}\frac{t_1(t_2-t_1)}{t_2}\right].
\end{equation}
Analogously, using the fact that the random variable $\beta_{t_2
T}$ is normally distributed with zero mean and variance
$t_2(T-t_2)/T$ and is independent of the random variable $H_T$, we
have:
\begin{equation}
\E\left[\exp\left(\im
y_2\xi_{t_2}\right)\right]=\sum^n_{j=0}\,p_j\exp\left[\kappa\sigma
h_j t_2-\frac{1}{2}\frac{t_2(T-t_2)}{T}y^2_2\right],
\end{equation}
where $p_j$ is the {\it a priori} probability that the random
variable $H_T$ takes the value $h_j$.

Thus, the term involving the two expectations in the equation
(\ref{e7.1a}) yields
\begin{align}
&\E\left[\exp\left(\im\,
y_2\,\xi_{t_2}\right)\right]\E\left[\exp\left(\im\,
y_1\left(\xi_{t_1}-\frac{t_1}{t_2}\,\xi_{t_2}\right)\right)\right]\nonumber\\
&=\sum^n_{j=0}p_j\exp\left(\im\,\sigma\, h_j\,t_2\,y_2\right)
\exp\left(-\frac{t_2(T-t_2)}{2
T}y_2^2-\frac{t_1(t_2-t_1)}{2t_2}y_1^2\right).
\end{align}
Hence, the Arrow-Debreu density (\ref{e7.1a}) can be written in
the form
\begin{eqnarray}
A(x_1,x_2)&=&\sum^{n}_{j=0}p_j\frac{1}{2\pi}\frac{1}{2\pi}\int^{\infty}_{-\infty}\int^{\infty}_{-\infty}\exp\left[-\im
              \left(x_1-\frac{t_1}{t_2}x_2\right)y_1-\frac{t_1(t_2-t_1)}{2t_2}y_1^2\right]\nn\\
          &\times&\exp\left[-\im(x_2-\sigma h_j\,
               t_2)y_2-\frac{t_2(T-t_2)}{2T}y^2_2\right]\rd y_1\rd y_2.
\end{eqnarray}
Here we have interchanged the sum with the integrals. Carrying out
the integration we are then led to the desired result.
\begin{prop}
Let $t_1$, $t_2$, and $T$ be fixed times, where $t_1\le t_2\le T$.
Consider the values of the information process $\{\xi_t\}$ at
$t_1$ and $t_2$. The associated bivariate intertemporal
Arrow-Debreu density is given by:
\begin{eqnarray}
A(x_1,x_2)&=&\frac{1}{2\pi}\sqrt{\frac{T}{t_1(T-t_2)(t_2-t_1)}}\
\exp\left[-\frac{t_2}{2\,t_1(t_2-t_1)}\left(x_1-\frac{t_1}{t_2}x_2\right)^2\right]\nn\\
&\times&\sum^n_{j=0}p_j\exp\left[-\frac{T}{2\,t_2(T-t_2)}(x_2-\sigma
h_j\,t_2)^2\right].
\end{eqnarray}
\end{prop}
The bivariate intertemporal Arrow-Debreu price enables us to
calculate the prices of options that depend on the values of the
information process at two distinct times. A multivariate
intertemporal Arrow-Debreu price can be similarly constructed.

\chapter{Information-based approach to interest rates and inflation}\label{chapIII}
\section{Overview}
In this chapter we apply the information-based framework to introduce a class of discrete-time models for interest rates and inflation. The key idea is that
market participants have at any time partial information about the
future values of macro-economic factors that influence
consumption, money supply, and other variables that determine
interest rates and price levels. We present a model for such
partial information, and show how it leads to a consistent
framework for the arbitrage-free dynamics of real and nominal
interest rates, price-indices, and index-linked securities.

We begin with a general model for discrete-time asset pricing. We
take a pricing kernel approach, which has the effect of building
in the arbitrage-free property, and providing the desired link to
economic equilibrium. We require that the pricing kernel should be
consistent with a pair of axioms, one giving the general
intertemporal relations for dividend-paying assets, and the other
relating to the existence of a money market asset. Instead of
directly assuming the existence of a previsible money-market
account, we make a somewhat weaker assumption, namely the
existence of an asset that offers a positive rate of return. It
can be deduced, however, that the assumption of the existence of a
positive-return asset is sufficient to imply the existence of a
previsible money-market account, once the intertemporal relations
implicit in the first axiom are taken into account.

The main result of Section \ref{sec7.1} is the derivation of a
general expression for the price process of a limited-liability
asset. This expression includes two terms, one being the familiar
discounted and risk-adjusted value of the dividend stream, and the
other characterising retained earnings. The vanishing of the
latter is shown in Proposition \ref{prop7.1} to be given by a
transversality condition, equation (\ref{2.3}). In particular, we
are able to show (under the conditions of Axioms A and B) that in
the case of a limited-liability asset with no permanently retained
earnings, the general form of the price process is given by the
ratio of a potential and the pricing kernel, as expressed in
equation (\ref{2.13}). In Section \ref{sec7.2} we consider the
per-period rate of return $\{\bar{r}_i\}$ offered by the positive
return asset, and we show in Proposition \ref{prop7.2} that there
exists a constant-value asset with limited liability such that the
associated dividend flow is given by $\{\bar{r}_i\}$. This result
is then used in Proposition \ref{prop7.3} to establish that the
pricing kernel admits a decomposition of the form (\ref{3.7}). In
Proposition \ref{prop7.4} we prove what then might be interpreted
as a converse to this result, thus giving us a procedure for
constructing examples of systems satisfying Axioms A and B. The
method involves the introduction of an increasing sequence that
converges to an integrable random variable. Given the sequence we
then construct an associated pricing kernel and positive-return
asset satisfying the intertemporal relations.

In Section \ref{sec7.3}  we introduce the nominal discount bond
system arising with the specification of a given pricing kernel,
and in Proposition \ref{prop7.5} we show that the discount bond
system admits a representation of the Flesaker-Hughston type. In
Section \ref{sec7.4} we consider the case when the
positive-return asset has a previsible price process, and hence
can be consistently interpreted (in a standard way) as a
money-market account, or ``risk-free" asset. The results of the
previous sections do not depend on this additional assumption. A
previsible money-market account has the structure of a series of
one-period discount-bond investments. Then in Proposition \ref{prop7.6}
we show under the conditions of Axioms A and B that there exists a
unique previsible money-market account. In other words, although we only assume the existence of a positive-return asset, we can then establish the existence of a money-market account asset.

In Section \ref{sec7.5} we outline a general approach to interest
rate modelling in the information-based framework. In Section
\ref{sec7.6} we are then able to propose a class of stochastic
models for the pricing of inflation-linked assets. The nominal and
real pricing kernels, in terms of which the consumer price index
can be expressed, are modelled by introducing a bivariate utility
function depending on (a) aggregate consumption, and (b) the
aggregate real liquidity benefit conferred by the money supply.
Consumption and money supply policies are chosen such that the
expected joint utility obtained over a specified time horizon is
maximised, subject to a budget constraint that takes into account
the ``value" of the liquidity benefit associated with the money
supply. For any choice of the bivariate utility function, the
resulting model determines a relation between the rate of
consumption, the price level, and the money supply. The model also
produces explicit expressions for the real and nominal pricing
kernels, and hence establishes a basis for the valuation of
inflation-linked securities.

\section{Asset pricing in discrete time}\label{sec7.1}
The development of asset-pricing theory in discrete time has been
pursued by many authors. In the context of interest rate
modelling, it is worth recalling that the first example of a fully
developed term-structure model where the initial discount function
is freely specifiable is that of Ho \& Lee 1986, in a
discrete-time setting. For our purposes it will be useful to
develop a general discrete-time scheme from first principles,
taking an axiomatic approach in the spirit of Hughston \&
Rafailidis (2005).

Let $\{t_i\}_{i=0,1,2,\ldots}$ denote a sequence of discrete
times, where $t_0$ represents the present and $t_{i+1}>t_i$ for
all $i\in\N_0$. We assume that the sequence $\{t_i\}$ is unbounded
in the sense that for any given time $T$ there exists a value of
$i$ such that $t_i>T$. We do not assume that the
elements of $\{t_i\}$ are equally spaced; for some applications,
however, we can consider the case where $t_n=n\tau$ for
all $n\in\mathbb{N}_0$ and for some unit of time
$\tau$.

Each asset is characterised by a pair of processes
$\{S_{t_i}\}_{i\ge 0}$ and $\{D_{t_i}\}_{i\ge 0}$ which we refer
to as the ``value process" and the ``dividend process",
respectively. We interpret $D_{t_i}$ as a random cash flow or
dividend paid to the owner of the asset at time $t_i$. Then
$S_{t_i}$ denotes the ``ex-dividend" value of the asset at $t_i$.
We can think of $S_{t_i}$ as the cash flow that would result if
the owner were to dispose of the asset at time $t_i$.

For simplicity, we shall frequently use an abbreviated notation,
and write $S_i=S_{t_i}$ and $D_i=D_{t_i}$. Thus $S_i$ denotes the
value of the asset at time $t_i$, and $D_i$ denotes the dividend
paid at time $t_i$. Both $S_i$ and $D_i$ are expressed in nominal
terms, i.e. in units of a fixed base currency. We use the term
``asset" in the broad sense here---the scheme is thus applicable
to any liquid financial position for which the values and cash
flows are well defined, and for which the principles of no
arbitrage are applicable.

The unfolding of random events in the economy will be represented
with the specification of a probability space $(\Omega,
\mathcal{F}, \mathbb{P})$ equipped with a filtration
$\{\mathcal{F}_i\}_{i\ge 0}$ which we call the ``market
filtration". For the moment we regard the market filtration as
given, but later we shall construct it explicitly. For each asset
we assume that the associated value and dividend processes are
adapted to $\{\F_i\}$. In what follows ${\mathbb{P}}$ is taken to
be the ``physical" or ``objective" probability measure; all
equalities and inequalities between random variables are to be
understood as holding almost surely with respect to $\mathbb{P}$.
For convenience we often write $\E_i[-]$ for $\E[-|\F_i]$.

In order to ensure the absence of arbitrage in the financial
markets and to establish intertemporal pricing relations, we
assume the existence of a strictly positive pricing kernel
$\{\pi_i\}_{i\ge 0}$, and make the following assumptions:
\vspace{.2cm}

\noindent {\bf Axiom~A}. {\it For any asset with associated value
process $\{S_i\}_{i\ge\infty}$ and dividend process $\{D_i\}_{i\ge
0}$, the process $\{M_i\}_{i\ge 0}$ defined by
\begin{equation}\label{2.1}
M_i=\pi_i S_i+\sum^{i}_{n=0}\pi_n D_n
\end{equation}
is a martingale, i.e. $\E[|M_i|]<\infty$ for all
$i\in\mathbb{N}_0$, and $\E[M_j|\F_i]=M_i$ for all $i\le j$. }

\vspace{0.2cm} \noindent {\bf Axiom~B}. {\it There exists a
strictly positive non-dividend-paying asset, with value process
$\{\bar{B}_i\}_{i\ge 0}$, that offers a strictly positive return,
i.e.~such that $\bar{B}_{i+1}>\bar{B}_i$ for all
$i\in\mathbb{N}_0$. We assume that the process $\{\bar{B}_i\}$ is
unbounded in the sense that for any $b\in\mathbb{R}$ there exists
a time $t_i$ such that $\bar{B}_i>b$.} \vspace{.2cm}

Given this axiomatic scheme, we proceed to explore its
consequences. The notation $\{\bar{B}_i\}$ is used in Axiom B to
distinguish the positive return asset from the previsible
money-market account asset $\{B_i\}$ that will be introduced
later; in particular, in Proposition \ref{prop7.6} it will be
shown that Axioms A and B imply the existence of a unique
money-market account asset.  We note that since the
positive-return asset is non-dividend paying, it follows from
Axiom A that $\{\pi_i \bar{B}_i\}$ is a martingale. Writing
$\bar{\rho}_i=\pi_i \bar{B}_i$, we have
$\pi_i=\bar{\rho}_i/\bar{B}_i$. Since $\{\bar{B}_i\}$ is assumed
to be strictly increasing, we see that $\{\pi_i\}$ is a
supermartingale. In fact, we have the somewhat stronger relation
$\E_i[\pi_j]<\pi_i$. Indeed, we note that
\begin{equation}
\E_i[\pi_j]=\E_i\left[\frac{\bar{\rho}_j}{\bar{B}_j}\right]<E_i\left[\frac{\bar{\rho}_j}{\bar{B}_i}\right]=\frac{E_i[\bar{\rho}_j]}{\bar{B}_i}=\frac{\bar{\rho}_i}{\bar{B}_i}=\pi_i.
\end{equation}
The significance of $\{\bar{\rho}_i\}$ is that it has the
interpretation of being the likelihood ratio appropriate for a
change of measure from the objective measure $\mathbb{P}$ to the
equivalent martingale measure $\mathbb{Q}$ characterised by the
property that non-dividend-paying assets when expressed in units
of the numeraire $\{\bar{B}_i\}$ are martingales.

We recall the definition of a potential. An adapted process
$\{x_i\}_{0\le i<\infty}$ on a probability space $(\Omega, \F,
\mathbb{P})$ with filtration $\{\F_i\}$ is said to be a potential
if $\{x_i\}$ is a non-negative supermartingale and
$\lim_{i\rightarrow\infty}\E[x_i]=0$. It is straightforward to
show that $\{\pi_i\}$ is a potential. We need to demonstrate that
given any $\epsilon>0$ we can find a time $t_j$ such
$\E[\pi_n]<\epsilon$ for all $n\ge j$. This follows from the
assumption that the positive-return asset price process
$\{\bar{B}_i\}$ is unbounded in the sense specified in Axiom B.
Thus given $\epsilon$ let us set $b=\bar{\rho}_0/\epsilon$. Now
given $b$ we can find a time $t_j$ such that $\bar{B}_{t_n}>b$ for
all $n\ge j$. But for that value of $t_j$ we have
\begin{equation}
\E[\pi_j]=\E\left[\frac{\bar{\rho}_j}{\bar{B}_j}\right]<\frac{\E[\bar{\rho}_j]}{b}=\frac{\bar{\rho}_0}{b}=\epsilon,
\end{equation}
and hence $\E[\pi_n]<\epsilon$ for all $n\ge j$. It follows that
\begin{equation}\label{3.5aa}
\lim_{i\rightarrow\infty}\E[\pi_i]=0.
\end{equation}

Next we recall the Doob decomposition for discrete-time
supermartingales (see, e.g., Meyer 1966, chapter 7). If $\{x_i\}$
is a supermartingale on a probability space
$(\Omega,\F,\mathbb{P})$ with filtration $\{\F_i\}$, then there
exists a martingale $\{y_i\}$ and a previsible increasing process
$\{a_i\}$ such that $x_i=y_i-a_i$ for all $i\ge 0$. By previsible,
we mean that $a_i$ is $\F_{i-1}$-measurable. The decomposition is
given explicitly by $a_0=0$ and
$a_i=a_{i-1}+x_{i-1}-\E_{i-1}[x_i]$ for $i\ge 1$.

It follows that the pricing kernel admits a decomposition of this
form, and that one can write
\begin{equation}\label{DoobPI}
\pi_i=Y_i-A_i,
\end{equation}
where $A_0=0$ and
\begin{equation}\label{DoobDecomp}
A_i=\sum^{i-1}_{n=0}\left(\pi_n-\E_n[\pi_{n+1}]\right)
\end{equation}
for $i\ge 1$; and where $Y_0=\pi_0$ and
\begin{equation}
Y_i=\sum^{i-1}_{n=0}\left(\pi_{n+1}-\E_n[\pi_{n+1}]\right)+\pi_0
\end{equation}
for $i\ge 1$. The Doob decomposition for $\{\pi_i\}$ has an
interesting expression in terms of discount bonds, which we shall
mention later, in Section \ref{sec7.4}.

In the case of a potential $\{x_i\}$ it can be shown (see, e.g.,
Gihman \& Skorohod 1979, chapter 1) that the limit
$a_{\infty}=\lim_{i\rightarrow\infty}a_i$ exists, and that
$x_i=\E_i[a_{\infty}]-a_i$. As a consequence, we conclude that the
pricing kernel admits a decomposition of the form
\begin{equation}
\pi_i=\E_i[A_{\infty}]-A_i,
\end{equation}
where $\{A_i\}$ is the previsible process defined by (\ref{DoobDecomp}).
With these facts in hand, we shall establish a useful result concerning the pricing of
limited-liability assets. By a limited-liability asset we mean an
asset such that $S_i\ge0$ and $D_i\ge0$ for all $i\in\mathbb{N}$.

\begin{prop}\label{prop7.1}
Let $\{S_i\}_{i\ge 0}$ and $\{D_i\}_{i\ge 0}$ be the value and
dividend processes associated with a limited-liability asset. Then
$\{S_i\}$ is of the form
\begin{equation}\label{2.2}
S_i=\frac{m_i}{\pi_i}+\frac{1}{\pi_i}\E_i\left[\sum^{\infty}_{n=i+1}\pi_n D_n\right],
\end{equation}
where $\{m_i\}$ is a non-negative martingale that vanishes if and
only if the following transversality condition holds:
\begin{equation}\label{2.3}
\lim_{j\rightarrow\infty}\E[\pi_j S_j]=0.
\end{equation}
\end{prop}

\noindent{\bf Proof}. It follows from Axiom A, as a consequence of
the martingale property, that
\begin{equation}\label{2.4}
\pi_i S_i+\sum^i_{n=0}\pi_n D_n=\E_i\left[\pi_j S_j+\sum^j_{n=0}\pi_n D_n\right]
\end{equation}
for all $i\le j$. Taking the limit $j\rightarrow\infty$ on the
right-hand side of this relation we have
\begin{equation}\label{2.5}
\pi_i S_i+\sum^i_{n=0}\pi_n D_n=\lim_{j\rightarrow\infty}\E_i[\pi_j S_j]+\lim_{j\rightarrow\infty}\E_i\left[\sum^j_{n=0}\pi_n D_n\right].
\end{equation}
Since $\pi_i D_i\ge 0$ for all $i\in\mathbb{N}_0$, it follows from
the conditional form of the monotone convergence theorem---see,
e.g., Steele 2001, Williams 1991---that
\begin{equation}\label{2.6}
\lim_{j\rightarrow\infty}\E_i\left[\sum^j_{n=0}\pi_n D_n\right]=\E_i\left[\lim_{j\rightarrow\infty}\sum^j_{n=0}\pi_n D_n\right],
\end{equation}
and hence that
\begin{equation}\label{2.7}
\pi_i S_i+\sum^i_{n=0}\pi_n D_n=\lim_{j\rightarrow\infty}\E_i[\pi_j S_j]+\E_i\left[\sum^{\infty}_{n=0}\pi_n D_n\right].
\end{equation}
Now let us define
\begin{equation}\label{2.8}
m_i=\lim_{j\rightarrow\infty}\E_i[\pi_j S_j].
\end{equation}
Then clearly $m_i\ge 0$ for all $i\in\mathbb{N}_0$. We see,
moreover, that $\{m_i\}_{i\ge 0}$ is a martingale, since
$m_i=M_i-\E_i[F_{\infty}]$, where $M_i$ is defined as in equation
(\ref{2.1}), and
\begin{equation}\label{2.9}
F_{\infty}=\sum^{\infty}_{n=0}\pi_n D_n.
\end{equation}
It is implicit in the axiomatic scheme that the sum
$\sum^{\infty}_{n=0}\pi_n D_n$ converges in the case of a
limited-liability asset. This follows as a consequence of the
martingale convergence theorem and Axiom A. Thus, writing equation
(\ref{2.7}) in the form
\begin{equation}
\pi_i S_i+\sum^i_{n=0}\pi_n D_n=m_i+\E_i\left[\sum^{\infty}_{n=0}\pi_n D_n\right],
\end{equation}
after some re-arrangement of terms we obtain
\begin{equation}\label{2.10}
\pi_i S_i=m_i+\E_i\left[\sum^{\infty}_{n=i+1}\pi_n D_n\right],
\end{equation}
and hence (\ref{2.2}), as required. On the other hand, by the
martingale property of $\{m_i\}$ we have $\E[m_i]=m_0$ and hence
\begin{equation}\label{2.11}
\E[m_i]=\lim_{j\rightarrow\infty}\E[\pi_j S_j]
\end{equation}
for all $i\in\N$. Thus since $m_i\ge 0$ we see that the transversality
condition (\ref{2.3}) holds if and only if $\{m_i\}=0$.\hfill$\Box$\\

The interpretation of the transversality condition is as follows.
For each $j\in\N_0$ the expectation $V_j=\E[\pi_j S_j]$ measures
the present value of an instrument that pays at $t_j$ an amount
equal to the proceeds of a liquidation of the asset with price
process $\{S_i\}_{i\ge 0}$. If $\lim_{j\rightarrow\infty}V_j=0$
then one can say that in the long term all of the value of the
asset will be dispersed in its dividends. On the other hand, if
some or all of the dividends are ``retained" indefinitely, then
$\{V_j\}$ will retain some value, even in the limit as $t_j$ goes
to infinity.

The following example may clarify this interpretation. Suppose
investors put \$100m of capital into a new company. The management
of the company deposits \$10m into a money market account. The
remaining \$90m is invested in an ordinary risky line of business,
the entire proceeds of which, after costs, are paid to
share-holders as dividends. Thus at time $t_i$ we have
$S_i=B_i+H_i$, where $B_i$ is a position in the money market
account initialised at \$10m, and where $H_i$ is the value of the
remaining dividend flow. Now $\{\pi_i B_i\}$ is a martingale, and
thus $\E[\pi_i B_i]=\$10{\text m}$ for all $i\in\N_0$, and
therefore $\lim_{i\rightarrow\infty}\E[\pi_i B_i]=\$10{\text m}$.
On the other hand $\lim_{i\rightarrow\infty}\E[\pi_i H_i]=0$; this
means that given any value $h$ we can find a time $T$ such that
$\E[\pi_i H_i]<h$ for all $t_i\ge T$.

There are other ways of ``retaining" funds than putting them into
a domestic money market account. For example, one could put the
\$10m into a foreign bank account; or one could invest it in
shares in a securities account, with a standing order that all
dividends should be immediately re-invested in further shares.
Thus if the investment is in a general ``dividend-retaining" asset
(such as a foreign bank account), then $\{m_i\}$ can in principal
be any non-negative martingale. The content of Proposition
\ref{prop7.1} is that any limited-liability investment can be
separated in a unique way into a growth component and an income
component.

In the case of a ``pure income" investment, i.e. in an asset for
which the transversality condition is satisfied, the price is
directly related to the future dividend flow, and we have
\begin{equation}\label{2.12}
S_i=\frac{1}{\pi_i}\E_i\left[\sum^{\infty}_{n=i+1}\pi_n
D_n\right].
\end{equation}
This is the so-called ``fundamental equation" which some authors
use directly as a basis for asset pricing theory---see, e.g.,
Cochrane 2005. Alternatively we can write (\ref{2.12}) in the form
\begin{equation}\label{2.13}
S_i=\frac{1}{\pi_i}(\E_i[F_{\infty}]-F_i),
\end{equation}
where
\begin{equation}\label{2.14}
F_i=\sum^i_{n=0}\pi_n D_n,\quad \textrm{and}\quad F_{\infty}=\lim_{i\rightarrow\infty}F_i.
\end{equation}
It is a straightforward exercise to show that the process $\{\pi_i
S_i\}$ is a potential. Clearly, $\{\E_i[F_{\infty}]-F_i\}$ is a
positive supermartingale, since $\{F_i\}$ is increasing; and by
the tower property and the monotone convergence theorem we have
$\lim_{i\rightarrow\infty}\E[\E_i[F_{\infty}]-F_i]=\E[F_{\infty}]-\lim_{i\rightarrow\infty}\E[F_i]=\E[F_{\infty}]-\E\left[\lim_{i\rightarrow\infty}F_i\right]=0$.
On the other hand, $\{\pi_i\}$ is also a potential, so we reach
the conclusion that in the case of an income generating asset the
price process can be expressed as a ratio of potentials, thus
giving us a discrete-time analogue of a result obtained by Rogers
1997. Indeed, the role of the concept of a potential as it appears
here is consistent with the continuous-time theories developed by
Flesaker \& Hughston 1996, Rogers 1997, Rutkowski 1997, Jin \&
Glasserman 2001, Hughston \& Rafailidis 2005, and others, where
essentially the same mathematical structures appear.
\section{Nominal pricing kernel}\label{sec7.2}
\label{sec:3} To proceed further we need to say more about the
relation between the pricing kernel $\{\pi_i\}$ and the
positive-return asset $\{\bar{B}_i\}$. To this end let us write
\begin{equation}\label{3.1}
\bar{r}_i=\frac{\bar{B}_i-\bar{B}_{i-1}}{\bar{B}_{i-1}}
\end{equation}
for the rate of return on the positive-return asset realised at
time $t_i$ on an investment made at time $t_{i-1}$. Since the time
interval $t_i-t_{i-1}$ is not necessarily small, there is no
specific reason to presume that the rate of return $\bar{r}_i$ is
already known at time $t_{i-1}$. This is consistent with the fact
that we have assumed that $\{\bar{B}_i\}$ is $\{\F_i\}$-adapted.
The notation $\bar{r}_i$ is used here to distinguish the rate of return on the positive-return asset from the rate of return $r_i$ on the money market account, which will be introduced in Section \ref{sec7.4}.

Next we present a simple argument to motivate the idea that there should exist an asset with constant value unity that pays a dividend stream given by $\{\bar{r}_i\}$. We consider the following portfolio strategy. The portfolio consists at any time of a certain number of units of the positive-return asset. Let $\phi_i$ denote the number of units, so that at time $t_i$ the (ex-dividend) value of the portfolio is given by $V_i=\phi_i\bar{B}_i$. Then in order to have $V_i=1$ for all $i\ge 0$ we set $\phi_i=1/\bar{B}_i$. Let $D_i$ denote the dividend paid out by the portfolio at time $t_i$. Then clearly if the portfolio value is to remain constant we must have $D_i=\phi_{i-1}\bar{B}_i-\phi_{i-1}\bar{B}_{i-1}$ for all $i\ge 1$. It follows immediately that $D_i=\bar{r}_i$, where $\bar{r}_i$ is given by (\ref{3.1}).

This shows that we can construct a portfolio with a constant value and with the desired cash flows. Now we need to show that such a system satisfies Axiom A.
\begin{prop}\label{prop7.2}
There exists an asset with constant nominal value $S_i=1$ for all
$i\in\N_0$, for which the associated cash flows are given by
$\{\bar{r}_i\}_{i\ge 1}$.
\end{prop}

\noindent{\bf Proof}. We need to verify
that the conditions of Axiom A are satisfied in the case for which
$S_i=1$ and $D_i=\bar{r}_i$ for $i\in\N_0$. In other words we need
to show that
\begin{equation}\label{3.2}
\pi_i=\E_i[\pi_j]+\E_i\left[\sum^j_{n=i+1}\pi_n \bar{r}_n\right]
\end{equation}
for all $i\le j$. The calculation proceeds as follows. We observe that
\begin{eqnarray}\label{3.3}
\E_i\left[\sum^j_{n=i+1}\pi_n \bar{r}_n\right]&=&\E_i\left[\sum^j_{n=i+1}\pi_n\frac{\bar{B}_n-\bar{B}_{n-1}}{\bar{B}_{n-1}}\right]\nonumber\\
&=&\E_i\left[\sum^j_{n=i+1}\frac{\bar{\rho}_n}{\bar{B}_n}\frac{\bar{B}_n-\bar{B}_{n-1}}{\bar{B}_{n-1}}\right]\nonumber\\
&=&\E_i\left[\sum^j_{n=i+1}\left(\frac{\bar{\rho}_n}{\bar{B}_{n-1}}-\frac{\bar{\rho}_n}{\bar{B}_n}\right)\right]\nonumber\\
&=&\E_i\left[\sum^j_{n=i+1}\left(\E_{n-1}\left[\frac{\bar{\rho}_n}{\bar{B}_{n-1}}\right]-\frac{\bar{\rho}_n}{\bar{B}_n}\right)\right],
\end{eqnarray}
the last step being achieved by use of the tower property. It
follows then by use of the martingale property of $\{\rho_n\}$
that:
\begin{eqnarray}
\E_i\left[\sum^j_{n=i+1}\pi_n \bar{r}_n\right]&=&\E_i\left[\sum^j_{n=i+1}\left(\frac{1}{\bar{B}_{n-1}}\E_{n-1}[\bar{\rho}_n]-\frac{\bar{\rho}_n}{\bar{B}_n}\right)\right]\nonumber\\
&=&\E_i\left[\sum^j_{n=i+1}\left(\frac{\bar{\rho}_{n-1}}{\bar{B}_{n-1}}-\frac{\bar{\rho}_n}{\bar{B}_n}\right)\right]\nonumber\\
&=&\E_i\left[\left(\frac{\bar{\rho}_i}{\bar{B}_i}-\frac{\bar{\rho}_{i+1}}{\bar{B}_{i+1}}\right)+\left(\frac{\bar{\rho}_{i+1}}{\bar{B}_{i+1}}-\frac{\bar{\rho}_{i+2}}{\bar{B}_{i+2}}\right)
+\ldots+\left(\frac{\bar{\rho}_{j-1}}{\bar{B}_{j-1}}-\frac{\bar{\rho}_j}{\bar{B}_j}\right)\right]\nonumber\\
&=&\E_i\left[\frac{\bar{\rho}_i}{\bar{B}_i}\right]-\E_i\left[\frac{\bar{\rho}_j}{\bar{B}_j}\right]\nonumber\\
&=&\pi_i-\E_i[\pi_j].
\end{eqnarray}
But that gives us (\ref{3.2}). \hfill$\Box$\\

The existence of the constant-value asset leads to an alternative decomposition of the pricing kernel, which can be described as follows.
\begin{prop}\label{prop7.3}
Let $\{\bar{B}_i\}$ be a positive-return asset satisfying the
conditions of Axiom B, and let $\{\bar{r}_i\}$ be its
rate-of-return process. Then the pricing kernel can be expressed
in the form $\pi_i=\E_i[G_{\infty}]-G_i$, where
$G_i=\sum^i_{n=1}\pi_n\bar{r}_n$ and
$G_{\infty}=\lim_{i\rightarrow\infty}G_i$.
\end{prop}

\noindent {\bf Proof}. First we remark that if an asset has
constant value then it satisfies the transversality condition
(\ref{2.3}). In particular, letting the constant be unity, we see
that the transversality condition reduces to
\begin{equation}\label{3.5a}
\lim_{i\rightarrow\infty}\E[\pi_i]=0,
\end{equation}
which is satisfied since $\{\pi_i\}$ is a potential. Next we show
that
\begin{equation}\label{3.5b}
\lim_{j\rightarrow\infty}\E_i[\pi_j]=0
\end{equation}
for all $i\in\N_0$. In particular, fixing $i$, we have
$\E\left[\E_i[\pi_j]\right]=\E[\pi_j]$ by the tower property, and
thus
\begin{equation}
\lim_{j\rightarrow\infty}\E\left[\E_i[\pi_j]\right]=0
\end{equation}
by virtue of (\ref{3.5a}). But $\E_i[\pi_j]<\pi_i$ for all $j>i$,
and $\E[\pi_i]<\infty$; hence by the dominated convergence theorem
we have
\begin{equation}
\lim_{j\rightarrow\infty}\E[\E_i[\pi_j]]=\E[\lim_{j\rightarrow\infty}\E_i[\pi_j]],
\end{equation}
from which the desired result (\ref{3.5b}) follows, since the
argument of the expectation is non-negative. As a consequence of
(\ref{3.5b}) it follows from (\ref{3.2}) that
\begin{equation}\label{3.6}
\pi_i=\lim_{j\rightarrow\infty}\E_i\left[\sum^j_{n=i+1}\pi_n
\bar{r}_n\right],
\end{equation}
and thus by the monotone convergence theorem we have
\begin{eqnarray}\label{3.7}
\pi_i&=&\E_i\left[\sum^{\infty}_{n=i+1}\pi_n \bar{r}_n\right]\nn\\
        &=&\E_i\left[\sum^{\infty}_{n=1}\pi_n \bar{r}_n\right]-\sum^{i}_{n=1}\pi_n
        \bar{r}_n\nn\\
        &=&\E_i\left[G_{\infty}\right]-G_i,
\end{eqnarray}
and that gives us the result of the proposition.\hfill$\Box$\\

We shall establish a converse to this result, which allows us to
construct a system satisfying Axioms A and B from any
strictly-increasing non-negative adapted process that converges,
providing a certain integrability condition holds.

\begin{prop}\label{prop7.4}
Let $\{G_i\}_{i\ge 0}$ be a strictly increasing adapted process
satisfying $G_0=0$, and $\E[G_{\infty}]<\infty$, where
$G_{\infty}=\lim_{i\rightarrow\infty}G_i$. Let the processes
$\{\pi_i\}$, $\{\bar{r}_i\}$, and $\{\bar{B}_i\}$, be defined by
$\pi_i=\E_i[G_{\infty}]-G_{i}$ for $i\ge 0$;
$\bar{r}_i=(G_i-G_{i-1})/\pi_i$ for $i\ge 1$;
$\bar{B}_i=\prod^i_{n=1}(1+\bar{r}_n)$ for $i\ge 1$, with
$\bar{B}_0=1$. Let the process $\{\bar{\rho}_i\}$ be defined by
$\bar{\rho}_i=\pi_i \bar{B}_i$ for $i\ge0$. Then
$\{\bar{\rho}_i\}$ is a martingale, and
$\lim_{j\rightarrow\infty}\bar{B}_j=\infty$. Thus $\{\pi_i\}$ and $\{\bar{B}_i\}$, as constructed, satisfy Axioms A and B.
\end{prop}

\noindent{\bf Proof}. Writing $g_i=G_i-G_{i-1}$ for $i\ge1$ we have
\begin{equation}\label{3.8}
\pi_i=\E_i[G_{\infty}]-G_i=\E_i\left[\sum^{\infty}_{n=i+1}g_n\right],
\end{equation}
and
\begin{eqnarray}\label{3.9}
\bar{B}_i=\prod^i_{n=1}(1+\bar{r}_n)=\prod^i_{n=1}\left(1+\frac{g_n}{\pi_n}\right)=\prod^i_{n=1}\left(\frac{\pi_n+g_n}{\pi_n}\right).
\end{eqnarray}
Hence, writing $\bar{\rho}_i=\pi_i \bar{B}_i$, we have
\begin{eqnarray}\label{3.10}
\bar{\rho}_i&=&\pi_i\prod^i_{n=1}\left(\frac{\pi_n+g_n}{\pi_n}\right)\nn\\
&=&(\pi_i+g_i)\prod^{i-1}_{n=1}\left(\frac{\pi_n+g_n}{\pi_n}\right),
\end{eqnarray}
and thus
\begin{equation}
\bar{\rho}_i=(\pi_i+g_i)\bar{B}_{i-1}=\frac{\pi_i+g_i}{\pi_{i-1}}\bar{\rho}_{i-1}.
\end{equation}
To show that $\{\bar{\rho}_i\}$ is a martingale it suffices to
verify for all $i\ge 1$ that $\E[\bar{\rho}_i]<\infty$ and that
$\E_{i-1}[\bar{\rho}_i]=\bar{\rho}_{i-1}$. In particular, if
$\E[\bar{\rho}_i]<\infty$ then the ``take out what is known rule''
applies, and by (\ref{3.8}) and (\ref{3.10}) we have
\begin{eqnarray}
\E_{i-1}[\bar{\rho}_i]&=&\E_{i-1}\left[\frac{\pi_i+g_i}{\pi_{i-1}}\bar{\rho}_{i-1}\right]\nn \\
             &=&\frac{\bar{\rho}_{i-1}}{\pi_{i-1}}\E_{i-1}\left[\pi_i+g_i\right]\nn\\
             &=&\frac{\bar{\rho}_{i-1}}{\pi_{i-1}}\E_{i-1}\left[\sum^{\infty}_{n=i}g_n\right]\nn\\
             &=&\frac{\bar{\rho}_{i-1}}{\pi_{i-1}}\left(\E_{i-1}\left[G_{\infty}\right]-G_{i-1}\right)\nn\\
             &=&\bar{\rho}_{i-1}.
\end{eqnarray}
Here, in going from the first to the second line we have used the
fact that $\E[\pi_i+g_i]<\infty$, together with the assumption
that $\E[\bar{\rho}_i]<\infty$. To verify that
$\E[\bar{\rho}_i]<\infty$ let us write
\begin{equation}
J^{\alpha}_{i-1}=\min\left[\frac{\bar{\rho}_{i-1}}{\pi_{i-1}},\alpha\right]
\end{equation}
for $\alpha\in\mathbb{N}_0$. Then by use of monotone convergence
and the tower property we have
\begin{eqnarray}
\E[\bar{\rho}_i]&=&\E\left[(\pi_i+g_i)\lim_{\alpha\rightarrow\infty}J^{\alpha}_{i-1}\right]\nn\\
        &=&\lim_{\alpha\rightarrow\infty}\E\left[(\pi_i+g_i)J^{\alpha}_{i-1}\right]\nn\\
        &=&\lim_{\alpha\rightarrow\infty}\E\left[\E_{i-1}\left[(\pi_i+g_i)J^{\alpha}_{i-1}\right]\right]\nn\\
        &=&\lim_{\alpha\rightarrow\infty}\E\left[J^{\alpha}_{i-1}\E_{i-1}\left[(\pi_i+g_i)\right]\right]\nn\\
        &\le&\E\left[\frac{\bar{\rho}_{i-1}}{\pi_{i-1}}\E_{i-1}[\pi_i+g_i]\right]\nn\\
        &=&\E[\bar{\rho}_{i-1}],
\end{eqnarray}
since
\begin{equation}
J^{\alpha}_{i-1}\le\frac{\bar{\rho}_{i-1}}{\pi_{i-1}}.
\end{equation}
Thus we see for all $i\ge 1$ that if
$\E\left[\bar{\rho}_{i-1}\right]<\infty$ then
$\E\left[\bar{\rho}_i\right]<\infty$. But $\bar{\rho}_0<\infty$ by
construction; hence by induction we deduce that
$\E\left[\bar{\rho}_i\right]<\infty$ for all $i\ge 0$.

 To show that $\lim_{j\rightarrow\infty}\{\bar{B}_j\}=\infty$ let us assume the contrary and show that this
 leads to a contradiction. Suppose, in particular, that there were to exist a number $b$ such that
 $\bar{B}_i<b$ for all $i\in\N_0$. Then for all $i\in\N_0$ we would have
\begin{equation}\label{3.13a}
\E\left[\frac{\bar{\rho}_i}{\bar{B}_i}\right]>\frac{1}{b}\E[\bar{\rho}_i]=\frac{\bar{\rho}_0}{b}.
\end{equation}
But by construction we know that $\lim_{i\rightarrow\infty}\E[\pi_i]=0$ and hence
\begin{equation}
\lim_{i\rightarrow\infty}\E\left[\frac{\bar{\rho}_i}{\bar{B}_i}\right]=0.
\end{equation}
Thus given any $\epsilon>0$ we can find a time $t_i$ such that
\begin{equation}\label{3.14}
\E\left[\frac{\bar{\rho}_i}{\bar{B}_i}\right]<\epsilon.
\end{equation}
But this is inconsistent with (\ref{3.13a}); and thus we conclude
that $\lim_{j\rightarrow\infty}\bar{B}_j=\infty$.
That completes the proof of Proposition \ref{prop7.4}.\hfill$\Box$\\
 \section{Nominal discount bonds}\label{sec7.3}
Now we proceed to consider the properties of nominal discount bonds.
By such an instrument we mean an asset that pays a single dividend
consisting of one  unit of domestic currency at some designated time
$t_j$. For the price $P_{ij}$ at time $t_i$ $(i<j)$ of a discount
bond that matures at time $t_j$ we thus have
\begin{equation}\label{3.15}
P_{ij}=\frac{1}{\pi_i}\E_i[\pi_j].
\end{equation}
Since $\pi_i>0$ for all $i\in\N$, and $\E_i[\pi_j]<\pi_i$ for all
$i<j$, it follows that $0<P_{ij}<1$ for all $i<j$. We observe, in
particular, that the associated interest rate $R_{ij}$ defined by
\begin{equation}\label{3.16}
P_{ij}=\frac{1}{1+R_{ij}}
\end{equation}
is strictly positive. Note that in our theory we regard a discount
bond as a ``dividend-paying" asset. Thus in the case of a discount
bond with maturity $t_j$ we have $P_{jj}=0$ and $D_j=1$. Usually
discount bonds are defined by setting $P_{jj}=1$ at maturity, with
$D_j=0$; but it is perhaps more logical to regard the bonds as
giving rise to a unit cash flow at maturity. We also note that the
definition of the discount bond system does not involve the
specific choice of the positive-return asset.

It is important to point out that in the present framework there
is no reason or need to model the dynamics of $\{P_{ij}\}$, or to
model the volatility structure of the discount bonds. Indeed, from
the present point of view this would be a little artificial. The
important issue, rather, is how to model the pricing kernel. Thus,
our scheme differs somewhat in spirit from the discrete-time
models discussed, e.g., in Heath {\it et al.} 1990, and
Filipovi\'c \& Zabczyk 2002.

As a simple example of a family of discrete-time interest rate
models admitting tractable formulae for the associated discount
bond price processes, suppose we set
\begin{equation}\label{3.17}
\pi_i=\alpha_i+\beta_i N_i
\end{equation}
where $\{\alpha_i\}$ and $\{\beta_i\}$ are strictly-positive,
strictly-decreasing deterministic sequences, satisfying
$\lim_{i\rightarrow\infty}\alpha_i=0$ and
$\lim_{i\rightarrow\infty}\beta_i=0$, and where $\{N_i\}$ is a
strictly positive martingale. Then by (\ref{3.15}) we have
\begin{eqnarray}\label{3.18}
P_{ij}=\frac{\alpha_j+\beta_j N_i}{\alpha_i+\beta_i N_i},
\end{eqnarray}
thus giving a family of ``rational" interest rate models. Note
that in a discrete-time setting we can produce classes of models
that have no immediate analogues in continuous time---for example,
we can let $\{N_i\}$ be the natural martingale associated with a
branching process.

Now we shall demonstrate that any discount bond system consistent
with our general scheme admits a representation of the
Flesaker-Hughston type. For accounts of the Flesaker-Hughston
theory see, e.g., Flesaker \& Hughston 1996, Rutkowski 1997, Hunt
\& Kennedy 2000, or Jin \& Glasserman 2001.

\begin{prop}\label{prop7.5}
Let $\{\pi_i\}$, $\{\bar{B}_i\}$, $\{P_{ij}\}$ satisfy the conditions of
{\rm Axioms A} and {\rm B}. Then there exists a family of positive
martingales $\{m_{in}\}_{0\le i\le n}$ indexed by $n\in\N$ such
that
\begin{equation}\label{3.19}
P_{ij}=\frac{\sum^{\infty}_{n=j+1}m_{in}}{\sum^{\infty}_{n=i+1}m_{in}}.
\end{equation}
\end{prop}

\noindent{\bf Proof}. We shall use the fact that $\pi_i$ can be written in the form
\begin{eqnarray}\label{3.20}
\pi_i&=&\E_i[G_{\infty}]-G_{i}\nonumber\\
       &=&\E_i\left[\sum^{\infty}_{n=1}g_n\right]-\sum^i_{n=1}g_n\nonumber\\
       &=&\E_i\left[\sum^{\infty}_{n=i+1}g_n\right],
\end{eqnarray}
where $g_i=G_i-G_{i-1}$ for each $i\ge 1$. Then $g_i>0$ for all
$i\ge 1$ since $\{G_i\}$ is a strictly increasing sequence. By the
monotone convergence theorem we have
\begin{equation}\label{3.21}
\pi_i=\sum^{\infty}_{n=i+1}\E_i[g_n]
\end{equation}
and
\begin{equation}\label{3.22}
\E_i[\pi_j]=\sum^{\infty}_{n=j+1}\E_i[g_n].
\end{equation}
For each $n\ge 1$ we define $m_{in}=\E_i[X_n]$. Then for each
$n\in\N$ we see that $\{m_{in}\}_{0\le i\le n}$ is a strictly
positive martingale, and (\ref{3.19}) follows
immediately.\hfill$\Box$
\section{Nominal money-market account}\label{sec7.4}
In the analysis presented so far we have assumed that the
positive-return process $\{\bar{B}_i\}$ is $\{F_i\}$-adapted, but
is not necessarily previsible. The point is that many of our
conclusions are valid under the weaker hypothesis of mere
adaptedness, as we have seen. There are also economic motivations
behind the use of the more general assumption. One can imagine
that the time sequence $\{t_i\}$ is in reality a ``course
graining" of a finer time sequence that includes the original
sequence as a sub-sequence. Then likewise one can imagine that
$\{\bar{B}_i\}$ is a sub-sequence of a finer process that assigns
a value to the positive-return asset at each time in the finer
time sequence. Finally, we can imagine that $\{\F_i\}$ is a
sub-filtration of a finer filtration based on the finer sequence.
In the case of a money market account, where the rate of interest
is set at the beginning of each short deposit period (say, one
day), we would like to regard the relevant value process as being
previsible with respect to the finer filtration, but merely
adapted with respect to the course-grained filtration.

Do positive-return assets, other than the standard previsible
money market account, actually exist in a discrete-time setting?
The following example gives an affirmative answer. In the setting
of the standard binomial model, in the case of a single period,
let $S_0$ denote the value at time $0$ of a risky asset, and let
$\{U,D\}$ denote its possible values at time $1$. Let $B_0$ and
$B_1$ denote the values at times 0 and 1 of a deterministic
money-market account. We assume that $B_1>B_0$ and $U>S_0
B_1/B_0>D$. A standard calculation shows that the risk-neutral
probabilities for $S_0\rightarrow U$ and $S_0\rightarrow D$ are
given by $p^*$ and $1-p^*$, where $p^*=(S_0 B_1/B_0-D)/(U-D)$. We
shall now construct a ``positive-return" asset, i.e. an asset with
initial value $\bar{S}_0$ and with possible values
$\{\bar{U},\bar{D}\}$ at time 1 such that $\bar{U}>\bar{S}_0$ and
$\bar{D}>\bar{S}_0$. Risk-neutral valuation implies that
$\bar{S}_0=(B_0/B_1)[p^*\bar{U}+(1-p^*)\bar{D}]$. Thus, given
$\bar{S}_0$, we can determine $\bar{U}$ in terms of $\bar{D}$. A
calculation then shows that if
$(B_1/B_0-p^*)/(1-p^*)>\bar{D}/\bar{S}_0>1$, then
$\bar{U}>\bar{S}_0$ and $\bar{D}>\bar{S}_0$, as desired. Thus, in
the one-period binomial model, for the given initial value
$\bar{S}_0$, we obtain a one-parameter family of positive-return
assets.

Let us consider now the special case where the positive-return asset is previsible.
Thus for $i\ge 1$ we assume that $B_i$ is $\F_{i-1}$-measurable and we drop the ``bar" over $B_i$ to signify the fact that we are now considering a money-market account. In that case we have
\begin{eqnarray}\label{3.23}
P_{i-1,i}&=&\frac{1}{\pi_{i-1}}\E_{i-1}[\pi_i]\nonumber\\
        &=&\frac{B_{i-1}}{\rho_{i-1}}\E_{i-1}\left[\frac{\rho_i}{B_i}\right]\nonumber\\
        &=&\frac{B_{i-1}}{B_i},
\end{eqnarray}
by virtue of the martingale property of $\{\rho_i\}$. Thus, in the
case of a money-market account we see that
\begin{equation}\label{3.24}
P_{i-1,i}=\frac{1}{1+r_i}.
\end{equation}
where $r_i=R_{i-1,i}$. In other words, the rate of return on the
money-market account is previsible, and is given by the one-period
simple discount factor associated with the discount bond that
matures at time $t_i$.

Reverting now to the general situation, it follows that if we are given a pricing kernel $\{\pi_i\}$ on a probability space $(\Omega,\F,\mathbb{P})$ with filtration
$\{\F_i\}$, and a system of assets satisfying Axioms A and B, then we can construct a plausible candidate for an
associated previsible money market account by setting $B_0=1$ and
defining
\begin{equation}\label{3.25}
B_i=(1+r_i)(1+r_{i-1})\cdots(1+r_1),
\end{equation}
for $i\ge 1$, where
\begin{equation}
r_i=\frac{\pi_{i-1}}{\E_{i-1}[\pi_i]}-1.
\end{equation}
We shall refer to the process $\{B_i\}$ thus constructed as the
``natural" money market account associated with the pricing kernel
$\{\pi_i\}$.

To justify this nomenclature, we need to verify that $\{B_i\}$, so
constructed, satisfies the conditions of Axioms A and B. To this
end, we make note of the following decomposition. Let $\{\pi_i\}$
be a positive supermartingale satisfying $\E_i[\pi_j]<\pi_i$ for
all $i<j$ and $\lim_{j\rightarrow\infty}[\pi_j]=0$. Then as an
identity we can write
\begin{equation}\label{3.26}
\pi_i=\frac{\rho_i}{B_i},
\end{equation}
where
\begin{equation}\label{3.27}
\rho_i=\frac{\pi_i}{\E_{i-1}[\pi_i]}\ \frac{\pi_{i-1}}{\E_{i-2}[\pi_{i-1]}}\cdots\frac{\pi_1}{\E_0[\pi_1]}\pi_0
\end{equation}
for $i\ge 0$, and
\begin{equation}\label{3.28}
B_i=\frac{\pi_{i-1}}{\E_{i-1}[\pi_i]}\ \frac{\pi_{i-2}}{\E_{i-2}[\pi_{i-1}]}\cdots\frac{\pi_1}{\E_1[\pi_2]}\ \frac{\pi_0}{\E_0[\pi_1]}
\end{equation}
for $i\ge 1$, with $B_0=1$. Thus, in this scheme we have
\begin{equation}\label{3.29}
\rho_i=\frac{\pi_i}{\E_{i-1}[\pi_i]}\rho_{i-1},
\end{equation}
with the initial condition $\rho_0=\pi_0$; and
\begin{equation}\label{3.30}
B_i=\frac{\pi_{i-1}}{\E_{i-1}[\pi_i]}B_{i-1},
\end{equation}
with the initial condition $B_0=1$. It is evident that
$\{\rho_i\}$ as thus defined is $\{\F_i\}$-adapted, and that
$\{B_i\}$ is previsible and strictly increasing. Making use of the
identity (\ref{3.30}) we are now in a position to establish the
following:

\begin{prop}\label{prop7.6}
Let $\{\pi_i\}$ be a non-negative supermartingale satisfying
$\E_i[\pi_j]<\pi_i$ for all $i<j\in\N_0$, and
$\lim_{i\rightarrow\infty}\E[\pi_i]=0$. Let $\{B_i\}$ be defined by
$B_0=1$ and $B_i=\prod^i_{n=1}(1+r_n)$ for $i\ge 1$, where
$1+r_i=\pi_{i-1}/\E_{i-1}[\pi_i]$, and set $\rho_i=\pi_i B_i$ for
$i\ge 0$. Then $\{\rho_i\}$ is a martingale, and the interest rate
system defined by $\{\pi_i\}$, $\{B_i\}$, and $\{P_{ij}\}$ satisfies
{\rm Axioms A} and {\rm B}.
\end{prop}

\noindent{\bf Proof}. To show that $\{\rho_i\}$ is a martingale it
suffices to verify for all $i\ge 1$ that $\E[\rho_i]<\infty$ and
that $\E_{i-1}[\rho_i]=\rho_{i-1}$. In particular, if
$\E[\rho_i]<\infty$ then the ``take out what is known rule" is
applicable, and by (\ref{3.29}) we have
\begin{equation}
\E_{i-1}[\rho_i]=\E_{i-1}\left[\frac{\pi_i}{\E_{i-1}[\pi_i]}\rho_{i-1}\right]=\rho_{i-1}.
\end{equation}
Thus to show that $\{\rho_i\}$ is a martingale all that remains is to verify that $\E[\rho_i]<\infty$. Let us write
\begin{equation}
J^{\alpha}_{i-1}=\min\left[\frac{\rho_{i-1}}{\E_{i-1}[\pi_i]},\alpha\right]
\end{equation}
for $\alpha\in\N_0$. Then by monotone convergence and the tower
property we have
\begin{eqnarray}
\E[\rho_i]&=&\E\left[\pi_i\lim_{\alpha\rightarrow\infty}J^{\alpha}_{i-1}\right]\\
               &=&\lim_{\alpha\rightarrow\infty}\E\left[\pi_i J^{\alpha}_{i-1}\right]\\
               &=&\lim_{\alpha\rightarrow\infty}\E\left[\E_{i-1}[\pi_i J^{\alpha}_{i-1}]\right].
\end{eqnarray}
But since $J^{\alpha}_{i-1}$ is bounded we can move this term
outside the inner conditional expectation to give
\begin{equation}
\E[\rho_i]=\lim_{\alpha\rightarrow\infty}\E\left[J^{\alpha}_{i-1}\E_{1-i}[\pi_i]\right]\le\E[\rho_{i-1}],
\end{equation}
since
\begin{equation}
J^{\alpha}_{i-1}\le\frac{\rho_{i-1}}{\E_{i-1}[\pi_i]}.
\end{equation}
Thus we see for all $i\ge 1$ that if $\E[\rho_{i-1}]<\infty$
then $\E[\rho_i]<\infty$. But $\rho_0<\infty$ by construction,
and hence by induction we deduce that $\E[\rho_i]<\infty$ for all $i\ge 0$.\hfill$\Box$\\

The martingale $\{\rho_i\}$ is the likelihood ratio process
appropriate for a change of measure from the objective measure
$\mathbb{P}$ to the equivalent martingale measure $\mathbb{Q}$
characterised by the property that non-dividend-paying assets are
martingales when expressed in units of the money-market account.
An interesting feature of Proposition \ref{prop7.6} is that no
integrability condition is required on $\{\rho_i\}$. In other
words, the natural previsible money market account defined by
(\ref{3.28}) ``automatically" satisfies the conditions of Axiom A.
For some purposes it may therefore be advantageous to incorporate
the existence of the natural money market account directly into
the axioms. Then instead of Axiom B we would have:

\vspace{0.4cm} \noindent {\bf Axiom~B$^{\ast}$}. {{\it There
exists a strictly-positive non-dividend paying asset, the
money-market account, with value process $\{B_i\}_{i\ge 0}$,
having the properties that $B_{i+1}>B_i$ for all $i\in\N_0$ and
that $B_i$ is $\F_{i-1}$-measurable for all $i\in\N$. We assume
that $\{B_i\}$ is unbounded in the sense that for any
$b\in\mathbb{R}$ there exists a time $t_i$ such that $B_i>b$.}}
\vspace{0.2cm}

The content of Proposition \ref{prop7.6} is that Axioms A and B
together imply Axiom B$^{\ast}$. As an exercise we shall establish
that the class of interest rate models satisfying Axioms ${\text
A}$ and ${\text B^{\ast}}$ is non-vacuus. In particular, suppose
we consider the ``rational" models defined by equations
(\ref{3.17}) and (\ref{3.18}) for some choice of the martingale
$\{N_i\}$. It is straightforward to see that the unique previsible
money market account in this model is given by $B_0=1$ and
\begin{equation}
B_i=\prod^i_{n=1}\frac{\alpha_{n-1}+\beta_{n-1}
N_{n-1}}{\alpha_n+\beta_n N_{n-1}}
\end{equation}
for $i\ge 1$. For $\{\rho_i\}$ we then have
\begin{equation}
\rho_i=\rho_0\prod^{i}_{n=1}\frac{\alpha_n+\beta_n
N_n}{\alpha_n+\beta_n N_{n-1}},
\end{equation}
where $\rho_0=\alpha_0+\beta_0 N_0$. But it is easy to check that
for each $i\ge 0$ the random variable $\rho_i$ is bounded;
therefore $\{\rho_i\}$ is a martingale, and the money market
account process $\{B_i\}$ satisfies the conditions of Axioms A and
B$^{\ast}$.

Now let us return to the Doob decomposition for $\{\pi_i\}$ given in formula (\ref{DoobPI}). Evidently, we have $\pi_i=\E_i[A_{\infty}]-A_i$, with
\begin{eqnarray}
A_i&=&\sum^{i-1}_{n=0}\left(\pi_n-\E_n[\pi_{n+1}]\right)\nn\\
    &=&\sum^{i-1}_{n=0}\pi_n\left(1-\frac{\E_n[\pi_{n+1}]}{\pi_n}\right)\nn\\
    &=&\sum^{i-1}_{n=0}\pi_n\left(1-P_{n,n+1}\right)\nn\\
    &=&\sum^{i-1}_{n=0}\pi_n r_{n+1}P_{n,n+1},
\end{eqnarray}
where $\{r_i\}$ is the previsible short rate process defined by
(\ref{3.24}). The pricing kernel can therefore be put in the form
\begin{equation}\label{modPK}
\pi_i=\E_i\left[\sum^{\infty}_{n=i}\pi_n r_{n+1}P_{n,n+1}\right].
\end{equation}
Comparing the Doob decomposition (\ref{modPK}) with the
alternative decomposition given by (\ref{3.7}), we thus deduce
that if we set
\begin{equation}
\bar{r}_i=\frac{r_i\pi_{i-1}P_{i-1,i}}{\pi_i}
\end{equation}
then we obtain a positive-return asset for which the corresponding
decomposition of the pricing kernel, as given by (\ref{3.7}), is
the Doob decomposition. On the other hand, since the money-market
account is a positive-return asset, by Proposition \ref{prop7.3}
we can also write
\begin{equation}
\pi_i=\E_i\left[\sum^{\infty}_{n=i+1}\pi_n r_n\right].
\end{equation}
As a consequence, we see that the price process of a pure income
asset can be written in the symmetrical form
\begin{equation}
S_i=\frac{\E_i\left[\sum^{\infty}_{n=i+1}\pi_n
D_n\right]}{\E_i\left[\sum^{\infty}_{n=i+1}\pi_n r_n\right]},
\end{equation}
where $\{D_n\}$ is the dividend process, and $\{r_n\}$ is the
short rate process.
\section{Information-based interest rate models}\label{sec7.5}
So far in the discussion we have regarded the pricing kernel
$\{\pi_i\}$ and the filtration $\{\F_i\}$ as being subject to an
exogenous specification. In order to develop the framework further
we need to make a more specific indication of how the pricing
kernel might be determined, and how information is made available
to market participants. To obtain a realistic model for
$\{\pi_i\}$ we need to develop the model in conjunction with a
theory of consumption, money supply, price level, inflation, real
interest rates, and information. We shall proceed in two steps.
First we consider a general ``reduced-form" model for nominal
interest rates, in which we model the filtration explicitly; then
in the next section we consider a more general ``structural''
model in which both the nominal and the real interest rate systems
are characterised.

Our reduced-form model for interest rates will be based on the
theory of $X$-factors, following the general line of the previous
chapters. Associated with each time $t_i$ we introduce a
collection of one or more random variables $X^{\alpha}_i$
$(\alpha=1,\ldots, m_i)$, where $m_i$ denotes the number of random
variables associated with time $t_i$. For each value of $n$, we
assume that the various random variables $X^{\alpha}_1,
X^{\alpha}_2,\ldots,X^{\alpha}_n$ are independent. We regard the
random variables $X^{\alpha}_n$ as being ``revealed" at time
$t_n$, and hence $\F_n$-measurable. More precisely, we shall
construct the filtration $\{\F_i\}$ in such a way that this
property holds. Intuitively, we can think of $X^{\alpha}_1,
X^{\alpha}_2,\ldots,X^{\alpha}_n$ as being the various independent
macroeconomic ``market factors" that determine cash flows at time
$t_n$.

Now let us consider how the filtration will be modelled. For each
$j\in\N_0$, at any time $t_i$ before $t_j$ only partial
information about the market factors $X^{\alpha}_j$ will be
available to market participants. We model this partial
information for each market factor $X^{\alpha}_j$ by defining a
discrete-time information process $\{\xi^{\alpha}_{t_i
t_j}\}_{0\le t_i\le t_j}$, setting
\begin{equation}
\xi^{\alpha}_{t_i t_j}=\sigma t_i X^{\alpha}_{j}+\beta^{\alpha}_{t_i t_j}.
\end{equation}
Here $\{\beta^{\alpha}_{t_i t_j}\}_{0\le t_i\le t_j}$ can, for
each value of $\alpha$, be thought of as an independent
discretised Brownian bridge. Thus, we consider a standard Brownian
motion starting at time zero and ending at time $t_j$, and sample
its values at the times $\{t_i\}_{i=0,\ldots,j}$. Let us write
$\xi^{\alpha}_{ij}=\xi^{\alpha}_{t_i t_j}$ and
$\beta^{\alpha}_{ij}=\beta^{\alpha}_{t_i t_j}$, in keeping with
our usual shorthand conventions for discrete-time modelling. Then
for each value of $\alpha$ we have $\E[\beta^{\alpha}_{ij}]=0$ and
\begin{equation}
\textrm{Cov}[\beta^{\alpha}_{ik},\beta^{\alpha}_{jk}]=\frac{t_i(t_k-t_{j})}{t_k}
\end{equation}
for $i\le j\le k$. We assume that the bridge processes are
independent of the $X$-factors (i.e., the macroeconomic factors);
and hence that the various information processes are independent
of one another. Finally, we assume that the market filtration is
generated collectively by the various information processes. For
each value of $k$ the sigma-algebra $\F_k$ is generated by the
random variables $\{\xi^{\alpha}_{ij}\}_{0\le i\le j\le k}$.

Thus, as in the earlier chapters, the filtration is not simply
``given", but rather is modelled explicitly. It is a
straightforward exercise to verify that, for each value of
$\alpha$, the process $\{\xi^{\alpha}_{ij}\}$ has the Markov
property. The proof follows the pattern of the continuous-time
argument. This has the implication that the conditional
expectation of a function of the market factors $X^{\alpha}_j$,
taken with respect to $\F_i$, can be reduced to a conditional
expectation with respect to the sigma-algebra
$\sigma(\xi^{\alpha}_{ij})$. That is to say, the history of the
process $\{\xi^{\alpha}_{nj}\}_{n=0,1,\ldots,i}$ can be neglected,
and only the most ``recent" information, $\xi^{\alpha}_{ij}$,
needs to be considered in taking the conditional expectation.

For example, in the case of a function of a single
$\F_j$-measurable market factor $X_j$, with the associated
information process $\{\xi_{nj}\}_{n=0,1,\ldots,j}$, we obtain:
\begin{equation}
\E[f(X_j)|\F_i]=\frac{\int^{\infty}_0
p(x)f(x)\exp\left[\frac{t_j}{t_j-t_i}\left(\sigma
x\xi_{ij}-\frac{1}{2}\sigma^2 x^2 t_i\right)\right]\rd x}
{\int^{\infty}_0 p(x)\exp\left[\frac{t_j}{t_j-t_i}\left(\sigma
x\xi_{ij}-\frac{1}{2}\sigma^2 x^2 t_i\right)\right]\rd x},
\end{equation}
for $i\le j$, where $p(x)$ denotes the {\it a priori} probability
density function for the random variable $X_j$.

In the formula above we have for convenience presented the result
in the case of a single $X$-factor represented by a continuous
random variable taking non-negative values; the extension to other
classes of random variables, and to collections of random
variables, is straightforward.

Now we are in a position to state how we propose to model the
pricing kernel. First, we shall assume that $\{\pi_i\}$ is adapted
to the market filtration $\{\F_i\}$. This is clearly a natural
assumption from an economic point of view, and is necessary for
the general consistency of the theory. This means that the random
variable $\pi_j$, for any fixed value of $j$, can be expressed as
a function of the totality of the available market information at
time $j$. In other words, $\pi_j$ is a function of the values
taken, between times $0$ and $j$, of the information processes
associated with the various market factors.

Next we make the simplifying assumption that $\pi_j$ (for any
fixed $j$) depends on the values of only a finite number of
information processes. This corresponds to the intuitive idea that
when we are pricing a contingent claim, there is a limit to the
amount of information we can consider.

But this implies that expectations of the form $\E_i[\pi_j]$, for
$i\le j$, can be computed explicitly. The point is that since
$\pi_j$ can be expressed as a function of a collection of
intertemporal information variables, the relevant conditional
expectations can be worked out in closed form by use of the
methods of Chapter 6. As a consequence, we are led to a system of
essentially tractable expressions for the resulting discount bond
prices and the previsible money market account. Thus we are left
only with the question of what is the correct functional form for
$\{\pi_i\}$, given the relevant market factors. If we simply
``propose" or ``guess" a form for $\{\pi_i\}$, then we have a
``reduced-form" or ``ad hoc" model. If we provide an economic
argument that leads to a specific form for $\{\pi_i\}$, then we
say that we have a ``structural" model.
\section{Models for inflation and index-linked securities}\label{sec7.6}
For a more complete picture we must regard the nominal interest
rate system as embedded in a larger system that takes into account
the various macroeconomic factors that inter-relate the money
supply, aggregate consumption, and the price level. We shall
present a simple model in this spirit that is consistent with the
information-based approach that we have been taking.

To this end we introduce the following quantities. We envisage a
closed economy with aggregate consumption $\{k_i\}_{i\ge 1}$. This
consumption takes place at discrete times, and $k_i$ denotes the
aggregate level of consumption, in units of goods and services,
taking place at time $t_i$. Let us write $\{M_i\}_{i\ge 0}$ for
the process corresponding to the nominal money supply, and
$\{C_i\}_{i\ge 0}$ for the process of the consumer price index
(the ``price level"). For convenience we can think of $\{k_i\}$
and $\{M_i\}$ both as being expressed on a {\it per capita} basis.
Hence these quantities can be regarded, respectively, as the
consumption and money balance associated with a representative
agent. We can therefore formulate the optimisation problem from
the perspective of the representative agent; but the role of the
agent here is to characterise the structure of the economy as a
whole.

We shall assume that at each time $t_i$ the agent receives a
benefit or service from the money balance maintained in the
economy; this will be given in nominal terms by $\lambda_i M_i$,
where $\lambda_i$ is the nominal liquidity benefit conferred to
the agent per unit of money ``carried" by the agent, and $M_i$ is
the money supply, expressed on a {\it per capita} basis, at that
time. The corresponding ``real" benefit (in units of goods and
services) provided by the money supply at time $t_i$ is defined by
the quantity
\begin{equation}
l_i=\frac{\lambda_i M_i}{C_i}.
\end{equation}
It follows from these definitions that we can think of
$\{\lambda_i\}$ as a kind of ``convenience yield" process
associated with the money supply. Rather in the way a country will
obtain a convenience yield (per barrel) from its oil reserves,
which can be expressed on a {\it per capita} basis, likewise an
economy derives a convenience yield (per unit of money) from its
money supply.

It is important to note that what matters in reality is the {\it
real benefit} of the money supply, which can be thought of
effectively as a flow of goods and services emanating from the
presence of the money supply. It is quite possible that the
``wealth" attributable to the face value of the money may in
totality be insignificant. For example, if the money supply
consists exclusively of notes issued by the government, and hence
takes the form of government debt, then the {\it per capita}
wealth associated with the face value of the notes is essentially
null, since the representative agent is also responsible
(ultimately) for a share of the government debt. Nevertheless, the
presence of the money supply confers an overall positive flow of
benefit to the agent. On the other hand, if the money supply
consists, say, of gold coins, or units of some other valuable
commodity, then the face value of the money supply will make a
positive contribution to overall wealth, as well as providing a
liquidity benefit.

Our goal is to obtain a consistent structural model for the
pricing kernel $\{\pi_i\}_{i\ge 0}$. We assume that the
representative agent gets utility both from consumption and from
the real benefit of the money supply in the spirit of Sidrauski 1969.
Let $U(x,y)$ be a standard
bivariate utility function $U:\R^+\times\R^+\rightarrow\R$,
satisfying $U_x>0$, $U_y>0$, $U_{xx}<0$, $U_{yy}<0$, and
$U_{xx}U_{yy}>(U_{xy})^2$. Then the objective of the
representative agent is to maximise an expression of the form
\begin{equation}
J=\E\left[\sum^{N}_{n=0}\e^{-\gamma t_n}U(k_n,l_n)\right]
\end{equation}
over the time horizon $[t_0,t_1,\ldots,t_N]$, where $\gamma$ is
the appropriate discount rate applicable to delayed gains in
utility. For simplicity of exposition we assume a constant
discount rate. The optimisation problem faced by the agent is
subject to the budget constraint
\begin{equation}
W=\E\left[\sum^N_{n=0}\pi_n(C_n k_n+\lambda_n M_n)\right].
\end{equation}
Here $W$ represent the total {\it per capita} wealth, in nominal
terms, available for consumption related expenditure over the
given time horizon. The agent can maintain a position in money,
and ``consume" the benefit of the money; or the money position can
be liquidated (in part, or in whole) to purchase consumption
goods. In any case, we must include the value of the benefit of
the money supply in the budget for the relevant period. In other
words, since the presence of the money supply ``adds value", we
need to recognise this value as a constituent of the budget. The
budget includes also any net initial funds available, together
with the value of any expected income (e.g., derivable from labour
or natural resources) over the relevant period.

The fact that the utility depends on the real benefit of the money
supply, whereas the budget depends on the nominal value of the
money supply, leads to a fundamental relationship between the
processes $\{k_i\}$, $\{M_i\}$, $\{C_i\}$, and $\{\lambda_i\}$.
Introducing a Lagrange multiplier $\mu$, after some re-arrangement
we obtain the associated unconstrained optimisation problem, for
which the objective is to maximise the following expression:
\begin{equation}
\E\left[\sum^N_{n=0}\e^{-\gamma t_n}U(k_n,l_n)-\mu\sum^N_{n=0}\pi_n C_n(k_n+l_n)\right].
\end{equation}
A straightforward argument then shows that the solution for the
optimal policy (if it exists) satisfies the first order conditions
\begin{equation}
U_x(k_n,l_n)=\mu\e^{\gamma t_n}\pi_n C_n\label{U1},
\end{equation}
and
\begin{equation}
U_y(k_n,l_n)=\mu\e^{\gamma t_n}\pi_n C_n\label{U2},
\end{equation}
for each value of $n$ in the relevant time frame, where $\mu$ is
determined by the budget constraint. As a consequence we obtain
the fundamental relation
\begin{equation}\label{fundrelation}
U_x\left(k_n,\lambda_n M_n/C_n\right)=U_y(k_n,\lambda_n M_n/C_n),
\end{equation}
which allows us to eliminate any one of the variables $k_n$,
$M_n$, $\lambda_n$, and $C_n$ in terms of the other three. In this
way, for a given level of consumption, money supply, and liquidity
benefit, we can work out the associated price level. Then by use
of (\ref{U1}), or equivalently (\ref{U2}), we can deduce the form
taken by the nominal pricing kernel, and hence the corresponding
interest rate system. We also obtain thereby an expression for the
``real" pricing kernel $\{\pi_i C_i\}$.

We shall take the view that aggregate consumption, the liquidity
benefit rate, and the money supply level are all determined
exogenously. In particular, in the information-based framework we
take these processes to be adapted to the market filtration, and
hence determined, at any given time, by the values of the
information variables upon which they depend. The theory outlined
above then shows how the values of the real and nominal pricing
kernels can be obtained, at each time, as functions of the
relevant information variables.

It will be useful to have an explicit example in mind, so let us
consider a standard ``log-separable" utility function of the form
\begin{equation}\label{logU}
U(x,y)=A\ln(x)+B\ln(y),
\end{equation}
where $A$ and $B$ are non-negative constants. From the fundamental
relation (\ref{fundrelation}) we immediately obtain
\begin{equation}
\frac{A}{k_n}=\frac{B}{l_n},
\end{equation}
and hence the equality
\begin{equation}
k_n C_n=\frac{A}{B}\lambda_{n}M_n.
\end{equation}
Thus, in the case of log-separable utility we see that the level
of consumption, in nominal terms, is always given by a fixed
proportion of the nominal liquidity benefit obtained from the
money supply. For any fixed values of $\lambda_n$ and $k_n$, we
note, for example, that an increase in the money supply leads to
an increase in the price level.

One observes that in the present framework we {\it derive} an
expression for the consumer price index process. This contrasts
somewhat with current well-known methodologies for pricing
inflation-linked securities---see, e.g., Hughston 1998, and Jarrow
\& Yildirim 2003---where the form of the consumer price index is
specified on an exogenous, essentially {\it ad hoc} basis.

The quantity $k_n C_n/M_n$ is commonly referred to as the
``velocity" of money. It measures, roughly speaking, the rate at
which money changes hands, as a percentage of the total money
supply, as a consequence of consumption. Evidently, in the case of
a log-separable utility (\ref{logU}), the velocity has a fixed
ratio to the liquidity benefit. This is a satisfying conclusion,
which shows that even with a relatively simple assumption about
the nature of the utility we are able to obtain an intuitively
natural relation between the velocity of money and the liquidity
benefit. In particular, if liquidity is increased, then a lower
money supply will be required to sustain a given level of nominal
consumption, and hence the velocity will be increased as well. The
situation when the velocity is constant leads to the so-called
``quantity" theory of money, which in the present approach arises
in the case of a representative agent with log-separable utility
and a constant liquidity benefit.

It is interesting to note that the results mentioned so far, in
connection with log-separable utility, are not too sensitive to
the choice of the discount rate $\gamma$, which does not enter
into the fundamental relation (\ref{fundrelation}). On the other
hand, $\gamma$ does enter into the expression for the nominal
pricing kernel; in particular, in the log-separable case we obtain
the following expression for the pricing kernel:
\begin{equation}
\pi_n=\frac{B\e^{-\gamma t_n}}{\mu\lambda_n M_n}.
\end{equation}
Hence, in the log-separable utility theory we can see explicitly
the relation between the nominal money supply and the term
structure of interest rates.

Let us consider now a contingent claim with the random nominal
payoff $H_j$ at time $t_j$. Then the value of the claim at time
$t_0$ in the log-separable utility model is given by the following
formula:
\begin{equation}
H_0=\lambda_0 M_0\e^{-\gamma t_j}\E\left[\frac{H_j}{\lambda_j M_j}\right].
\end{equation}
One can evidently see two different influences on the value of
$H_0$. First one has the discount factor; but equally importantly
one sees the effect of the money supply. For a given level of the
liquidity (i.e., for constant $\lambda_j$), an increase in the
likely money supply at time $t_j$ will reduce the value of $H_0$.
This example illustrates how market perceptions of the direction
of future monetary policy can potentially affect the valuation of
contingent claims in a fundamental way. In particular, the value
of the money supply $M_j$ at time $t_j$ will be given as a
function of the best available information at that time concerning
future random factors affecting the economy. The question of how
best to model the money supply process $\{M_i\}$ takes us, to some
extent, outside of the realm of pure mathematical finance, and
more into the territory of macroeconomics and, ultimately,
political economics. Nevertheless, it is gratifying and perhaps
surprising that we can have come as far as we have.

\newpage
\addcontentsline{toc}{chapter}{\protect
    \numberline{}{References}}

\pagestyle{fancy} \fancyhf{}
\renewcommand{\chaptermark}[1]{\markboth{}}
\renewcommand{\sectionmark}[1]{\markright{}}
\fancyhead[RO]{\sffamily\small \thepage}
\fancyhead[LO]{\sffamily\small References}
\fancyhead[LE]{\sffamily\small \thepage}
\fancyhead[RE]{\sffamily\small \leftmark}
\fancypagestyle{plain}{\fancyhf{}\renewcommand{\headrulewidth}{0pt}\renewcommand{\footrulewidth}{0pt}}

\mbox{}\newline {\LARGE {\bf References}}
\\
\begin{description}
%

\bibitem[1]{back} K.~Back (1992) Insider trading in continuous time,
{\em Review of Financial Studies} \textbf{5}, 387-407.

\bibitem[2]{back2} K.~Back \& S.~Baruch (2004) Information in securities
markets: Kyle meets Glosten and Milgrom, {\em Econometrica}
\textbf{72}, 433-465.

\bibitem[3]{balltorous} C.~A.~Ball \& W.~N.~Torous (1983) Bond price dynamics and options, {\em Journal of Financial and Quantitative Analysis}
\textbf{18}, 517-531.

\bibitem[4]{baxter-rennie} M.~W.~Baxter \& A.~J.~O.~Rennie (1996)
{\em Financial calculus: an introduction to derivative pricing}.
Cambridge University Press, Cambridge.


\bibitem[5]{bielecki1}T.~R.~Bielecki, M.~Jeanblanc \&
M.~Rutkowski (2004) Modelling and valuation of credit risk,
Bressanone Lectures 2003, in {\em Stochastic Methods in Finance},
edited by M.~Fritelli \& W.~Runggaldier. Springer, Berlin.

\bibitem[6]{bielecki} T.~R.~Bielecki \& M.~Rutkowski (2002)
{\em Credit risk: modelling, valuation and hedging}. Berlin,
Springer.

\bibitem[7]{bjork} T.~Bj\"ork (2004)
{\em Arbitrage theory in continuous time}, second edition. Oxford
University Press, Oxford.

\bibitem[8]{black} F.~Black \& J.~C.~Cox (1976) Valuing corporate securities: some effects of bond indenture provisions, {\em Journal of Finance} \textbf{31}, 351-367.


\bibitem[9]{bhm2} D.~C.~Brody, L.~P.~Hughston \& A.~Macrina (2006)
Information-based asset pricing. Imperial College London and
King's College London working paper. \\
Downloadable at:
www.mth.kcl.ac.uk/research/finmath/publications.html.

\bibitem[10]{bhm1} D.~C.~Brody, L.~P.~Hughston \& A.~Macrina (2007) Beyond
hazard rates: a new framework for credit-risk modelling, in {\em
Advances in Mathematical Finance, Festschrift volume in honour of
Dilip Madan}, edited by R.~Elliott, M.~Fu, R.~Jarrow \& J.-Y.~Yen.
Birkh\"auser Basel \& Springer, Berlin. Downloadable at:\\
www.defaultrisk.com.

\bibitem[11]{bj} R.~S.~Bucy \& P.~D.~Joseph (1968) {\em Filtering for
stochastic processes with applications to guidance}. Interscience
Publishers, New York.

\bibitem[12]{CGM} P.~Carr, H.~Geman \& D.~Madan (2001) Pricing and hedging in incomplete markets,
{\em Journal of Financial Economics} \textbf{62}, 131-167.

\bibitem[13]{cetin} U.~Cetin, R.~Jarrow, P.~Protter \& Y.~Yildirim (2004)
Modelling Credit Risk with Partial Information, {\em Annals of
Applied Probability} \textbf{14}, 1167-1172.

\bibitem[14]{chen-filipovic} L.~Chen \& D.~Filipovi\'c (2005)
A simple model for credit migration and spread curves, {\em
Finance and Stochastics} \textbf{9}, 211-231.

\bibitem[15]{cochrane} J.~H.~Cochrane (2005) {\em Asset pricing}. Princeton University
Press, Princeton.

\bibitem[16]{davis1} M.~H.~A.~Davis (2004) Complete-market models of
stochastic volatility, {\em Proceedings of the Royal Society
London} A\textbf{460}, 11-26.

\bibitem[17]{davis-lo} M.~H.~A.~Davis \& V.~Lo (2001) Infectious defaults,
{\em Quantitative Finance} \textbf{1}, 382-387.

\bibitem[18]{davis2} M.~H.~A.~Davis \& S.~I.~Marcus (1981) An introduction
to nonlinear filtering, in {\em Stochastic systems: The
mathematics of filtering and identification and application},
edited by M.~Hazewinkel \& J.~C.~Willems. D.~Reidel, Dordrecht.

\bibitem[19]{duffie0} D.~Duffie (2001)
{\em Dynamic Asset Pricing Theory}, third edition. Princeton
University Press, Princeton.

\bibitem[20]{duffie1} D.~Duffie \& N.~Garleanu (2001) Risk and
valuation of collateralised debt obligations, {\em Financial
Analyst's Journal} \textbf{57}, 41-59.

\bibitem[21]{duffie2} D.~Duffie \& D.~Lando (2001) Term
structure of credit spreads with incomplete accounting
information, {\em Econometrica} \textbf{69}, 633-664.

\bibitem[22]{duffie4} D.~Duffie, M.~Schroder \& C.~Skidas (1996)
Recursive valuation of defaultable securities and the timing of
resolution of uncertainty, {\em Annals of Applied Probability}
\textbf{6}, 1075-1090.

\bibitem[23]{duffie5} D.~Duffie \& K.~J.~Singleton (1999) Modelling
term structures of defaultable bonds, {\em Review of Financial
Studies} \textbf{12}, 687-720.

\bibitem[24]{duffie6} D.~Duffie \& K.~J.~Singleton (2003)
{\em Credit risk: pricing, measurement and management}. Princeton
University Press, Princeton.

\bibitem[25]{elizalde1} A.~Elizalde (2003) Credit risk models: I
Default Correlation in Intensity Models, II Structural Models, III
Reconciliation Structural--Reduced Models, IV Understanding and
pricing CDOs. CEMFI \& UPNA working paper, downloadable at: \\
www.abelelizalde.com.

\bibitem[26]{flesaker} B.~Flesaker, L.~P.~Hughston, L.~Schreiber \&
L.~Sprung (1994) Taking all the credit, {\em Risk} \textbf{7},
104-108.

\bibitem[27]{flesaker_hughston} B.~Flesaker \& L.~P.~Hughston (1996) Positive interest, {\em Risk} \textbf{9}, 46-49.

\bibitem[28]{frey} R.~Frey \& A.~J.~McNeil (2003) Modelling dependent
defaults, {\em Journal of Risk} \textbf{6}, 59-92.

\bibitem[29]{filipovic} D.~Filipovi\'c \& J.~ Zabczyk (2002) Markovian term structure  models in discrete time, {\em Annals of Applied Probability} \textbf{12},
710-729.

\bibitem[30]{FWY} H.~F\"ollmer, C.~T.~Wu \& M.~Yor (1999) Canonical decomposition of linear transformations of two independent Brownian motions motivated by models of insider trading, {\em Stochastic Processes and their Applications}
\textbf{84}, 137-164.

\bibitem[31]{FKK} M.~Fujisaki, G.~Kallianpur \& H.~Kunita (1972) Stochastic differential equations for the nonlinear filtering problem, {\em Osaka Journal of Mathematics} \textbf{9}, 19-40.

\bibitem[32]{GKR} H.~Geman, N.~E.~Karoui \& J.~C.~Rochet (1995) Changes of numeraire, changes of probability measure and option pricing, {\em Journal of Applied Probability}
\textbf{32}, 443-458.

\bibitem[33]{giesecke1} K.~Giesecke (2004) Correlated default with
incomplete information, {\em Journal of Banking and Finance}
\textbf{28}, 1521-1545.

\bibitem[34]{giesecke2} K.~Giesecke \& L.~R.~Goldberg (2004)
Sequential default and incomplete information {\em Journal of
Risk} \textbf{7}, 1-26.

\bibitem[35]{gih} I.~I.~Gihman \& A.~V.~Skorohod (1979) {\em The theory of stochastic processes III}. Springer, Berlin.

\bibitem[36]{guo} X.~Guo, R.~A.~Jarrow \& Y.~Zeng (2005) Information
reduction in credit risk modelling. Cornell working paper.

\bibitem[37]{hjm} D.~Heath, R.~Jarrow \& A.~Morton (1990) Bond pricing and the term structure of interest rates: a discrete time approximation, {\em Journal of Finance and Quantitative Analysis} \textbf{25}, 419-440.

\bibitem[38]{heston} S.~L.~Heston (1993) A closed-form solution for
options with stochastic volatility with applications to bond and
currency options, {\em Review of Financial Studies} \textbf{6},
327-343.

\bibitem[39]{hilberink} B.~Hilberink \& C.~Rogers (2002) Optimal
capital structure and endogenous default {\em Finance and
Stochastics} \textbf{6}, 237-263.

\bibitem[40]{holee} T.~S.~Y.~Ho \& S.~B.~Lee (1986) Term-structure movements and pricing interest rate contingent claims, {\em Journal of Finance and Quantitative Analysis} \textbf{41}, 1011-1029.

\bibitem[41]{hughston1} L.~P.~Hughston (1998) Inflation derivatives. Merrill Lynch and King's College London report, with added note (2004), downloadable at: \\ www.mth.ac.uk/research/finmath/publications.html.

\bibitem[42]{hughston-macrina} L.~P.~Hughston \& A.~Macrina (2006)
Information, inflation, and interest. King's College London
working paper, downloadable at: \\
www.mth.kcl.ac.uk/research/finmath/publications.html.

\bibitem[43]{hughston_rafailidis} L.~P.~Hughston \& A.~Rafailidis  (2005) A chaotic approach to interest rate modelling, {\em Finance and Stochastics} \textbf{9}, 43-65.

\bibitem[44]{hughston} L.~P.~Hughston \& S.~M.~Turnbull (2000) Credit
derivatives made simple, {\em Journal of Risk} \textbf{13}, 36-43.

\bibitem[45]{hughston2} L.~P.~Hughston \& S.~M.~Turnbull (2001) Credit
risk: constructing the basic building blocks {\em Economic Notes}
\textbf{30}, 281-292.

\bibitem[46]{hull1} J.~Hull \& A.~White (2004a) Valuation of a CDO
and an $n^{\rm th}$ to default CDS without Monte Carlo simulation,
{\em Journal of Derivatives} \textbf{12}, 8-23.

\bibitem[47]{hull3} J.~Hull \& A.~White (2004b) Merton's model,
credit risk and volatility skews, {\em Journal of Credit Risk}
\textbf{1}, 1-27.

\bibitem[48]{hunt_kennedy} P.~J.~Hunt \& J.~E. Kennedy (2004) {\em Financial derivatives in theory and practice}, revised edition. Wiley, Chichester.

\bibitem[49]{jarrow1} R.~A.~Jarrow, D.~Lando \& S.~M.~Turnbull
(1997) A Markov model for the term structure of credit risk
spreads, {\em Review of Financial Studies} \textbf{10}, 481-528.

\bibitem[50]{jarrow2} R.~A.~Jarrow \& P.~Protter (2004) Structural
versus reduced form models: a new information based perspective,
{\em Journal of Investment Management} \textbf{2}, 1-10.

\bibitem[51]{jarrow3} R.~A.~Jarrow \& S.~M.~Turnbull (1995) Pricing
derivatives on financial securities subject to credit risk, {\em
Journal of Finance} \textbf{50}, 53-85.


\bibitem[52]{jarrow5a} R.~A.~Jarrow \& Y.~Yildirim (2003) Pricing treasury inflation protected securities and related derivatives using an HJM model,
{\em Journal of Financial and Quantitative Analysis} \textbf{38},
409.

\bibitem[53]{jarrow5} R.~A.~Jarrow \& F.~Yu (2001) Counterparty
risk and the pricing of defaultable securities, {\em Journal of
Finance} \textbf{56}, 1765-1799.

\bibitem[54]{jeanblanc} M.~Jeanblanc \& M.~Rutkowski (2000)
Modelling of default risk: an overview, in {\em Mathematical
finance: theory and practice}. Higher Education Press, Beijing.

\bibitem[55]{jin_glasserman} Y.~Jin \& P.~Glasserman (2001) Equilibrium positive interest rates: a unified view, {\em Review of Financial Studies} \textbf{14}, 187-214.

\bibitem[56]{ks} G.~Kallianpur \& C.~Striebel (1968) Estimation of
stochastic systems: Arbitrary system process with additive white
noise observation errors, {\em Annals of Mathematics and
Statistcs} \textbf{39}, 785.

\bibitem[57]{karatzas-Shreve} I.~Karatzas \& S.~Shreve (1991) {\em Brownian
motion and stochastic calculus}, second edition. Springer, Berlin.

\bibitem[58]{karatzas-Shreve2} I.~Karatzas \& S.~Shreve (1998) {\em Methods of mathematical
finance}. Springer, Berlin.


\bibitem[59]{kusuoka} S.~Kusuoka (1999) A remark on default risk
models, {\em Advances in Mathematical Economics} \textbf{1},
69-82.

\bibitem[60]{lando1} D.~Lando (1994) Three essays on contingent claims pricing. Ph.D. thesis, Cornell
University.

\bibitem[61]{lando1} D.~Lando (1998) On Cox processes and credit risky
securities, {\em Review of Derivatives Research} \textbf{2},
99-120.

\bibitem[62]{lando2} D.~Lando (2004) {\em Credit risk modelling}.
Princeton University Press, Princeton.

\bibitem[63]{leland} H.~E.~Leland \& K.~B.~Toft (1996) Optimal
capital structure, endogenous bankruptcy and the term structure of
credit spreads, {\em Journal of Finance} \textbf{50}, 789-819.

\bibitem[64]{ls} R.~S.~Liptser \& A.~N.~Shiryaev (2000) {\em Statistics
of random processes}, Volumes I \& II, second edition. Springer,
Berlin.

\bibitem[65]{madan} D.~Madan \& H.~Unal (1998) Pricing the risks of
default, {\em Review of Derivatives Research} \textbf{2}, 121-160.

\bibitem[66]{merton} R.~Merton (1974) On the pricing of corporate debt:
the risk structure of interest rates, {\em Journal of Finance}
\textbf{29}, 449-470.

\bibitem[67]{meyer} P.~A.~Meyer (1966) {\em Probability and
potentials}. Blaisdell Publishing Company, Waltham, Massachusetts.

\bibitem[68]{musiela-rutkowski} M.~Musiela, M. Rutkowski \& A.
Lichnewsky (2004) {\em Martingale methods in financial modelling
(Stochastic Modelling and Applied Probability )}, Springer,
Berlin.

\bibitem[69]{ohara} M.~O'Hara (1995) {\em Market Microstructure Theory}. Blackwell Publishers, Cambridge, Massachusetts.


\bibitem[70]{protter} P.~Protter (2004) {\em Stochastic integration and
differential equations}, second edition. Springer, Berlin.

\bibitem[71]{rogers} L.~C.~G.~Rogers (1997) The potential approach to the term structure of interest rate and foreign exchange rates,
{\em Mathematical Finance} \textbf{7}, 157-176.

\bibitem[72]{rutkowski} M.~Rutkowski (1997) A note on the Flesaker-Hughston model of the term structure of interest rates,
{\em Applied Mathematical Finance} \textbf{4}, 151-163.

\bibitem[73]{rutkowski-yu} M.~Rutkowski \& N.~Yu (2005) On the
Brody-Hughston-Macrina approach to modelling of defaultable term
structure. University of New South Wales working paper, downloadable at: \\
www.maths.unsw.edu.au/statistics/pubs/statspreprints2005.html.

\bibitem[74]{schonbucher} P.~J.~Sch\"{o}nbucher (2003) {\em Credit
derivatives pricing models}. Wiley, New York.

\bibitem[75]{shiryaev} A.~N.~Shiryaev (1999) {\em Essentials of stochastic finance}. World Scientific, Singapore.

\bibitem[76]{sidrauski} M.~Sidrauski (1969) Rational choice and patterns of growth,
{\em Journal of Political Economy} \textbf{4}, 575-585.

\bibitem[77]{steele} J.~M.~Steele (2001) {\em Stochastic calculus and financial applications}. Springer, Berlin.


\bibitem[78]{williams} D.~Williams (1991) {\em Probability with martingales}. Cambridge Mathematical Textbooks, Cambridge.


\bibitem[79]{yor1} M.~Yor (1992) {\em Some aspects of Brownian
motion}, Part I. Birkh\"auser, Basel.

\bibitem[80]{yor2} M.~Yor (1996) {\em Some aspects of Brownian
motion}, Part II. Birkh\"auser, Basel.
\end{description}

\end{document}